\documentstyle[12pt,epsf,a4wide]{article} 

\newcommand{\nd}[1]{/\hspace{-0.5em} #1}
\begin{document}

\begin{titlepage} 

\begin{centering} 

\begin{flushright} 
hep-th/0111275 \\
\end{flushright} 

\vspace{0.2in} 

{\Large {\bf String Cosmology}}

\vspace{0.2in} 

{\bf Nick E. Mavromatos}

\vspace{0.2in} 

King's College London, Department of Physics, \\
Theoretical Physics, Strand, London WC2R 2LS, U.K.

\vspace{0.2in} 

{\bf Abstract}
  
\vspace{0.1in} 

\end{centering} 

{\small ``Old'' String Theory is a theory of  
one-dimensional extended objects,
 whose vibrations correspond to excitations of various target-space 
field modes
including gravity. 
It is for this reason that strings 
present the first, up to
now, mathematically consistent framework where quantum gravity is unified with
the rest of the fundamental interactions in nature.  
In these lectures I will give an 
introduction to low-energy Effective Target-Space 
Actions
derived from conformal invariance conditions of the underlying sigma models in
string theory.
In this context, I shall discuss cosmology, 
emphasizing the role of the dilaton field in
inducing inflationary scenaria and in general expanding string universes. 
Specifically, I shall 
analyse some exact solutions of string theory with a linear dilaton, and
discuss their role in inducing expanding Robertson-Walker Universes. I will
mention briefly pre-Big-Bang scenaria of String Cosmology, in which
the dilaton plays a crucial role. In view of 
recent claims on experimental
evidence (from diverse astrophysical sources) on the existence of cosmic
acceleration in the universe today, with a positive non-zero cosmological
constant (de Sitter type), I shall also  
discuss difficulties of incorporating such
Universes with eternal acceleration in  the context 
of critical string theory, and present scenaria for a graceful exit from such 
a phase.}

\vspace{0.6in} 

\begin{centering}

{\it Lectures presented at the First Aegean Summer School on Cosmology,
Karlovassi (Samos), Greece, September 21-19 2001.}

\end{centering} 

\end{titlepage}

\section{Introduction}

Our way of thinking towards an understanding of the 
fundamental forces in nature, 
as well as of the structure of matter and that of space time,
has evolved over the last decades of the previous 
century from that of using point-like structures as the basic 
constituents of matter, 
to that employing one-dimensional extended objects (strings~\cite{green}),
and, recently (from the mid 90's), higher-dimensional 
domain-wall like solitonic 
objects, called (Dirichlet) (mem)branes~\cite{membranes}.

The passage from point-like fundamental constituents to 
strings, in the mid 1980's, has already revolutionarized our view
of space time and of the unification of fundamental interactions 
in Nature, including gravity. Although in the framework of point-like 
field theories, the uncontrollable ultraviolet (short-distance)
divergencies of quantum gravity prevented the development 
of a mathematically 
consistent unifying theory of all known interactions in Nature,
the discovery of one-dimensional fundamental constituents of 
matter and space-time, called {\it strings}, which were in 
principle free from such divergenecies, opened up the way 
for a mathematically cosistent way of incorporating quantum gravity 
on an equal footing with the rest of the interactions. 
The existence of a {\it minimum} length $\ell_s$ in string theory,
in such a way that the quantum uncertainty principle
between position $X$ and momenta $P$:  $\Delta X \Delta P \ge \hbar$, 
of point-like quantum mechanics is replaced by:
$\Delta X \ge \ell_s, \quad \Delta X \Delta P \ge \hbar + {\cal O}(\ell_s^2)\Delta P^2 + \dots$, revolutionaized the way we looked at 
the structure of space time at such small scales.
The unification of gravitaional interactions with the rest
is achieved in this framework if one identifies the string scale 
$\ell_s$ with the Planck scale, $\ell_P=10^{-35}$ m, 
where gravitational interactions 
are expected to set in. The concept of space time,  
as we preceive it, breaks down beyond the string (Planck) scale,
and thus there is a fundamental short-distance cutoff built-in in the 
theory, which results in its {\it finiteness}.

The cost, however, for such an achievement, was that 
mathematical consistency implied a {\it higher-dimensional}
target space time, in which the strings propagate. 
This immediately lead the physicists to try and determine
the correct vacuum configurations of string theory 
which would result into a four-dimensional 
Universe, i.e. a Universe with four dimensions being 
``large'' compared to the 
gravitational scale, the Planck length, $10^{-35} $ m, 
with the extra dimensions compactified on Planckian size 
manifolds. 
Unfortunately such consistent ground states are not unique,
and there is a huge degeneracy among such string vacua, 
the lifitng of which is still an important unresolved 
problem in string physics. 

In the last half of the 1990's the discovery of 
{\it string dualities}, i.e. discrete stringy 
(non-perturbative) gauge symmetries linking various 
string theories, 
showed another interesting possibility, which 
could contribute
significantly towards the elimination of the 
huge degeneracy problem of the string vacua.
Namely, many string theories were found to be {\it dual}
to each other
in the sense of exhibiting invariances of their physical spectra
of excitations under the action of such discrete  
symmetries. In fact, by virtue of such dualities
one could argue that there exist a sort of {\it unification
of string theories}, in which all the known string theories
(type IIA, type IIB, 
$SO(32)/Z_2$, Heterotic $E_8 \times E_8$,
type I), together with 11 dimensional supergavity 
(living in one-dimension higher than the critical dimension of superstrings) 
can be all connected with string dualities, so that one may view
them as {\it low energy limits} of a mysterious larger theory,
termed $M$-theory~\cite{membranes}, whose precise 
dynamics is still not known.  

A crucial r\^ole in such string dualities is played by domain walls,
stringy solitons, which can be derived from ordinary strings
upon the application of such dualities. Such extended higher-dimensional
objects are also excitations of this mysterious M-theory, and they are on 
a completely equal footing with their one dimensional (stringy) 
counterparts. 

In this framework one could discuss cosmology. 
The latter is nothing other but a theory of the 
gravitational field, in which the Universe is treated as a whole.
As such, string or M-theory theory, which includes the gravitational
field 
in its spectrum of excitations, seems the appropriate 
framework for providing analyses on issues of the Early Universe
Cosmology, such as the nature of the initial singularity
(Big Bang), the inflationary phase and graceful exit from it {\it etc}, 
which conventional local field theories cannot give 
a reliable answer to.
It is the purpose of this lectures to provide a
very brief, but hopefully comprehensive discussion, 
on String Cosmology. We use the terminoloy string cosmology here
to discuss Cosmology based on one-dimensional
fundamental constituents (strings). Cosmology may also 
be discussed from the more modern point of view of 
membrane structures in M-theory, mentioned above, but this 
will not be covered in these lectures. Other lecturers
in the School will discuss this issue.

The structure of the lectures is as follows:
in the first lecture we shall introduce the layman
into the subject of string effective actions, and 
discuss how 
equations of motion of the various low-energy modes of 
strings are associated with fundamental 
consistency properties (conformal invariance) 
of the underlying string theory. 
In the second lecture we shall discuss various scenaria
for String Cosmology, together with their physical
consequences. Specifically I will discuss how expanding 
and inflationary (de Sitter) Universes are incorporated
in string theory, with emphasis on describing new 
fatures, not characterizing conventional point-like
cosmologies. 
Finally, in the third lecture we shall
speculate on ways of 
providing possible resolution to 
various theoretical challenges for string theory,
especially in view of recent astrophysical evidence
of a current-era acceleration of our Universe. 
In this respect we shall discuss the application 
of the so-called {\it non-critical} (Liouville) 
string theory to cosmology, as a way of going 
{\it off equilibrium} in a string-theory setting,
in analogy with the use of non-equilibrium field 
theories in conventional point-like cosmological 
field theories of the Early Universe.

\section{Lecture 1: Introduction to String Effective Actions} 

\subsection{World-sheet String formalism}

\begin{figure}[htb]
\epsfxsize=4in
\begin{center}
\epsffile{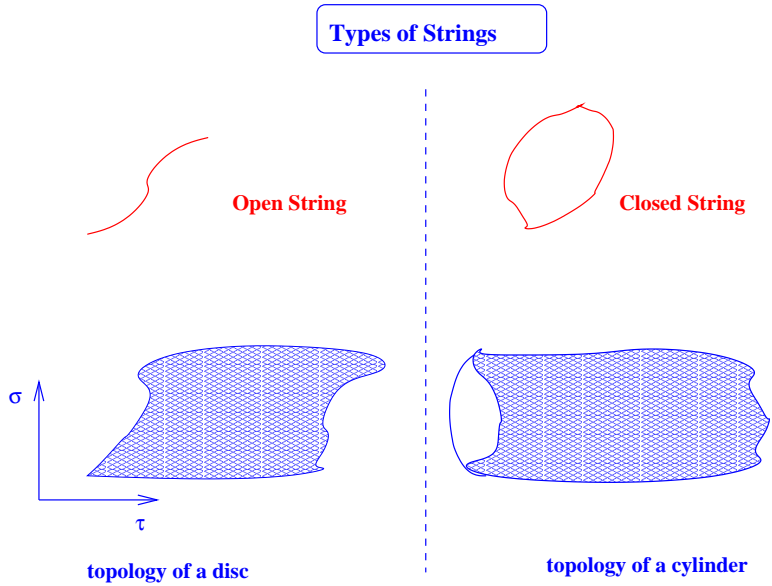}
\end{center} 
\caption[]{Types of strings and the associated world-sheets
swept as the string propagates through a (higher-dimensional) 
target space time. In the closed-string case, which incorporates
gravity, the point-like low-energy field theory limit 
is obtained by shrinking the size of the external strings
(at the tips of the cylinder) to zero, thereby obtaining 
the topology of a punctured sphere} 
\label{fig:stringtypes}
\end{figure}

\begin{figure}[htb]
\epsfxsize=4in 
\begin{center}
\epsffile{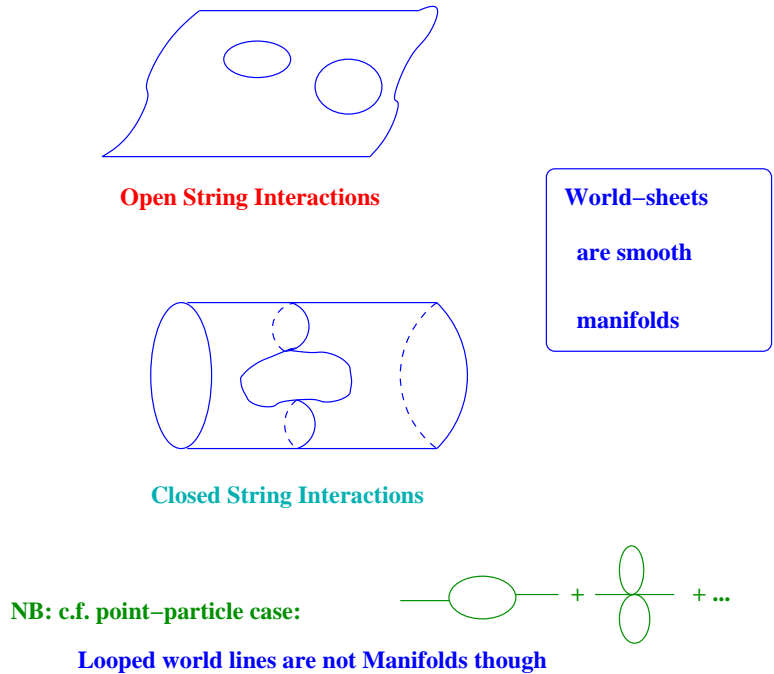}
\end{center} 
\caption[]{Quantum String Interactions are represented
by higher-topologies on the associated world-sheets. 
The two-dimensional nature of the string world-sheet,
which makes it a smooth manifold, should be contrasted 
with the point-particle world-line case, where loops
are not manifolds.}
\label{fig:genus}
\end{figure}

In this lectures the terminology ``string theory'' will be restricted to 
the ``old'' concept of {\it one (spatial) dimensional extended objects},
propagating in target-space times of dimensions higher than four,
specifically 26 for Bosonic strings and 10 for Super(symmetric)strings.
There are in general two types of such objects, as illustrated 
in a self-explanatory way in figure \ref{fig:stringtypes}: {\it open strings}
and {\it closed strings}. In the first quantized formalism, one 
is interested in the propagation of such extended objects in 
a background space time. By direct extension of the concept of
a point-particle, the motion of a string as it glides through spacetime
is described by the {\it world sheet}, a two dimensional Riemann surface
which is swept by the extended object during its propagation 
through spacetime. The world-sheet is a direct extension of the 
concept of the {\it world line} in the case of a point particle. 
The important formal difference of the string case, as compared with the 
particle one, is the fact that {\it quantum corrections}, i.e. string loops,
are incorporated in a smooth and straightforward manner 
in the case of string theory by means of summing over Riemann
surfaces with non-trivial  {\it topologies} (``genus'') (c.f. figure \ref{fig:genus}). This is allowed because in two (world-sheet) dimensions 
one is allowed to discuss loop corrections in a way compatible 
with a (two-dimensional) smooth manifold concept, in contrast to the 
point-particle one-dimensional case, where a loop correction 
on the world-line (c.f. figure \ref{fig:genus}) 
cannot be described in a smooth way, given that a particle loop 
does not constitute a manifold. 
The `somooth-manifold' property of quantum fluctuating world sheets 
is essential in analysing target-space 
quantum corrections within a first-quantization framework,
which cannot be done in the one-dimensional particle case. 
Specifically, as we shall discuss later on, 
by considering the propagation of a stringy extended object
in a curved target space time manifold of 
higher-dimensionality (26 for Bosoninc or 10 for Superstrings),
one will be able of arriving at consistency conditions on the 
background geometry, which are, in turn, interpreted as 
equations of motion derived from an effective low-energy 
action constituting the 
local field theory limit of strings. Summation over genera
will describe quantum fluctuations about classical ground states
of the strings described by world-sheet with the topology of the 
sphere (for closed strings) or disc (for open strings).

To begin our discussion we first consider the propagation
of a Bosonic string in a flat target space of space-time dimensionality $D$,
which will be determined dynamically below by means of certain 
mathematical self-consistency conditions. 
From a first quantization view point, 
such a propagation is described by considering the following 
world-sheet two-dimensional action:
\begin{equation}
{\cal S}_\sigma = -\frac{{\cal T}}{2}\int_\Sigma d^2\sigma \sqrt{\gamma}
\gamma^{\alpha\beta}\eta_{MN}\partial_\alpha X^M \partial_\beta X^N~,
\qquad \alpha, \beta =\sigma, \tau 
\label{smodelaction}
\end{equation}
where $\gamma^{\alpha\beta}$ is the world-sheet
metric, and $X^M(\sigma,\tau),~D=0, \dots D-1$ 
denote a mapping from the world-sheet $\Sigma$
to a target space manifold of dimensionality $D$,
of flat Minkowski metric $\eta_{MN},~M,N=0, \dots D-1$.
The world-sheet zero modes of the $\sigma$-model fields $X^M$
are therefore the spacetime coordinates, the $0$ index indicating the 
(Minkowski) time. The action (\ref{smodelaction}) is related to the 
invariant world-sheet area, 
in direct extension of the point-particle case, where 
a particle sweeps out a world line as it glides through space time, and 
hence its action is proportional to a section of an invariant curve.
The quantity ${\cal T}$ is the {\it string tension}, which 
from a target-space viewpoint is a dimensionful parameter with dimensions
of [length]$^{-2}$. One then denotes 
\begin{equation}
{\cal T}=\frac{1}{2\pi \alpha'} 
\label{reggeslope}
\end{equation}
where $\alpha'$ is the Regge slope. This notation is a result of the 
original idea for which string theory was invented, namely to explain
hadron physics, and in particular the linear dependence of the 
various hadron resonances of (toal) spin $J$ vs Energy, the slope of which
was identified with the Regge slope $\sqrt{\alpha '}$.

The dynamical world-sheet theory based on (\ref{smodelaction})
is a {\it constrained theory}. This follows from invariances,
which are: (i) the reparametrizations of the world-sheet 
\begin{equation} 
\left(\sigma, \tau \right) \rightarrow \left(\sigma', \tau' \right)~,
\label{gencoord}
\end{equation}
playing the r\^ole of general coordinate tranasformations 
in the two-dimensional world-sheet manifold, and (ii) Weyl invariance,
i.e. invariance of the theory under local 
conformal rescalings of the metric :
\begin{equation} 
\gamma_{\alpha\beta} \rightarrow e^{\varphi(\sigma,\tau)}\gamma_{\alpha\beta}~,
\label{weyl}
\end{equation} 
where $\varphi(\sigma,\tau)$ is a function of $\sigma,\tau$. 
It should be noted that in two-dimensions the conformal group is 
infinite dimensional, in contrast to its finite nature in all higher 
dimensions. It is generated by the Virasoro algebra as we shall discuss 
later, and plays a crucial r\^ole for the quantum consistency 
of the theory (\ref{smodelaction}), with  important 
restrictions on the nature of the 
target-space time manifold in which the string propagates. 

The symmetry under (i) and (ii) above allows one to fix the 
world-sheet metric into the form:
\begin{equation} 
\gamma_{\alpha\beta}=e^{\rho(\sigma,\tau)}{\widehat \gamma}_{\alpha\beta}
\label{fiducial}
\end{equation}
where ${\widehat \gamma}_{\alpha\beta}$ is a {\it fiducial} (fixed)
metric on the world-sheet. As far as the two-dimensional 
gravity (world-sheet) theory is concerned, 
the choice (\ref{fiducial}) 
is, in a sense, a ``gauge choice''; this is the reason why 
the ansatz is commonly called 
{\it a conformal gauge}. For most practical purposes
the metric 
${\widehat \gamma}_{\alpha\beta}$ is taken to be flat $\eta_{\alpha\beta}$
(plane). However formally this is not quite correct in general, as
it depends on the kind of string theory considered. 
For open strings, whose classical (tree-level) propagation
implies world-sheets with the topology of a disc, the fiducial metric
is that of a disc, i.e. a manifold with boundary.
On the other hand, for closed strings, whose classical (tree-level) propagation
implies world-sheets with the topology of a sphere (punctured), for point-like 
excitations, or cylinder, for stringy excitations, the fiducial metric
is taken to be that of a sphere or cylinder. In particular, 
in the case of low-energy limit of strings, which implies that the 
external strings have been shrunk to zero size, and hence they are 
punctures for all practical purposes, the spherical topology of the 
fiducial geometry implies an Euler characteristic 
\begin{eqnarray}
&~& \chi ={\rm Euler~characteristic}=2-{\rm no.~of~holes}-2\times{\rm no.~of~handles}= \nonumber \\  
&~& 2= \frac{1}{4\pi}\int_{{\widehat \Sigma}=S^{(2)}}\sqrt{\widehat \gamma}{\widehat R}^{(2)}
\label{euler}
\end{eqnarray} 
where ${\widehat R}^{(2)}$ is the two-dimensional curvature. 
On the other hand, if one used naively a planar fiducial metric,
which as mentioned earlier, in many respect is sufficient, 
such topological properties as (\ref{euler}), would be obscured. 
The importance of (\ref{euler}) will become obvious later on,
when we discuss quantum target-space 
string corrections (string loops), as opposed to 
$\sigma$-model loops, i.e. world-sheet theory 
quantum corrections, which will be discussed immediately below. 

Under the gauge choice (\ref{fiducial}) the string equations
(i.e. the equations of motion of the fields $X^M$) read: 
\begin{equation} 
\left(\frac{\partial^2}{\partial \tau^2} -  
\frac{\partial^2}{\partial \sigma^2}
\right)X^M =0~, \qquad ({\rm wave~equation})
\label{waveeq}
\end{equation} 
and are supplemented with the {\it constraint } equations
arising from vanishing variations with respect to 
the world-sheet metric field $\gamma_{\alpha\beta}$ (which 
should be first varied and then be constrained in the gauge
(\ref{fiducial}) ):
\begin{equation} 
\frac{\delta {\cal S}_\sigma }{\delta \gamma_{\alpha\beta}}=0
\label{constraint}
\end{equation} 
The constraint (\ref{constraint}) is nothing other than the vanishing
of the stress-energy tensor $T_{\alpha\beta}$ 
of the two-dimensional (world-sheet)
field theory, defined as:
\begin{equation} 
T_{\alpha\beta} \equiv -\frac{2}{{\cal T}\sqrt{\gamma}}\frac{\delta {\cal S}_\sigma }{\delta \gamma^{\alpha\beta}}
\label{stress}
\end{equation} 
The above equations (\ref{waveeq}),(\ref{constraint}) take their simplest 
form if one uses {\it light-cone coordinates} on the world-sheet:
\begin{equation}
\sigma^{\pm} = \tau \pm \sigma
\label{lightcone}
\end{equation}
Indeed, in this system of coordinates (\ref{constraint}) becomes:
\begin{eqnarray} 
T_{\pm\pm} &=& \partial_{\pm}X^M \partial_{\pm}X^N \eta_{MN}~=0, 
\nonumber \\
T_{+-} &=& 0 \qquad ({\rm trace~of~stress~tensor})
\label{stresslightcone}
\end{eqnarray} 
The vanishing of the trace of the stress tensor of the 
world-sheet theory implies an important {\it symmetry}, that 
of {\bf CONFORMAL INVARIANCE}. The maintainance of this
classical symmetry at a quantum level is essential for the consistency
of the theory, given that above we have used this classical symmetry
in order to make the choice (\ref{fiducial}). 
In the next subsection we shall turn to a 
rather detailed discussion on  
the implications of the requirement of 
conformal symmetry (which in 
two-dimensions implies an underlying {\it infinite dimensional}
(Virasoro) symmetry) 
at a quantum $\sigma$-model level. 

Before doing this we simply mention 
that, in order to understand the existence of an infinite number of 
conserved quantities, leading to an infinite-dimensional 
symmetry, in the case of conformal 
symmetry in two space-time dimensions, it suffices to  notice that 
the conservation of the stress tensor $T_{\alpha\beta}$, which is a
consequence of two-dimensional reparametrization invariance, 
in light-cone coordinates reads: $\partial_{-}T_{++} + \partial_+T_{-+}=0$.
In view of $T_{-+}=T_{+-}=0$, then, this 
implies $\partial_{-}T_{++}=0$. 
If $f(\sigma^+)$ is {\it an arbitrary} function 
of $\sigma^+$, so that $\partial_{-}f=0$, then 
the current $fT_{++}$ is conserved, and hence the spatial integral 
$Q_f \equiv \int d\sigma f(\sigma^+) T_{++}$ is a {\it conserved 
charge}. The arbitrariness of $f$ implies therefore an infinity of 
conserved charges. Clearly the argument above holds only
in two dimensions. In higher dimensions the conformal symmetry 
is finite dimensional. 

\subsection{Conformal Invariance and Critical Dimension of Strings} 

In this subsection we shall discuss the way by which 
conformal invariance is maintained at a quantum $\sigma$-model level.
First of all we should distinguish the quantum $\sigma$-model
level, which pertains to quantising the fields $X^M$ of the 
$\sigma$-model (in, say, a path integral) at a {\it fixed} world-sheet
topology, but integrating over world-sheet metrics (geometries),
from the quantum target-space level, at which one also summs
up world-sheet topologies (string loops).

The requirement of vanishing of the trace of the world-sheet stress tensor 
at a quantum $\sigma$-model level implies important restrictions on the 
structure of the target space-time of string theory. 
The first important restriction concerns the {\it dimensionality} 
of target space time. There are various ways in which one can see this.
In this lectures we shall follow the covariant path integral 
quantization, which is most relevant for our purposes.
For details on other methods we refer the interested reader
in the literature~\cite{green}.

Consider the free field-theory world-sheet action, describing propagation 
of a free string in a flat target space time (\ref{smodelaction}). 
\begin{equation} 
{\cal S}_\sigma[\gamma, X] =-\frac{1}{4\pi\alpha '}
\int_\Sigma d^2\sigma \partial^\alpha X^M \partial^\beta X^N \eta_{MN} \gamma_{\alpha\beta}\sqrt{\gamma}
\end{equation} 
To quantize in a covariant path-integral way the above world-sheet
action one considers the partition function at a {\it fixed}
world-sheet topology (genus): 
\begin{equation} 
Z=\int {\cal D}\gamma {\cal D}X e^{-i{\cal S}_\sigma [\gamma, X]}
\label{pathintegral}
\end{equation} 
Formally one should analyticaly continue to a Euclidean world sheet
and go back to the Minkowskian signature world-sheet theory
only at the end of the computations. This will be understood
in what follows. 

We now concentrate on the integration over geometries on the world-sheet,
${\cal D}\gamma$. This integral is over 
three independent world-sheet metric 
components~\footnote{we work in the light-cone 
coordinate system, whose choice is allowed by
postulating invariance under general coordinate transformations
of the two-dimensional quantum gravity theory. Notice that in two-dimensions
gravity is a renormalizable theory so the quantum path integral over
world-sheet metrics is rigorously well defined.}:
$\gamma_{++}(\sigma,\tau)$, $\gamma_{--}(\sigma,\tau)$, 
$\gamma_{+-}(\sigma,\tau)$. An important r\^ole is played by anomalies,
i.e. the potential breakdown of certain symmetries at a quantum
world-sheet level, which result in the impossibility 
of preserving all of the apparent 
classical symmetries of (\ref{pathintegral}).

As we mentioned earlier there are three `gauge invariances'
of the action (\ref{smodelaction}), two reparametrizations of the 
world-sheet coordinates and a Weyl rescaling. 
Locally we can use these symmetries to fix the gauge (\ref{fiducial}).
For simplicity, in what follows, and given that we shall
work only at a fixed lowest topology on the world sheet, 
we shall consider the 
case of flat fiducial metrics; 
however, the precise discussion on disc (spherical 
geometries) in case of open (closed strings) 
should be kept in the back of the reader's
mind as the appropriate procedure when one sums up genera.

In this case the covariant gauge reads:
\begin{equation} 
\gamma_{\alpha\beta}=e^{\rho(\sigma,\tau)}{\eta}_{\alpha\beta}
\label{fiducialflat}
\end{equation}
In light-cone coordinates then, the condition (\ref{fiducialflat})
implies: 
\begin{equation}
0=\gamma_{++}=\gamma_{--} 
\label{condition}
\end{equation} 
Under reparametrizations $\sigma^{\pm} \rightarrow \sigma^{\pm} + \xi^{\pm}$ 
the world-sheet metric components in (\ref{condition}) transform as:
\begin{equation} 
\delta \gamma_{++} = 2 \nabla_+\xi_+ ~ ; \qquad 
\delta \gamma_{--} = 2 \nabla_-\xi_- ~.
\label{reparam}
\end{equation}
where $\nabla_{\alpha}$ denotes covariant world-sheet derivative,
with respect to the metric $\gamma$. To maintain (\ref{condition}) 
one should constraint the variations (\ref{reparam}) to {\it vanish}. 

Such conditions are implemented 
in the path integral (\ref{pathintegral}) by means 
of insertion of the identity:
\begin{equation} 
1 = \int {\cal D}g(\sigma,\tau) \delta(\gamma_{++}^g)\delta(\gamma_{--}^g)
{\rm det}\left(\frac{\delta \gamma_{++}^g}{\delta g}\right)
{\rm det}\left(\frac{\delta \gamma_{--}^g}{\delta g}\right)
\label{identity}
\end{equation}
where ${\cal D}g$ denotes integration over the group $G$ of reparametrizations
of the string world-sheet, and $\gamma^g$ denotes the world-sheet metric
into which $\gamma$ is transformed under the action of $G$. The determinants
${\rm det}\left( \dots \right)$ 
appearing in (\ref{identity}) are due to the gauge 
fixing procedure (\ref{fiducialflat}). We then have: 
\begin{equation} 
Z= \int {\cal D}g(\sigma,\tau) 
\int {\cal D}\gamma {\cal D}X e^{-{\cal S}_\sigma [\gamma, X]} 
\delta(\gamma_{++}^g)\delta(\gamma_{--}^g)
{\rm det}\left(\frac{\delta \gamma_{++}^g}{\delta g}\right)
{\rm det}\left(\frac{\delta \gamma_{--}^g}{\delta g}\right)
\label{fullpathintegral}
\end{equation} 
Reparametrization invariance
implies that ${\cal S}_\sigma [\gamma, X]={\cal S}_\sigma [\gamma^g, X]$,
i.e. that the integrand of the path integral depends on $\gamma, g$ 
{\it only through} $\gamma^g$. Making a change of variables from $\gamma,g$
to $g$ and $\gamma' \equiv \gamma^g$, and discarding the ${\cal D}g$
intergation, which can be performed trivially yielding an irrelevant
constant proportionality (normalization) factor, one arrives at: 
\begin{eqnarray} 
&~& Z= \int {\cal D}\gamma^g {\cal D}X 
e^{-{\cal S}_\sigma [\gamma^g, X]} 
\delta(\gamma_{++}^g)\delta(\gamma_{--}^g)
{\rm det}\left(\frac{\delta \gamma_{++}^g}{\delta g}\right)
{\rm det}\left(\frac{\delta \gamma_{--}^g}{\delta g}\right) = \nonumber \\
&~& \int {\cal D}\gamma_{+-}^g {\cal D}X e^{-{\cal S}_\sigma[\gamma^g, X] }
{\rm det}\left(\frac{\delta \gamma_{++}^g}{\delta g}\right)|_{\gamma_{++}=0}
{\rm det}\left(\frac{\delta \gamma_{--}^g}{\delta g}\right)|_{\gamma_{--}=0}
\label{fullpathintegral2}
\end{eqnarray} 
The integration over $\gamma_{+-}^g$ is equivalent 
to an integration over the function $\rho(\sigma,\tau)$ 
(c.f. (\ref{fiducialflat})). 
The determinants in the last expression can be expressed in terms of a
set of 
`reparametrization ghost fields of Fadeev-Popov type', 
$\{ c^{\pm}, b_{\pm\pm}\}$, 
of Grassmann 
statistics: 
\begin{eqnarray} 
{\rm det}\left(\frac{\delta \gamma_{++}^g}{\delta g}\right)|_{\gamma_{++}=0}
&=&\int {\cal D}c^-(\sigma,\tau){\cal D}b_{--}(\sigma,\tau)
e^{-\frac{1}{\pi}\int _\Sigma d^2\sigma c^-\nabla_+b_{--}}~, \nonumber \\
{\rm det}\left(\frac{\delta \gamma_{--}^g}{\delta g}\right)|_{\gamma_{--}=0}
&=&\int {\cal D}c^+(\sigma,\tau){\cal D}b_{++}(\sigma,\tau)
e^{-\frac{1}{\pi}\int _\Sigma d^2\sigma c^+\nabla_-b_{++}}~.
\label{fpghostdet}
\end{eqnarray}
Hence one should have as a final result:
\begin{equation}
Z=\int {\cal D}\rho(\sigma,\tau) \int {\cal D}X(\sigma,\tau) {\cal D}c(\sigma,\tau) {\cal D}b(\sigma,\tau)e^{-{\cal S}_{\rm total}[c,b,X]}~,
\end{equation}
where ${\cal S}_{\rm total} = {\cal S}_\sigma + {\cal S}_{\rm ghost}$,
with 
\begin{equation}
{\cal S}_{\rm ghost} =\frac{1}{2\pi}\int d^2\sigma \sqrt{\gamma}
\gamma^{\alpha\beta}c^\gamma \nabla_\alpha b_{\beta\gamma}
\label{ghostaction}
\end{equation}
the action for the Fadeev-Popov ghost fields, written
in a covariant form for completeness. The $c^\gamma$ ghost field  
is a contravariant vector, while the ghost field $b_{\beta\gamma}$ 
is a symmetric traceless tensor. Both fields $b,c$ are
of course anticommuting (Grassmann) variables, as mentioned previously.

\noindent {\it Quantization of the Ghost Sector.}

We now proceed to discuss in some detail the quantization 
of the ghost sector of theory, which has crucial implications
for the dimensionality of the target space. 
{}From (\ref{ghostaction}), the stress tensor of the ghost sector
$T_{\alpha\beta}^{\rm ghost} 
\equiv -\frac{2\pi}{\sqrt{\gamma}}\frac{\delta {\cal S}_{\rm ghost} }{\delta \gamma_{\alpha\beta}}$ (imposing the conformal gauge fixing (\ref{fiducialflat})
at the end) reads:
\begin{equation} 
T_{\alpha\beta}^{\rm ghost}=\frac{1}{2}c^\gamma
\nabla_{(\alpha}b_{\beta)\gamma} + 
\nabla_{(\alpha}c^\gamma b_{\beta)\gamma} - {\rm trace}
\label{ghoststress}
\end{equation}
In the light-cone coordinate system the only non-trivial 
components of $T^{\rm ghost}$ are: $T_{++}^{\rm ghost}, T_{--}^{\rm ghost}$:
\begin{eqnarray}
T_{++}^{\rm ghost}&=&\frac{1}{2}c^+\partial_+b_{++} + (\partial_+c^+) 
b_{++}~, \nonumber \\
T_{--}^{\rm ghost}&=&\frac{1}{2}c^-\partial_-b_{--} 
+ (\partial_-c^-) b_{--}~.
\label{lightconeghoststress}
\end{eqnarray}
Canonical quantization of ghost fields imply the following 
anticommutation relation~\cite{green}:
\begin{eqnarray} 
&~& \{ b_{++} (\sigma,\tau), 
c^+(\sigma ', \tau)\}=2\pi \delta(\sigma-\sigma')~, 
\nonumber \\
&~& \{ b_{--} (\sigma,\tau), c^-(\sigma ', \tau)\}=2\pi \delta(\sigma-\sigma')
\label{ghostquant}
\end{eqnarray}
In what follows, for simplicity, 
we concentrate on the open string
case. Comments on the closed strings will be made 
where appropriate. The interested reader can find details
on this case in the literature~\cite{green}. 
In terms of ghost-field oscillation modes: 
\begin{eqnarray} 
c^+ = \sum_{-\infty}^{+\infty} c_n e^{-i n (\tau + \sigma)}~, \nonumber \\
c^- = \sum_{-\infty}^{+\infty} c_n e^{-i n (\tau - \sigma)}~, \nonumber \\
b_{++} = \sum_{-\infty}^{+\infty} b_n e^{-i n (\tau + \sigma)}~, \nonumber \\
b_{--} = \sum_{-\infty}^{+\infty} b_n e^{-i n (\tau - \sigma)}~,
\label{modes}
\end{eqnarray} 
one has the following anticommutation relations:
\begin{eqnarray} 
\{ c_n, b_m \} =\delta_{m+n}~, \qquad \{ c_n, c_m \} = \{ b_n, b_m \} =0
\label{modealgebra}
\end{eqnarray}
Using the Fourier modes of $T^{\rm ghost}$ at $\tau=0$: 
\begin{equation}
{\cal L}_m^{\rm ghost} =\frac{1}{\pi}\int_{-\pi}^{\pi}e^{im\sigma}
T_{++}^{\rm ghost}
\label{virasorogen}
\end{equation}
we have:
\begin{equation} 
{\cal L}_m^{\rm ghost} = \sum_{n=-\infty}^{\infty}
[m(J-1)-n]b_{m+n}c_{-n}
\end{equation}
where $J$ is the conformal spin of the field $b$, with 
$1-J$ that of the field $c$. 

\noindent {\bf [NB1}: {\it For completeness we note that conformal dimensions
are defined as follows (open string case for definiteness): consider a local operator on the world sheet
${\cal F}(\sigma,\tau)$. Set $\sigma=0$ (or $\sigma=\pi$, the 
position of the boundaries of the open string) and study ${\cal F}(0,\tau)
\equiv {\cal F}(\tau)$. Then, ${\cal F}(\tau)$ is defined to have 
conformal  dimension (or `spin') $J$ if and only if,
under an arbitrary change of variables} $\tau \to \tau'(\tau)$, 
${\cal F}(\tau)$ {\it transforms as}: 
\begin{equation} 
{\cal F}'(\tau ') = \left(\frac{d \tau }{d \tau'}\right)^J {\cal F}(\tau)
\end{equation}

{\it The operators ${\cal L}_m^{\rm ghost}$ in (\ref{virasorogen})
are the {\it generators} of the infinite-dimensional 
Virasoro algebra. The action of ${\cal L}_m$ on ${\cal F}$ 
is}:
\begin{equation}
\left[ {\cal L}_m, {\cal F}(\tau)\right]=e^{im\tau}
\left(-i\frac{d}{d\tau} + mJ\right){\cal F}(\tau)
\end{equation}
{\it or in terms of modes}:
\begin{equation} 
\left[{\cal L}_m, {\cal F} \right]=[m(J-1)-n]{\cal F}_{m+n}
\end{equation}

{\it Note for completeness that for {\it closed strings} there is a second set 
of ghost Virasoro generators}.{\bf ]}

The Virasoro algebra of ${\cal L}_m^{\rm ghost}$ 
is defined by the respective commutation relations:
\begin{equation} 
\left[ {\cal L}_m^{\rm ghost}, {\cal L}_n^{\rm ghost} \right]=
(m-n){\cal L}_{m+n}^{\rm ghost} + {\cal A}(m)^{\rm ghost}\delta_{m+n} 
\label{ghostvirasoro}
\end{equation}
where the second term on the right-hand-side is a 
``conformal anomaly term'', indicating the breakdown of conformal 
symmetry at a quantum $\sigma$-model level. It can be calculated to be:
\begin{equation} 
{\cal A}(m)^{\rm ghost}=\frac{1}{12}[1-3(2J-1)^2]m^3+\frac{1}{6}m
\end{equation} 

\noindent {\bf [NB2:} {\it The 
easiest way to evaluate the anomaly is to look at 
specific matrix elements, e.g.} : ${\cal A}(1)^{\rm ghost} = 
\langle 0|\left[{\cal L}_1^{\rm ghost}, {\cal L}_{-1}^{\rm ghost} \right]|0\rangle $. {\bf ]}

The ghost field $b$ has $J=2$, so that the anomaly in the ghost sector is:
\begin{equation} 
{\cal A}(m)^{\rm ghost} = \frac{1}{6}(m-13m^3)
\label{ghostanomaly}
\end{equation} 

Similar quantization conditions characterize the matter 
sector of the $\sigma$-model (\ref{smodelaction}),
pertaining to the fields/coordinates $X^M$.
We shall not do the analysis here. The interested reader is referred
for details and results in the literature~\cite{green}. 
Adding such ghost and matter contributions, 
the total conformal anomaly (for a $D$-dimensional target space time) 
is~\cite{green} is found as follows: first 
we note that the Virasoro generators 
corresponding to ${\cal S}_{\rm total}={\cal S}_\sigma + {\cal S}_{\rm ghost}$,
are 
the Fourier modes ${\cal L}_m =\frac{1}{\pi}\int_{-\pi}^{\pi}
d\sigma e^{im\sigma}T_{++}^{\rm total}$, where 
$T_{++}^{\rm total} =-\frac{2\pi}{\gamma}\frac{\delta {\cal S}_{\rm total}}{\delta \gamma_{++}}|_{\gamma_{++}=0}$, and ${\cal L}_m= {\cal L}_m^{\rm matter}
+ {\cal L}_m^{\rm ghost} - a\delta_m$, and we have shifted the 
definition of ${\cal L}_0$ (related to the Hamiltonian of the string)
so that the zeroth-order Virasoro constraint is ${\cal L}_0 =0$.
Then, following similar mode expansions for the matter sector, 
as 
those of the ghost sector outlined above, 
one arrives at the total conformal anomaly:
\begin{equation}
{\cal A}(m) = \frac{D}{12}(m^3 - m) + \frac{1}{6}(m-m^3)+2am
\label{anomalytotal}
\end{equation}
where $D$ is the target-space dimensionality (corresponding to 
the contributions from $D$ 
$\sigma$-model ``mater'' fields $X^M$).  From (\ref{anomalytotal})
one observes that the anomaly {\bf VANISHES}, and thus conformal 
world-sheet symmetry is a good symmetry at a quantum $\sigma$-model level, 
as required for mathematical self-consistency of the theory,
if and only if: 
\begin{equation}
                    D_c=26 ~\qquad ({\rm Bosonic~String})~,
\qquad a=1
\end{equation}

For fermionic (supersymmetric strings (cf below)) 
the critical space-time dimension is $D_c=10$.

\subsection{Some Hints towards Supersymmetric Strings} 

So far we have examined Bosonic strings. Supersymmetric 
strings are more relevant for particle phenomenology,
because as we shall discuss now, do not suffer from vacuum 
instabilities like the bosonic counterparts, which are known 
to contain in their spectrum tachyons (negative mass squared modes).
Moreover such theories are capable of incorporating
fermionic target-space backgrounds. 

There are two ways to include fermionic backgrounds in a 
$\sigma$-model  string theory, and thus to achieve target-space
{\it supersymmetry}:

\noindent {\bf (1)} The first one is to supersymemtrize the world-sheet
theory by introducing fermionic partners $\psi^M (\sigma, \tau)$ 
to the $X^M(\sigma,\tau)$ fields. 
There are two kinds of fermions that can be introduced, depending 
on their boundary conditions (b.c.) on a circle, so that the world-sheet 
fermion action is invariant under periodic identification on a 
cylinder $\sigma \to \sigma + 2\pi$:
\begin{eqnarray}
&~&\psi^M(\sigma=0) = - \psi^M (\sigma=2\pi)~~~ {\rm antiperiodic~b.c.:~Neveu-Schwarz~(NS)}, \nonumber \\
&~&\psi^M(\sigma=0) =  \psi^M (\sigma=2\pi)~~~~~{\rm periodic~b.c.:~Ramond~(R)}
\end{eqnarray} 
As a result of the presence
of these extra degrees of freedom, 
world-sheet supersymmetry leads to a reduction of the 
critical target-space dimension, for which the conformal 
anomaly is absent, from 26 to 10 (i.e. the critical 
target-space dimensionality of a superstring is 10). 

A world-sheet supersymmetric $\sigma$-model does not have manifest supersymmetry in 
target space; the latter is obtained after appropriate spectrum 
projection
(Goddard, Scherk and Olive~\cite{green}). 

\noindent {\bf (2)} The second way of introducing fermionic backgrounds
in string theory is to have bosonic world sheets but with manifest 
target-space Supersymmetry (Green and Schwarz). 

The two methods are equivalent, as far as target-space Supersymmetry 
is concerned. 

\noindent {\it Features of Supersymmetric Strings.}

\begin{itemize} 

\item{(i)} The tachyonic instabilities in the spectrum,
which plagued the Bosonic string, are absent in the 
supersymmetric string case. This stability of the 
superstring vacuum is one of the most important 
arguments in favour of (target-space) 
supersymmetry from the point of view of string theory.

\item{(ii)} From a world-sheet viewpoint, in the Neveu-Schwarz-Ramond 
formulation of fermionic strings, the world-sheet action becomes a curved 
two-dimensional locally supersymmetric theory (world-sheet supergravity 
theory).

\item{(iii)} Target Supersymmetry is broken in general when one 
considers strings at finite temperatures, obtained upon appropriate
compactification of the target-space coordinate. In general, however,
the breaking of target supersymmetry at zero temperature, so as to 
make contact with realistic phenomenologies, is an open issue 
at present, despite considerable effort and the existence of many scenaria. 

\end{itemize} 

\subsection{Kaluza-Klein Compactification} 

The fact that the target space-time dimensionality of strings
turns out to be higher than four implies the need for 
{\it compactification} of the extra dimensions. 

Compactification means that the ground state of string theory has the form:
\begin{equation}
      {\cal M}^{(4)} \otimes {\cal K} 
\end{equation}
where ${\cal M}^{(4)}$ is a four-dimensional non-compact 
manifold (assumed Minkowski, but in fact it can be 
any other space time encountered in four-dimensional general relativity, 
priovided it satisfies certain consistency conditions 
to be discussed below), and ${\cal K}$ is a compact manifold,
six dimensional in the case of superstrings, or 22 dimensional
in the case of (unstable) Bosonic strings. 

In ``old'' (conventional) string theory~\cite{green}, 
the ``size'' of the extra dimensions is assumed Planckian, something 
which in the 
modern brane version is not necessarily true. For our purposes 
in these Lectures we shall restrict ourselves 
to the ``old'' string theory approach to compactification. 

Consider a 26-(or 10-)dimensional metric on 
${\cal M}^{(4)} \otimes {\cal K}$,  $g_{MN}$, and let 
$g_{\mu\nu} \in {\cal M}^{(4)}$, and $g_{ij} \in {\cal K}$. 

{} From a four-dimensional point of view $g_{ij}$ appear 
as {\it massless spin-one particles}, i.e. massless gauge bosons. 
This is the central point of Kaluza-Klein (KK) approach.
Such particles appear if a suitable subgroup of 
the underlying 
ten-dimensional general covariance is {\it left unbroken}
under compactification to ${\cal M}^{(4)} \otimes {\cal K} $. 
Let us see this in some detail. 

Consider a general coordinate transformation on the manifold ${\cal K}$:
\begin{equation}
    y^k \to y^k + \epsilon V^k (y^j)
\label{coord}
\end{equation}
where $\epsilon$ is a small parameter, and $V^k$ a vector field. 
In the passive frame, the corresponding 
change of the metric tensor $g_{ij}$ is:
\begin{equation} 
\delta g_{ij} = \epsilon \left(\nabla_i V_j + \nabla_j V_i \right)
\end{equation}
where $\nabla_i$ is the gravitational covariant derivative. The 
metrc on ${\cal K}$ is therefore {\it invariant} if $V^k$ obeys
a Killing-vector equation:
\begin{equation} 
\nabla_i V_j + \nabla_j V_i =0
\end{equation} 
Thus, the coordinate transformation (\ref{coord}), generated by the 
Killing vector $V^k$, is a symmetry of any generally-covariant equation
for the metric of ${\cal K}$. 
More generally, if one studies an equation 
involving a coupled system of the metric with some other matter fields (e.g. gauge fields {\it etc.}), then one obtains a symmetry if $V^k$ can be combined with a suitable transformation of the matter fields that leaves their 
expectation values invariant.

Consider the case in which one has several Killing vector 
fields $V^i_a, a=1, \dots N$, generating a Lie algebra ${\cal H}$ 
of some kind: 
\begin{equation} 
\left[ V_a^i\partial_i, V_b^j\partial_j \right]=f_{abc}V_c^k\partial_k
\end{equation}
where $f_{abc}$ are the corresponding structure 
constants of the Lie algebra that generates a symmetry group 
${\cal H}$ on ${\cal K}$. 

Consider the transformation 
\begin{equation}
\left( x^\mu, y^k \right) \rightarrow \left(x^\mu, y^k + 
\sum_{a} \epsilon_a V_a^k \right)
\label{gencoor}
\end{equation} 
In the general case one may consider non-constant 
$\epsilon_a = \epsilon_a (x^\mu)$ on ${\cal M}^{(4)}$. 

At long wavelengths, 
which are of interest to any low-energy observer, 
only massless modes are important. Therefore, the transformation
(\ref{gencoor}) will be a symmetry 
of the theory compactified on ${\cal M}^{(4)} \otimes {\cal K}$.
From the point of view of the four-dimensional 
effective low-energy theory the transformations (\ref{gencoor}) 
will look like ${\cal M}^{(4)}$-dependent {\it local gauge 
transformations} with gauge group ${\cal H}$. 
The effective four-dimensional theory will therefore 
have {\it massless gauge bosons} given by the 
ansatz:
\begin{equation} 
g_{\mu j} = \sum_{a}A_\mu^a(x^\nu) V_{ja}
(y^k)
\label{gaugefields}
\end{equation}
where $A_\mu^a (x^\nu)$ are the massless gauge fields
that appear in ${\cal M}^{(4)}$. This follows from the fact that 
under (\ref{gencoor}), the fields $A_\mu^a$ in (\ref{gaugefields}) 
transform as ordinary gauge fieldds: $\delta A_\mu^a = \partial_\mu 
\epsilon^a + f^{abc}\epsilon_b A_{\mu c} $. 

An interesting question arises at this point as to what symmetry groups 
can arise via KK compactification. This is equivalent to asking 
what symmetry groups an $n$-dimensional manifold can have. 

We consider for completeness the case where ${\cal K}$ has dimension $n$,
which is kept general at this point. 
The most general answer to the above question is complicated. 
An interesting question, of phenomenological interest, is for which 
$n$ one can get the standard model group 
$SU(3) \otimes SU(2) \otimes U(1)$. It can be shown~\cite{green}
that this happens for $n=7$ which {it is not}
the case of string theory (superstrings), since in that case $n=6$. 
This is what put off people's interest in the traditional
KK compactification, which was instead replaced by the
heterortic string construction, which we 
shall not analyse here~\cite{green}. 
On the other hand, it should be mentioned that in the modern version 
of string theory, involving branes, KK modes play an important r\^ole again.
For more details we refer the reader to the lectures on brane 
theory in this School. 

\subsection{Strings in Background Fields} 

So far we have dealt with flat Minkowski target space times.
In general strings may be formulated in curved space times,
and, in general, in the presence of non-trivial background fields. 
In this case conformal invariance conditions
of the underlying $\sigma$-model theory become equivalent,
as we shall discuss below, to equations of motion of the various
target-space background fields. 

The lowest lying energy multiplet in superstring theory
consists (in its bosonic part) 
of {\it massless} states of gravitons $g_{MN}$ (spin two
traceless and symmetric tensor field),
dilaton $\Phi$ (scalar, spin 0) and antisymmetric tensor $B_{MN}$ 
field~\footnote{In the Bosonic states the lowest lying energy state (vacuum) 
is 
tachyonic, and the above multiplet occurs at the next level.}.
Target space supersymmetry, of course, implies the existence
of the supersymmetric (fermionic) partners of the states in this multiplet.  
In this section we shall discuss the formalism, and its physical consequences,
for string propagation in the bosonic part of the massless superstring 
multiplet, starting from graviton backgrounds, which are discussed next. 

\subsubsection{Formulation of Strings in Curved Space times-Graviton Backgrounds}

The corresponding $\sigma$-model action, describing the propagation
of a string in a space time with metric $g_{MN}$ reads:
\begin{equation} 
{\cal S}_\sigma = \frac{1}{4\pi \alpha'}\int _\Sigma 
d^2\sigma \sqrt{\gamma}
\gamma^{\alpha\beta} g_{MN}(X^P(\sigma,\tau))\partial_\alpha X^M \partial_\beta X^N~,
\qquad \alpha, \beta =\sigma, \tau 
\label{smodelactioncurved}
\end{equation}
One expands around a flat target space time $g_{MN}=\eta_{MN} + 
h_{MN}(X)$. For $|h_{MN}(X)| \ll 1$ one may expand in 
Fourier series: 
\begin{equation} 
   h_{MN}(X)_ = \int \frac{d^Dk}{(2\pi)^D}e^{i k_M X^M}{\tilde h}_{MN}(k) 
\end{equation} 
in which case the $\sigma$-model action becomes schematically:
\begin{eqnarray}
&~& {\cal S}_\sigma = S^* + \frac{1}{4\pi \alpha'} 
\int _\Sigma d^2\sigma \sqrt{\gamma}\gamma^{\alpha\beta}
\partial_\alpha X^M \partial_\beta X^N \int \frac{d^Dk}{(2\pi)^D}
e^{ik_M X^M}{\tilde h}_{MN}(k) \equiv \nonumber \\
&~& S^* + g^i \int _\Sigma d^2\sigma V_i 
\label{expansion} 
\end{eqnarray} 
where $S^*$ is the flat space-time action (\ref{smodelaction}), 
and one has the correspondence 
$g^i \leftarrow\rightarrow {\tilde h}_{MN}(k)$, 
$V_i \leftarrow\rightarrow 
\sqrt{\gamma} \gamma^{\alpha\beta} \partial_\alpha X^M \partial_\beta X^N
e^{i k_P X^P}$, and $\sum_{i} \leftarrow\rightarrow \int 
\frac{d^Dk}{(2\pi)^D}$.

It should be stressed that implementing a 
Fourier expansion necessitates an expansion in the neighborhood
of the Minkowski space time, so as to be able to define 
plane waves appropriately. For generic space times one may consider 
an expansion about an appropriate conformal (fixed point)
$\sigma$-model action $S^*$, as in the last line of
the right-hand-side of (\ref{expansion}), 
but in this case the set of background fields/$\sigma$-model couplings $\{ g^i \}$ is found as follows: 
consider $g_{MN}=g_{MN}^* + h_{MN}(X)$, where $g_{MN}^*$ a conformal 
(fixed-point) non-flat metric, and $h_{MN}(X)$ an expansion around it.
Then, 
\begin{eqnarray} 
&~& {\cal S}=\frac{1}{4\pi\alpha'} \int_\Sigma d^2\sigma \partial_\alpha X^M 
\partial_\beta X^N g_{MN}\gamma^{\alpha\beta}\sqrt{\gamma} = \nonumber \\
&~& {\cal S}^* + \frac{1}{4\pi\alpha'}\int_\Sigma d^2 \sigma 
h_{MN}(X)\partial_\alpha X^M \partial_\beta 
X^N \gamma^{\alpha\beta}\sqrt{\gamma} = \nonumber \\
&~& {\cal S}^* + \frac{1}{4\pi\alpha'}\int_\Sigma d^2\sigma \int d^Dy 
\sqrt{g^*(y)}\delta^{(D)}\left(y^M - X^M(\sigma,\tau)\right) \cdot \nonumber \\
&~& h_{MN}(y)
\partial_\alpha X^M \partial _\beta X^N \gamma^{\alpha\beta}\sqrt{\gamma} 
\equiv \nonumber \\
&~& {\cal S}^* + g^i \int_\sigma d^2 \sigma V_i 
\end{eqnarray} 
where $g^i \leftarrow\rightarrow \{ h_{MN}(y) \}$, 
$V_i \leftarrow\rightarrow \delta^{(D)}\left(y^M - X^M(\sigma,\tau)\right)
\partial_\alpha X^M \partial_\beta X^N \gamma^{\alpha\beta}\sqrt{\gamma}$,
and $\sum_{i} \leftarrow\rightarrow \int d^Dy \sqrt{g^*(y)}$. 
As the reader must have noticed, for general backgrounds one 
pulls out the world-sheet zero mode of $X^M$ appropriately, 
which defines the target-space coordinates, and integrates over it,
thereby determining the (infinite dimensional) set 
of $\sigma$-model couplings.

\subsubsection{Other Backgrounds} 

We continue our discussion on formulating string
propagation in non-trivial 
backgrounds, in the first-quantized formalism, 
by studying next antisymmetric tensor and dilatons.

\noindent {\it Antisymmetric Tensor Background}

The antisymmetric tensor backgrounds $B_{MN}$ 
are spin one, antisymmetric tensor fields
$B_{MN}=-B_{NM}$. There is an {\it Abelian
gauge symmery} which characterizes the corresponding scattering 
amplitudes (with antisymmetric tensors as external particles),
\begin{equation} 
B_{MN} \rightarrow B_{MN} + \partial_{[M}\Lambda_{N]}
\label{abelian}
\end{equation}
which implies that the corresponding low-energy 
effective action, which reproduces the scattering amplitudes,
will depend only through the field strength of $B_{MN}$:
$H_{MNP}=\partial_{[M}B_{NP]}$.

In a $\sigma$-model action the pertinent deformation
has the form:
\begin{equation}
\frac{1}{4\pi \alpha'}\int_\Sigma d^2\sigma B_{MN}V^{(B)MN} =
\frac{1}{4\pi \alpha'} \int_\Sigma d^2\sigma B_{MN}
\epsilon^{\alpha\beta}\partial_\alpha X^M \partial_\beta X^N
\label{antisymm}
\end{equation}
where $\epsilon^{\alpha\beta}$ is the contravariant antisymmetric
symbol.  

\noindent {\bf [NB3}:{\it due to its presence there is no explicit
$\sqrt{\gamma}$ factor in (\ref{antisymm}), as this is incorporated
in the contravariant $\epsilon$-symbol}. {\bf ]} 

\vspace{0.2in} 

\noindent {\it Dilaton Backgrounds and the String Coupling }

The dilaton $\Phi (X)$ is a spin-0 mode of the massless superstring 
multiplet, which in a $\sigma$-model framework couples to the 
world-sheet scalar curvature $R^{(2)}(\sigma\tau)$:
\begin{equation} 
{\cal S}_\sigma = \frac{1}{4\pi \alpha'}\int_\Sigma d^2\sigma \sqrt{\gamma}
\gamma^{\alpha\beta} \partial_\alpha X^M \partial_\beta X_M 
+ \frac{1}{4\pi} \int_\Sigma d^2\sigma \sqrt{\gamma} \Phi (X)R^{(2)}(\sigma,\tau)
\label{dilaton}
\end{equation}
Notice that in the dilaton term there is no $\alpha'$ factor, which mplies that in a (perturbative) series expansion in terms of $\alpha'$ the 
dilaton couplings are of higher order as compared with the graviton
and antisymmetric tensor backgrounds.

An important r\^ole of the dilaton is that it determines (via its vacuum expectation value)
the strength of the string interactions, the {\it string coupling}:
\begin{equation} 
{\rm String~Coupling}~g_s = e^{\langle \Phi \rangle }
\label{stringcoupling}
\end{equation}
where $< \dots >=\int {\cal D}X e^{{\cal S}_\sigma }$ 
is computed with respect to the string path integral
for the $\sigma$-model propgating in the background under
consideration.

\begin{figure}[htb]
\epsfxsize=4in
\begin{center}
\epsffile{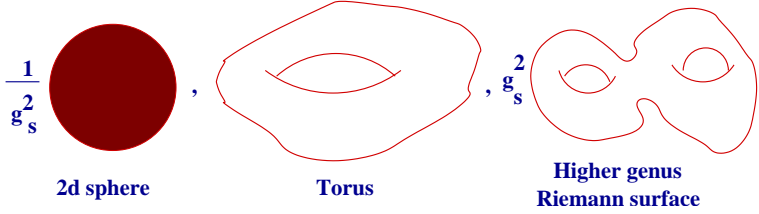}
\end{center}
\caption[]{The string coupling $g_s=e^{<\Phi>}$ as a string-loop
counting parameter. The loop expansion parameter is $g_s^{-\chi}$, where
$\chi$ is the Euler characteristic of the manifold. 
For the sphere one has $\chi_S=2$,
for a torus (flat) $\chi_T=0$ {\it etc.}. Such weights are depicted explicitly in the figure}
\label{fig:stringloops}
\end{figure}

The string coupling is a string-loop counting parameter (c.f. figure
\ref{fig:stringloops}). This can be seen easily by first recalling 
the index theorem (\ref{euler}) 
that connects a geometrical world-sheet
quantity like the curvature $R^{(2)}$ to a topological quantity,
the Euler characteristic $\chi$, which counts the genus of the surface:
\begin{equation}
\chi = 2-{\rm no.~of~holes}-2\times{\rm no.~of~handles} = 
\frac{1}{4\pi}\int_{\Sigma}\sqrt{\gamma}{R}^{(2)}
\label{euler2}
\end{equation} 
Consider the $\sigma$-model deformation (\ref{dilaton}) and split 
the dilaton into a classical (world-sheet coordinate independent)
part $<\Phi >$ and 
a quantum part \\ 
$\varphi\equiv:\Phi:$, where $: \dots :$ denotes appropriate
normal ordering of the corresponding operators:
$\Phi = <\Phi> + \varphi (\sigma,\tau)$. Using 
(\ref{euler2}),(\ref{stringcoupling}),    
we can 
then write for the $\sigma$-model partition function summed 
over surfaces of genus $\chi$: 
\begin{eqnarray} 
&~& Z = \sum_{\chi}\int {\cal D}X e^{-{\cal S}^{\rm rest} 
- \chi <\Phi > - \frac{1}{4\pi} 
\int_\Sigma d^2 \sigma \sqrt{\gamma} \varphi R^{(2)}} = \nonumber \\
&~& \sum_{\chi} g_s^{-\chi} \int {\cal D}X e^{-{\cal S}^{\rm rest} - \frac{1}{4\pi} 
\int_\Sigma d^2 \sigma \sqrt{\gamma} \varphi R^{(2)}}
\label{strringlooppart} 
\end{eqnarray}
where $S^{\rm rest}$ denotes a $\sigma$-model action involving the rest of the 
background deformations except the dilaton.  
For a sphere, $\chi =2$ (for a disc (open strings) $\chi=1$), 
for torus (one string loop) $\chi =0$, etc.  By normalizing
the higher-loop contributions to the sphere, then, one gets
the string-loop series depicted in \ref{fig:stringloops},
with a clear interpretation of the quantity (\ref{stringcoupling})  
as a string-loop counting parameter.

\subsection{Conformal Invariance and Background Fields} 

The presence of $\sigma$-model ``deformations'' $g^i \int_\Sigma V_i$ 
imply in general deviations from conformal invariance on the world sheet.
To ensure conformal invariance we must impose certain conditions 
on the couplings $g^i$. Such conditions, and their implications 
will be studied in this section. As we shall see, the conformal 
invariance conditions are equivalent to equations of motion
for the target-space background $g^i$ which are derived 
from a target-space string effective action. This action 
constitutes the low-energy (field-theory) limit of strings and 
will be the main topic of these lectures. String cosmology,
which we shall discuss in the second and third lectures, will be based
on such string effective actions.
 
To start with, let us consider a deformed $\sigma$-model action
\begin{equation}
{\cal S}={\cal S}^* + g^i\int_\Sigma d^2\sigma V_i 
\label{deformed}
\end{equation} 
which, as we have discussed above, 
describes propagation (in a first quantized formalism)
of a string in backgrounds $\{ g^i \}= \{ g_{MN}, \Phi, B_{MN}, \dots \}$. 

The partition function of the deformed string may be expanded
in an (infinite) series in powers of $g^i$ (assumed weak):
\begin{eqnarray} 
&~& Z[g] = \int {\cal D}\rho {\cal D}X e^{-{\cal S}^* - g^i\int_\Sigma d^2\sigma V_i }  = \nonumber \\
&~& \sum_{{i_i}}\int_\Sigma \dots \int_\Sigma \langle V_{i_1} \dots V_{i_N}
\rangle _*g^{i_1} \dots g^{i_N} d^2 \sigma_1 \dots d^2\sigma_N 
\end{eqnarray}
where $\langle \dots \rangle _* =\int {\cal D}\rho {\cal D}X e^{-{\cal S}^*}$.
We work in the conformal gauge (\ref{fiducial}), and thus the mode $\rho$ 
is whatever is left from the integration over world-sheet geometries.
In conformal (`critical') string theory
the quantities $\langle V_{i_1} \dots V_{i_N}
\rangle _* $ are nothing other
than the string scattering amplitudes (defining the 
on-shell S-matrix elements)
for the modes
corresponding to $\{ g^i \}$. It must be stressed that critical 
string theory is by definition a theory of the S-matrix, and hence 
this imposes a severe restriction on the appropriate 
backgrounds.
Namely, as we shall discuss in Lecture 3, appropriate
string backgrounds are those which can admit asymptotic states,
and hence well-defined on-shell S-matrix 
elements~\footnote{Eternally accelerating string Universe backgrounds, for instance, 
which will be the topic of disucssion in the last part 
of our lectures, are incompatible with critical string theory,
precisely because of this, namely in such backgrounds
one cannot define appropriate asymptotic pure quantum states. 
We shall discuss how such problems may be overcome in the last 
part of the lectures.}.

As a two-dimensional quantum field theory, the model (\ref{deformed})
suffers from world-sheet 
ultraviolet (short-distance) divergences, which should not
be confused with target-space ultraviolet infinities. 
Such world-sheet
infinities arise from short-distance regions \\
${\rm lim}_{\sigma_1 \to \sigma_2} \langle V_{i_1}(\sigma_1) \dots 
V_{i_2}(\sigma_2) V_{i_3}(\sigma_3) \dots V_{i_N}(\sigma_N)\rangle _* $,
and they
are responsible for the breaking of the conformal invariance
at a quantum level, because they require regularization, and regularisation
implies the existence of a length (short-ditance) cutoff. The presence of 
such length cutoff regulators break the local (and global) scale invariance
in general. Below we shall seek conditions under which the conformal 
invariance is restored. 

To this end, we first observe that, according to the 
general case of renormalizable 
quantum field theories, one of which is 
the $\sigma$-model two-dimensional theory (\ref{deformed}), 
such infinities may be absorbed in a 
renormalization of the string couplings. 
To this end, one adds appropriate
counterterms in the $\sigma$-model action, which have the same form as the
original (bare) deformations, but they are renormalization-group 
scale dependendent. 
Therefore their effect is to `renormalise' the couplings 
$g_i \to g_i^R ({\rm ln}\mu)$, where $\mu$ is a world-sheet renormalization
group scale. 

The scale defines the $\beta$-functions of the theory:
\begin{equation} 
\beta^i \equiv \frac{dg_R^i}{d{\rm ln}\mu}=\sum_{{i_n}}C^i_{i_1 \dots i_n}g_R^{i_1}\dots g_R^{i_n}
\label{beta}
\end{equation}
One can show in general that the 
(2d-gravitational) trace $\Theta \equiv T_{\alpha\beta}\gamma^{\alpha\beta}$ 
of the world-sheet stress tensor in such a renormalized theory
can be expressed as:
\begin{equation} 
\langle \Theta \rangle = c~R^{(2)} + \beta^i \langle V_i \rangle 
\label{trace}
\end{equation}
where $c$ is the conformal anomaly of the world-sheet theory, and 
$R^{(2)}$ is the world-sheet curvature. In the case of strings
living in their critical dimension, the total conformal anomaly 
$c$, when Fadeev-Popov contributions are taken into account
vanishes, as we have seen in the beginning of this lecture. 
Thus to ensure conformal invariance in the presence 
of background fields $g^i$, i.e. 
$\langle \Theta \rangle =0$ one must {\it impose} 
\begin{equation}
\beta^i =0 
\label{betacond}
\end{equation}
These are the conformal invariance conditions, which in 
view of (\ref{beta}) imply restrictions on the background fields $g^i$. 

A few comments are important at this point before we embark on a discussion
on the physical implications for the target-space theory 
of the conditions (\ref{betacond}).
The comments concern the geometry of the `space of coupling constants
$\{ g^i \}$', so called moduli space of strings, or {\it string theory space}.
As discussed first by Zamolodchikov~\cite{zam}, such a space
is a metric space, with the metric being provided by the 
two-point functions of vertex operators $V_i$ in the deformed theory,
\begin{equation}
{\cal G}_{ij} = z^2 {\bar z}^2\langle V_i (z, {\bar z})V_i(0,0)\rangle _g
\label{zammetric}
\end{equation}
where $z, {\bar z}$ are complex coordinate of a Euclidean world sheet,
which is necessary for convergence  
of our path integral formalism.
The notation $\langle \dots \rangle _g$ denotes path integral 
with respect to the deformed $\sigma$-model action (\ref{deformed})
in the background $\{ g^i\}$.   
The metric (\ref{zammetric}) acts as a raising and lowering 
indices operator in $g^i$-space. 

An important property of the stringy $\sigma$-model $\beta$-functions
is the fact that the `covariant' $\beta$-functions, defined as
$\beta_i = {\cal G}_{ij}\beta^j$, when expanded in powers
of $g^i$ have coefficients completely symmetric under permutation
of their indices, i.e. 
\begin{equation}
\beta_i = {\cal G}_{ij}\beta^j = \sum_{{i_n}}c_{i_1 i_2 \dots i_n}g^{i_2} \dots g^{i_n}
\label{covariantbeta}
\end{equation}
with $c_{i_1 i_2 \dots i_n}$ totally symmetric in the indices $i_j$.
This can be proven by using specific properties of the 
world-sheet renormalization group~\cite{mms}. Such totally
symmetric coefficients are associated with dual string scattering 
amplitudes, as we shall demonstrate explicitly later on.

What (\ref{covariantbeta}) implies 
is a gradient flow property of the stringy $\beta$-functions,
namely that 
\begin{equation}
\frac{\delta C[g]}{\delta g^i} = {\cal G}_{ij}\beta^j 
\label{flow}
\end{equation}
where $C[g]$ is a target-space space-time integrated functional of 
the fields $g^i(y)$. 

Notice that the conformal invariance conditions (\ref{beta}) 
are then equivalent to equations of motion obtained from this 
functional $C[g]$, which thus 
plays the r\^ole of a target-space 
effective action functional for 
the low-energy dynamics of string theory.

An important note should be made at this point, concerning
the r\^ole of target-space diffeomorphism invariance in 
stringy $\sigma$-models. As a result of this invariance, which is 
a crucial target-space symmetry, that makes contact 
with general relativity in the target manifold, the conformal 
invariance conditions (\ref{beta}) in the case of strings
are slightly modified by terms which express precisely 
the change of the background couplings $g^i$ under 
general coordinate diffeomorphisms in target space $\delta g^i$:
\begin{equation}
{\widehat \beta}^i = \beta^i + \delta g^i =0
\label{diffbeta}
\end{equation}
in other words conformal invariance in $\sigma$-models
implies the vanishing of the modified $\beta$-functions,
i.e. it is valid up to general coordinate diffeomorphism terms. 
This modification plays an important r\^ole in ensuring the 
compatibility of the solutions with general coordinate invariance
of the target manifold. The modified $\beta$-functions 
${\widehat \beta}^i$ are known in the string literature as
{\it Weyl anomaly coefficients}~\cite{green}. 
In fact, for the 
stringy $\sigma$-model case, they appear in the 
expression (\ref{trace}), in place of the ordinary $\beta^i$. 

\subsection{General Methods for Computing $\beta$-functions}

In general there are two kinds of perturbative 
expansions in $\sigma$-model theory. 

\begin{itemize} 

\item{(I)} {\bf Weak Coupling} $g^i$-{\bf expansion}: in which one 
assumes weak deformations of conformal $\sigma$-model actions,
with $g^i$ small enough so as a perturbative series expansion 
in powers of $g^i$ suffices. Usually in this method one deals
with Fourier modes (cf below) of background deformations, and 
hence the results are available in target-momentum space; this is 
appropriate when one considers scattering 
amplitudes of strings. 

\item{(II)} $\alpha '$-{\bf Regge slope expansion}: in which one 
considers an expansion of the partition function and correlation functions
of $\sigma$-models in powers of $\alpha '$. Given that the Regge slope 
has dimensions of [length]$^2$, such expansions imply (in Fourier space)
appropriate derivative expansions of the string effective actions.
It is the second expansion that will be directly relevant for our
Cosmological considerations. The Regge slope expansion 
preserves general covariance explicitly.

\end{itemize}

It should be stressed that physically the two methods of expansion
are completely equivalent. Formally though, as we have mentioned, 
the various methods
may have advantages and disadvantages, compared to each other,
dependending on the physical problem at hand. For instance when one 
deals with weak fields, then the first method seems appropriate.
In field theory limit of strings, on the other hand, where by definition
we are interested in low-energies compared with the string 
(Planckian $\sim 10^{19}$ GeV) scale, then the second expansion is 
more relevant. Moreover it is this method that allows 
configuration-space general covariant expressions for the 
effective action in arbitrary space-time backgrounds, in which 
momentum space may not always be a well-defined concept. 

Before we turn into an explicit discussion on 
string effective actions we consider it as instructive to 
discuss, thorugh a simple but quite generic example, 
the connection of conformal invariance conditions to 
string scattering amplitudes through the first method.

\subsubsection{String Amplitudes and World-Sheet Renormalization Group}

A generic structure of a renormalization-group  $\beta$-function
in powers of the renormalized couplings $g^i(t)$ is:
\begin{equation}
\beta^i = \frac{dg^i}{dt} =y_i g^i + \alpha^i_{jk}g^j g^k + \gamma^i_{jk\ell}
g^j g^k g^\ell + \dots~, 
\qquad t={\rm ln}\mu
\label{betacomplete}
\end{equation}
where $y_i$ are the anomalous dimensions, and no summation 
over the index $i$ is implied in the first term. Summation
over repeated indices in the other terms is implied as usual.
The bare cuplings are the ones for which $t=0$, $g^i(0) \equiv g_0^i$.
The perturbative solution of (\ref{betacomplete}), order by order in 
a power series in $g^i$, is: 
\begin{itemize} 
\item{First~Order:} 
\begin{equation} 
g^i(t) =e^{y_i t}g^i(0).
\label{1ord}
\end{equation} 
\item{Second~Order} 
\begin{equation}
g^i(t)=e^{y_it}g^i(0) + \delta ^i_{jk}g^j(0)g^k(0)~,
\label{2ord}
\end{equation} 
with ${\dot \delta}^i_{jk} \equiv \frac{d}{dt}\delta^i_{jk} = 
\alpha^i_{jk}e^{y_jt}e^{y_kt} + y_i \delta^i_{jk}; \quad \delta^i_{jk}(0)=0$,
from which :
\begin{equation}
\delta^i_{jk}(t) =\left(e^{(y_j+ y_k)t} -e^{y_it}\right)
\frac{\alpha^i_{jk}}{y_j + y_k - y_i}
\label{2ordaux}
\end{equation} 

\end{itemize} 

\noindent and so on. Notice from the expression for the 
second order terms 
the resemblance of the 
anomalous-dimension denominators with 
``energy denominators'' in scattering amplitudes. As we shall discuss
below this is not a coincidence; it is a highly non-trivial 
property of string renormalization group to have a close connection
with string scattering amplitudes. We shall explain this through 
a simplified but quite instructive, and in many respects generic, excample,
that of an open Bosonic string in a tachyonic background~\cite{klebanov}.

\noindent {\it Open Strings in Tachyonic Backgrounds:~Weak Field Expansion} 

The $\sigma$-model action, for an open string propagating in 
flat space time in a tachyon background $T(X)$, is:
\begin{equation} 
{\cal S}_{\rm open} =\frac{1}{4\pi }\int dx dy~\eta^{\alpha\beta}
\partial_\alpha X^M \partial_\beta X^N~\eta_{MN} 
+ \int_{-\infty}^{+\infty} \frac{dx}{a} \int d^{26}k {\tilde T}(k)e^{ik_M X^M}
\label{opensmodel}
\end{equation}
where we work in units of $\alpha'=1$, and $a$ is a length scale, which 
will play the r\^ole of a short-distance cut-off scale.
Notice that the world-sheet is taken here to be the upper half plane
for simplicity. The open string interactions occur at the world-sheet
boundary, and this is expressed by the fact that the 
tachyonic background term is over the real $x$ axis. 

We apply the background field method for quantization, according to which 
we split the fields $X^M = X_0^M + \xi^M$, where $X_0^M$ satisfies the 
classical equations of motion, and varies slow with respect to the 
cut-off scale  $a$. The effective action is defined as 
$S_{\rm eff}[X_0] = -{\rm ln}W[X_0]$, 
where $W[X_0]$ is the partition function of the $\sigma$-model 
(\ref{opensmodel}): 
\begin{equation}
W[X_0]= \int {\cal D}\xi 
e^{-\frac{1}{4\pi}\int_{y>0} dx dy~\eta^{\alpha\beta}
\partial_\alpha X^M \partial_\beta X^N~\eta_{MN}}~
e^{-\int_{-\infty}^{+\infty}\frac{dx}{a}\int dk 
{\tilde T}(k)e^{ik \cdot X_0 }e^{ik \cdot \xi }}
\label{part}
\end{equation}
where $dk \equiv \frac{d^{26} k}{(2\pi)^{26}}$ is the target 
momentum space
integration. 
Using the free-feld contraction, with the scale $a$ 
as a short-distance regulator,
\begin{equation}
\langle \xi(x_1) \xi(\_2) \rangle _* = -2{\rm ln}\left(|x_1 - x_2| + a\right)
\label{freefield}
\end{equation}
where * denotes free-field $\sigma$-model action (in flat target space), 
and expanding the $\sigma$-model partition function 
in powers of ${\tilde T}(k)$, we obtain
the folowing results,
order by order in the weak-field (tachyon) expansion: 

\noindent {\bf Linear order in } ${\tilde T}(k)$: to this order,
the 
partition function $W[X_0]$ becomes:
\begin{eqnarray} 
&~& W[X_0]^{(1)} = -\int_{-\infty}^{+\infty}\frac{dx}{a} 
\int dk {\tilde T}(k) e^{i k \cdot X_0}\langle 
e^{i k \cdot \xi}\rangle _*  = \nonumber \\
&~& -\int_{-\infty}^{+\infty} dx \int dk a^{k^2 -1}{\tilde T}(k) 
e^{i k \cdot X_0}
\end{eqnarray} 
where we used the free-field contraction (\ref{freefield}). 
The scale $a$-dependence may be absorbed in a 
renormalization of the coupling ${\tilde T}(k)$ :
\begin{equation}
{\tilde T}_R(k) \equiv a^{k^2 -1}{\tilde T}(k)
\label{renormtach}
\end{equation}  
Comparing with (\ref{1ord}), we observe that one may identify
$a=e^{-t}$, $t$ the renormalization-group (RG) scale, from which 
one obtains the $\beta$-function:
\begin{equation} 
\beta^T(k) = -\frac{d {\tilde T}_R(k)}{d {\rm ln}a} = -(k^2 -1){\tilde T}_R(k)
\label{tachyonnormbeta}
\end{equation}
Comparison with (\ref{betacomplete}), then, indicates that the 
anomalous dimension is $k^2 -1$.

The conformal invariance conditions (\ref{betacond}) 
amount to the vanishing of the $\beta$-function, which thus turns out
to be equivalent to the {\it on-shell} condition for tachyons:
\begin{equation} 
-(k^2-1){\tilde T}_R(k) =0 \rightarrow k^2=1
\label{onshell}
\end{equation} 
This is the first important indication that the conformal invariance
conditions of the stringy $\sigma$-model imply important 
restrictions for the dynamics of the background over which 
it propagates.

Less trivial consequences for the background become apparent if one
examines the next order in the expansion in powers of ${\tilde T}(k)$.

\noindent {\bf Quadratic Order in } ${\tilde T}(k)$: 
to this order the partition function $W[X_0]$ reads: 
\begin{eqnarray} 
&~& W[X_0]^{(2)} = \int_{-\infty}^{+\infty}\frac{dx_1}{a}
\int_{-\infty}^{+\infty}\frac{dx_2}{a} \int dk_1 {\tilde T}(k_1)
\int dk_2 {\tilde T}(k_2) \cdot \nonumber \\ 
&~& e^{ik_1 \cdot X_0(x_1) + i k_2 \cdot X_0(x_2)}
\langle  e^{ik_1 \cdot \xi(x_1) + i k_2 \cdot \xi(x_2)} \rangle _* 
\end{eqnarray}
Since $X_0$ varies slowly, one may expands a l\'a Taylor:
$X_0(x_2) = X_0(x_1) + (x_2 - x_1)X_0'(x_1) + \dots \simeq X_0(x_1)$
to a good approximation. 

Implementing the free-field contraction (\ref{freefield}),
and performing straightforward algebraic manipulations,
we arrive at integrals of the form:
\begin{equation} 
a^{k_1^2 + k_2^2 -2} \int _{-\infty}^{x_1} 
dx_2 (x_1 - x_2 - a)^{2k_1 \cdot k_2}
=-a^{(k_1 + k_2)^2 -1}\frac{1}{2k_1 \cdot k_2}
\end{equation}
The integral converges for 
\begin{equation} 
2k_1 \cdot k_2 - 1 < 0
\label{conv}
\end{equation} 
Absorbing the 
scale dependence in renormalized tachyons, as before, one obtains
to second order:
\begin{equation} 
{\tilde T}_R(k) =a^{k^2-1}\left( {\tilde T}_R(k) + \int dk_1 \int dk_2 
\frac{{\tilde T}_R(k_1) {\tilde T}_R(k_2)}{2k_1 \cdot k_2 + 1}\delta^{(26)}(k_1 + k_2 - k) \right)
\label{secondorder}
\end{equation}
Comparing (\ref{2ord}), (\ref{2ordaux}) 
with (\ref{secondorder}), we observe that we are
missing the term $a^{-y_1 - y_2}=a^{k_1^2 + k_2^2 - 2}$. To the order we 
are working, this discrepancy can be justified as folows:
removing the cut-off, i.e. going to a non-trivial fixed point 
$t \to \infty$, and taking into account the convergence region
(\ref{conv}), which implies $y_1 + y_2 < y$, with $y=k^2 -1 $ the anomalous
dimension, we observe that in the regime $t \to \infty$ the 
missing term is negligible compared with the one which is present, and thus the above 
computation is consistent with the generic renormalization 
group analysis, near a non-trivial fixed point.
One then defines the $\beta$-functions of the theory, 
away from a fixed point (in the entire (target) momentum space) 
by analytic continuation. 

Comparing the above results with (\ref{2ord}),(\ref{2ordaux})
we then find that to second order:
\begin{eqnarray}
&~& 2 k_1\cdot k_2 + 1 = y_1 + y_2 - y, \qquad (y_i = 1 - k_i^2), \nonumber \\
&~& \alpha^{k}_{k_1k_2}=-\delta^{(26)}(k_1 + k_2 - k)
\label{secondordresult}
\end{eqnarray}

\begin{figure}[htb]
\epsfxsize=4in
\begin{center}
\epsffile{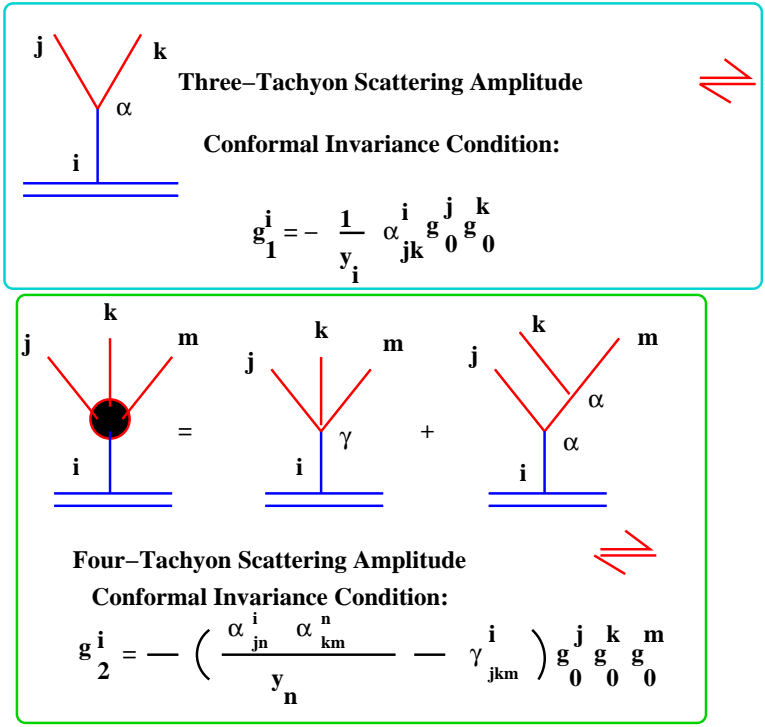}
\end{center}
\caption[]{Schematic representation of the equivalence of 
conformal invariance conditions (vanishing of world-sheet
renormalization group $\beta$-functions) and on-shell
string scattering amplitudes in the case of an open string 
in a tachyonic background.}
\label{fig:tachyonamplitude}
\end{figure}

The corresponding conformal invariance condition can be found by 
iterating the one at previous order as follows:
the first order result yields $y_i g_0^i =0$; to second order
we write for the coupling $g^i = g_0^i + g_1^i$, which then,
on account of the vanishing of the $\beta$-function
$\beta^i = y_i g^i + \alpha^i_{jk}g^j g^k + \dots =0$, yields:
\begin{equation} 
g_1 ^i = -\frac{1}{y_i}\alpha^i_{jk}g_0^j g_0^k 
\end{equation} 
The situation is depicted in figure \ref{fig:tachyonamplitude}.
It represents  a three-tachyon scattering amplitude, with two external
legs set on-shell, and with one propagator pole at $y_i=0$.
If one sets this third leg on shell two, then the residue of the pole is 
the three-on-shell tachyon scattering amplitude. 

\noindent {\bf Higher orders in} ${\tilde T}_R(k)$: at the next level
one obtains a highly non-trivial demonstration of the 
above-mentioned equivalence between conformal invariance conditions
and on-shell $S$-matrix elements. We shall not give details here,
as these can be found 
by the interested reader in the literature~\cite{klebanov}.
Below we shall only outline the results. 
Schematically the situation is depicted in fig. \ref{fig:tachyonamplitude}.

Following a similar treatment as before, but encountering 
signficantly more complex mathematical manipulations, 
one obtains as a solution of the conformal invariance 
conditions to this order:
\begin{eqnarray} 
&~& g_2^i = \frac{1}{y_i} \left( \frac{\alpha^i_{jm}\alpha^m_{k\ell}}{y_m}
- \gamma^i_{jk\ell}\right)g_0^j g_0^k g_0^\ell~, \nonumber \\
&~& \gamma^i_{jk\ell} g_0^j g_0^k g_0^\ell = 
\left( -{\cal D}^i_{jk\ell} +  \frac{2\alpha^i_{jm}\alpha^m_{k\ell}}{y_j + 
y_m - y_i}\right)g_0^j g_0^k g_0^\ell 
\end{eqnarray}
where, in the tachyonic background open string case, the {\it contact}
terms of the graph are:
${\cal D}^k_{k_1k_2k_3}= \frac{\delta^{(26)}(k_1 + k_2 + k_3 - k)}{1 + B + C}$
$~_3F_2(1,~ -1-B-C,~ -C,~-1-A-B-C,~-B-C;~1)$, with $~_3F_2$ denoting
a hypergeometric function, and 
$A=2k_1 \cdot k_2$, $B=2k_1 \cdot k_3$, $C=2k_1 \cdot k_3$,
and $2+A+B+C =y_j + y_k + y_\ell - y_i$, with $y_i$ the anomalous dimensions
defined above.

This completes the demonstration on the equivalence of the conformal 
invariance conditions of a stringy $\sigma$-model with string 
scattering amplitudes. As we have discussed above such amplitudes can be 
reproduced by a target space diffeomorphism invariant effective action. 
The form of this action can be most easily obtained if one 
folows the second method 
of perturbative expansion for computing the $\beta$-functions,
the so-called Regge-slope $\alpha'$ expansion, which from now 
on we shall restrict ourselves upon. For simplicity, 
in these lectures we shall restrict ourselves to ${\cal O}(\alpha ')$ 
in this expansion. This will be sufficient for our cosmological 
considerations. Some comments on higher orders will be made where appropriate.

\subsubsection{Regge-slope ($\alpha'$) expansion: 
${\cal O}(\alpha')$-Weyl anomaly 
Coefficients}

The second method of perturbative $\sigma$-model expansion, 
which we shall make use of in the context of the present lectures, 
consists of expanding the partition function, correlation functions
and $\beta$-functions in powers of $\alpha'$, or rather in
the dimensionless quantity $\alpha ' k^2$, where $k^M$ is a 
target momentum contravariant vector (for open strings 
the expansion is actually made in powers of $\sqrt{\alpha'} k$).
The Regge slope $\alpha'$-expansion is independent of 
the $g^i$-expansion, studied above, but formally 
it is equivalent to that, in the sense that the exact 
expressions (resummed to all orders) of the pertinent $\sigma$-model 
partition function in both expansion methods contain the same physical 
information. In practice,
the $\alpha'$ expansion is appropriate if one is interested, as we are 
in the cosmological context of these lectures, in 
long-wavelength (compared
to Planck scales) effective actions. In such a case the first few orders
in the $\alpha'$ expansion (actually up to and including ${\cal O}(\alpha')$ )
will suffice to provide an adequate description of the observed Universe,
as we shall discuss in Lecture 2. 

In these lectures we shall not discuss in detail the 
very intersting techniques underlying the $\alpha'$-expansion 
of $\sigma$-model renormalization-group analysis. The interested
reader may find details on this in the vast literature~\cite{green}. 
For our purposes here, we shall merely quote the results 
for the ${\cal O}(\alpha')$ Weyl anomaly coefficients
for Bosonic (or better the bosonic part of) $\sigma$-model backgrounds
of graviton, antisymmetric tensor and dilaton fields. 

For such backgrounds in the Bosonic string case 
(for definiteness) we have: 

\begin{itemize} 

\item{{\bf Graviton:}} For the Weyl anomaly coeffcient of the 
graviton background one has: 
\begin{equation} 
{\widehat \beta}^g_{MN} = \alpha' \left(
R_{MN} - \frac{1}{4}H_{M}^{PQ}H_{NPQ} + 2\nabla_{(M}\partial_{N)} \Phi
\right)
\label{gravwac}
\end{equation} 
where the last part (depending on $\Phi$) may be attributed to the 
differomorphism $\delta g^i$ part of the Weyl anomaly coefficient.

\item{{\bf Antisymmetric Tensor:}} For the antisymmetric tensor backgrounds
one finds: 
\begin{equation} 
{\widehat \beta}^B_{MN} = \frac{\alpha'}{2} \left(
-\nabla_P H^P_{MN} + 2(\partial _P \Phi)H^P_{MN} \right)
\label{antisymwac}
\end{equation} 
where again the dilaton ($\Phi$) dependent part is attributed to 
target-space diffeomorphism parts.

\item{{\bf Dilaton Fields:}} For dilaton fields it is convenient, for 
reasons that will become clear below, to define 
a Weyl anomaly coefficient with the  (target-space) gravitational 
trace of graviton Weyl anomaly coefficient subtracted:
\begin{eqnarray}
&~& {\tilde {\widehat \beta}}^\Phi = {\widehat \beta}^\Phi - 
\frac{1}{4}g^{MN}\beta_{MN}^g
= \nonumber \\
&~& \frac{\alpha'}{4} \left(-4 (\partial_M \Phi)^2 + 4 \nabla^2 \Phi + R - \frac{1}{12} 
H_{MNP}^2 - \frac{2(D-26)}{3\alpha'} \right)
\label{dilatonwac}
\end{eqnarray} 

Notice in the last expression (\ref{dilatonwac}) that the 
appearance of the scalar curvature is an exclusive consequence of 
the presence of the trace of the graviton Weyl anomaly coefficient
in ${\tilde {\widehat \beta}}^\Phi$. The dilaton Weyl anomaly
coefficient, to ${\cal O}(\alpha ')$ \emph{does not depend} on
the target-space curvature, only on derivatives of the dilaton field. 
Moreover, we also notice that to zeroth order in $\alpha'$, the 
dilaton Weyl anomaly coefficient does depend on the 
conformal anomaly $D-26$, which is absent for critical dimension 
strings. This term, if present, would act as an exponential
dilaton potential (or equivalently vacuum energy ). 
In the critical dimension $D_c=26$ (for bosonic strings) 
is absent. 
We shall come back to this important issue in our 
third lecture, when we discuss the issue of cosmological constant
in the context of string theory. For superstrings the 
$D-26$ term is replaced by $D-10$, and the vacuum energy term
is absent for the case of critical superstring space-time 
dimension $D_c=10$.

\end{itemize} 

We now notice that, as can be shown straightforwardly, the 
{\bf vanising} of the above 
expressions (i.e. the conformal invariance conditions (\ref{diffbeta})
for this set of background fields) corresponds to 
{\bf equations of motion} of a low-energy ${\cal O}(\alpha ')$ 
target-space effective action:
\begin{equation} 
I_{\rm eff} = -\frac{1}{2\kappa^2}\int d^DX \sqrt{g} e^{-2\Phi} 
\left( R + 4(\partial_M \Phi)^2 - \frac{1}{12}H_{MNP}^2 
- \frac{2(D-26)}{3\alpha'} + \dots \right)
\label{sea}
\end{equation} 
where $\kappa^2$ is the Gravitational constant 
in $D$ target space time dimensions
(related appropriately to 
the Planck (or string) mass scale $M_s$ ).   

In fact, as mentioned earlier in the context of 
$g^i$ weak field expansion, it can also be shown
explicitly within the $\alpha '$ expansion~\cite{osborn}, 
that the above-mentioned Weyl anomaly coefficients 
${\widehat \beta}^i$ are gradient flows in $g^i$ space
of $I_{\rm eff}$:
\begin{equation}
     \frac{\delta I_{\rm eff}}{\delta g^i}={\cal G}_{ij}{\widehat \beta}^j
\end{equation}
where, up to appropriate field redefinitions, which are irrelevant 
from the point of view of scattering amplitudes, as they leave them 
invariant,
the 
function ${\cal G}_{ij}$ coincides with the 
Zamolodchikov metric (\ref{zammetric}).  

\subsubsection{World-Sheet Renormalizability Constraints on the ${\widehat \beta}$-functions}

The world-sheet renormalizability of the $\sigma$-model 
action, deformed by background fields $g^i$, i.e. the fact that this
two-dimensional theory has ultraviolet divergencies which can be 
absorbed in appropriate redefinition of its coupling/fields $g^i$, 
without the necessity for introducing new types of interactions
that do not exist in the bare theory, 
is expressed simply in terms of the renormnalization-group 
scale invariance of the components of the world-sheet 
stress tensor of the theory:
\begin{equation}
\frac{d}{d{\rm ln}\mu}T_{\alpha\beta}=0~, \qquad \alpha,\beta =\sigma, \tau
\label{tracerginv} 
\end{equation}
where ${\rm ln}\mu$ is the renormalization-group scale.

Equations of the type (\ref{tracerginv}) 
implies severe constraints
among the ${\widehat \beta}$-functions which,
after some elegant $\sigma$-model 
renormalization-group analysis,  
are expressed 
by means of the 
Curci-Paffuti equation~\cite{curci}. To order $\alpha '$
this equation reads: 
\begin{equation} 
\nabla_N {\widehat \beta}^\Phi = 2g^{MP}e^{-2\Phi}\nabla_N \left(e^{-2\Phi}{\widehat \beta}^g_{MP} \right) + {\cal O}({\widehat \beta}^B)
\label{cpeq}
\end{equation}
An immediate consequence of this equation is that 
{\it not all} of the ${\widehat \beta}^i=0$ equations are independent.
In particular, at a fixed point of the renormalization-group 
on the world-sheet, for which ${\widehat \beta}_{MN}^g 
= {\widehat \beta}_{MN}^B =  0$, one obtains from (\ref{cpeq}) 
that the dilaton Weyl anomaly coefficient is constant, not necessarily 
zero. In the particular case of strings in Bosonic massless backgrounds, 
for instance, this constant is simply the conformal anomaly 
$D-26$ (Bosonic Strings) or $D-10$ (Superstrings).

When discussing equations of motion the Curci-Paffuti constraint
(\ref{cpeq}) should always be taken into account. Although the constraint
may seem trivial in case one is interested in solutions 
of the confornal invariance conditions (\ref{diffbeta}), 
${\widehat \beta}^i=0$, this is not the case when one encounters non-trivial
${\widehat \beta}^i \ne 0$  away from the fixed points of the renormalization
group on the world sheet. Such a situation (non-critical Strings) 
may be of interest in 
non-equilibrium cosmological situations, and we shall discuss it briefly
in Lecture 3.

\subsubsection{A note about ``Frames''}

The action (\ref{sea}) is derived in the so-called $\sigma$-model frame,
because it is derived directly from expressions
obtained in $\sigma$-model renormalization-group ananlysis. 
Such a terminology {\it should not be confused} with the 
general coordinate frames in general relativity. 
The terminology ``frame'' here is used to mean a 
given background metric configuration. 
In string theory, the perturbative string S-matrix elements
(scattering amplitudes) are invariant under {\it local
redefnition of the background fields} $g^i$ (``equivalence
theorem''), which simply
corresponds to a particular renormalization-group scheme choice. 

The $\sigma$-model frame metric corresponds to one such 
configuration. One may {\it redefine} the metric field
so as to pass to an effective action, where the curvature
scalar term in the action will have the 
standard (from the point of view of General relativity) 
coefficient $1/\kappa^2$, 
{\it without} the dilaton conformal factor $e^{-2\Phi}$ 
in front. In other words, it will have the canonically
normalized Einstein action form. 
Such a ``frame'', termed {\it Einstein } (or ``physical'') {\it frame},
is obtained upon redefining the $\sigma$-model background metric as
follows (the superscript $E$ denotes quantities in the Einstein frame):
\begin{equation}
g_{MN} \rightarrow g_{MN}^E = e^{-\frac{4}{D-2}\Phi }g_{MN} 
\label{einstein}
\end{equation}
In this frame, then, the effective action (\ref{sea}) 
acquires, as mentioned already, its canonical Einstein form, as far as 
the gravitational parts are concerned:
\begin{eqnarray} 
&~& I_{\rm eff}^E = -\frac{1}{2\kappa^2}\int d^DX \sqrt{g^E} 
\left( R^E - \frac{4}{D-2}(\partial_M \Phi)^2  
- \right. \nonumber \\ 
&~& \left. \frac{1}{12}e^{-\frac{8}{D-2}\Phi}H_{MNP}^2 - 
e^{\frac{4}{D-2}\Phi}\frac{2(D-26)}{3\alpha'} 
+ \dots \right)
\label{esea}
\end{eqnarray} 
where the $\dots$ denote higher-order terms, as well as other fields,
such as 
gauge-boson terms (in the case of heterotic string) {\it etc}.
Notice the change of relative sign between the curvature and dilaton 
kinetic terms in the Einstein frame. 

From the point of view of discussing physical low-energy applications
of string theory, such as cosmological models based on strings, 
the Einstein frame is the ``physical'' one, where the astrophysical
observations are made. This will always be understood
when we discuss string cosmology in Lectures 2 and 3. 

\subsubsection{Higher orders in $\alpha'$ }

Corrections to General Relativity occur at the next order in $\alpha '$,
at which one can show, for instance, that the graviton 
$\beta$-function 
has the form (ignoring the contributions from 
other backgrounds for simplicity): 
\begin{equation}
\beta_{MN}^g (X^P) = -\alpha ' \left( R_{MN} + \frac{\alpha '}{2}
R_{MKLP}{R_N}^{KLP}\right) 
\label{higherorder}
\end{equation} 
The higher-curvature terms will result in corrections to the 
Einstein term in the target-space effective action. 
Such action terms 
have some ambiguities concerning their coefficients, since the scattering
$S$-matrix elements one derives from an effective action 
correspond to more than one set of these coefficients
(the equivalence theorem, mentioned earlier in the Lecture). 
The amplitudes are invariant under ocal redefinitions 
of the graviton field (in this case): $g_{MN} \to g_{MN} + c \alpha' R_{MN}$,
where $c$ a constant coefficient. 
Such redefinitions  affect the higher order in $\alpha '$ 
terms of the target-space effective action, in such a way that one 
can always cast it in the Gauss-Bonnet
(gravitational ghost-free) combination: 
\begin{equation} 
{\cal S}=\int d^Dx \sqrt{g} \left(R + \alpha \alpha' \left(R_{MNRP}^2 
- 4R_{MN}^2 + R^2 \right) + \dots \right)
\label{gaussbonnet}
\end{equation} 
where the coefficient $\alpha$ is determined 
by comparison with string tree amplitudes. 
It is found to be: $\alpha=1/4$ (Bosonic String), $\alpha =0$ 
(Superstring type II), and $\alpha=1/8$ (Heterotic string). 

The fact that stringy higher-order corrections 
to the low-energy effective actions of string theories are free from 
gravitational ghosts, in the sense that the effective action can always
be cast, under local field redefinitions, in the ghost-free
Gauss-Bonnet combination, is consistent with the unitarity of 
the underlying string theory. 

The higher-order corrections to Einstein's general relativity 
are in principle an infinite series of terms, which 
become stronger at high energies (short distances).
From a cosmological viewpoint, 
the higher-curvature terms may thus have effects 
at very early stages of our Universe, but such effects
are negligible at redshifts $z \sim 1 $ and lower, 
where we shall concetrate most of our discussion 
in these lectures. One should notice that the 
presence of higher-curvature correction terms 
of Einstein's general relativity leads some times
to highly non-trivial effects. For instance, one may have 
black hole solutions with (secondary) dilaton hair~\cite{kanti} 
in such models, which do not exist in standard Einstein's 
relativity. Such objects may play a r\^ole in the Early Universe.

\section{Lecture 2: String Cosmology} 

\subsection{An Expanding Universe in String Theory and the 
r\^ole of the Dilaton Background}

As has already been discussed in the cosmology lectures 
in this School,
the Observed Universe is, to a good approximation, {\it homogeneous}
and {\it isotropic}. From the point of view of string theory,
therefore, one is interested in describing the propagation 
of strings in such homogeneous backgrounds, i.e. space-time 
geometries whose metric tensors depend {\it only on time}, and thus
have {\it no spatial} dependence. 

As we have discussed in the previous lecture, conformal invariance
conditions (\ref{diffbeta}) 
of the associated $\sigma$-model will imply target-space 
equations of motion for the background fields, 
which will determine the dynamics. This is, in general terms, what 
{\it String Cosmology} is about. The pertinent dynamics will be described
by means of string effective actions for the various (time dependent
only) modes. Of course, this is a first order approximation.
{\it Spatial Inhomogeneities} can be incorporated by allowing spatial 
dependence of the various $\sigma$-model couplings/background fields. 

It is the purpose of this part of the lectures to discuss how one can 
incorporate {\it expanding Universe} scenaria in the above string context. 
We shall start with the simplest scenario, that of a {\it linearly 
expanding non-accelerating Universe}. Subsequently we shall discuss
more complicated models, including {\it inflationary scenaria} 
in string theory, and mechanisms for {\it graceful exit} from it.
Due to lack of time, the discussion 
will be relatively brief. For more details,
the interested reader will be referred to the literature,
which is vast, and still growing. 
In the last two lectures I will try 
to give whatever details, and technical aspects, I believe are essential 
for introducing the layman into the subject of string cosmology 
and make him/her understand 
the various subtleties involved. It should be stressed that 
string cosmology is not a physically well established subject,
and part of the third lecture will be devoted to discussing 
open issues, motivated by recent astrophysical observations
on the possibility of a currently accelerating phase of the Unvierse, 
and the way such issues can be tackled in the 
framework of string theory.

In all stringy cosmological scenaria of expanding Universes
that we shall examine here the {\it dilaton} plays a crucial r\^ole,
as being directly responsible for providing consistent
time-dependent backgrounds in string theory. This is an
important feature which differentiates string cosmology from 
conventional one (this feature is, of course, 
in addition to the fact that
the target-space dimensionality of string theory is higher than four).

We commence our discussion by considering the 
$\sigma$-model action of a string propagating 
in time-dependent backgrounds of graviton
$g_{MN}$, antisymmetric tensor $B_{MN}$ and 
dilatons $\Phi$. Although given in Lecture 1,
for completeness we give again the action
explicitly (in this subsection 
we work in units of $\alpha '=2$ (closed strings) for convenience, and 
we follow the normalization of \cite{aben} for the dilaton field,
which implies that the dilaton field here equals twice the dilaton 
field in the previous section):
\begin{eqnarray} 
&~& {\cal S}_\sigma = \int _\Sigma \frac{1}{4\pi}
d^2\sigma \left( \sqrt{\gamma} \gamma^{\alpha\beta}g_{MN}(X^0) \partial_\alpha X^M \partial_\beta
X^N +  \right. \nonumber \\
&~& \left. B_{MN}(X^0) \epsilon^{\alpha\beta}
\partial_\alpha X^M \partial_\beta X^N + 
\frac{\sqrt{\gamma}}{2}\Phi (X^0) R^{(2)} \right)
\label{smodelcosmol}
\end{eqnarray} 
where $M,N = 0, \dots, D-1$, 
and $X^0$ denotes the target time. The reader is required
to remember that the dilaton coupling is of one order in $\alpha '$ higher
than the rest of the terms in (\ref{smodelcosmol}). 
As already mentioned, 
the time dependence of the backgrounds is appropriate for a discussion of 
isotropic and homogneoeus cosmological solutions
of the conformal invariance conditions (\ref{diffbeta}), which we now turn to.

\subsubsection{Linear Dilaton Background Conformal Field Theory} 

Consider the $\sigma$-model background~\cite{aben}:
\begin{eqnarray} 
g_{MN} = \eta_{mn},~ B_{MN}=0,~ \Phi = -2QX^0~, \qquad Q={\rm const} 
\label{ldb}
\end{eqnarray}
in which the dilaton is growing linearly with the target time. 
We observe that this is an exact solution of the $\sigma$-model 
conformal invariance conditions (\ref{diffbeta}), (\ref{cpeq}), 
for the Weyl anomaly coefficients, which, for the problem at hand, 
and to ${\cal O}(\alpha ')$ 
are 
given by (\ref{gravwac}), (\ref{antisymwac}),
(\ref{dilatonwac}). 
Hence it is an acceptable background in string theory. 

Let us describe the basic features of this conformal theory. We wish to determine first the central charge (conformal anomaly). To this end we need to 
compute the world-sheet stress tensor~\cite{green}.  
As we have discussed in Lecture 1, the latter is defined
by the response (\ref{stress}) of the world-sheet action (\ref{smodelcosmol}) 
to a variation of the world-sheet metric.
The presence of the dilaton term results in the following form:
 \begin{equation} 
T_{zz} = -\frac{1}{2}\partial_z X^M \partial_z X^N g_{MN} (X^0) + 
Q\partial_z^2 X^0 
\label{stresscosmol} 
\end{equation} 
where $z$ is the complexified world-sheet coordinate (we work in a 
Euclidean world-sheet, appropriate for the convergence of the 
$\sigma$-model path integral formalism we adopt here). 

\noindent {\bf [ NB4}: {\it For completeness we sketch below the derivation of 
the $Q\partial_z^2 X^0$  term in (\ref{stresscosmol}). This comes
from varying the world-sheet curvature/dilaton term with respect to 
the world-sheet metric, and setting at the end $\gamma^{\alpha\beta}=
\delta^{\alpha\beta}$ (for Euclidean world sheets):
$\frac{\delta }{\delta \gamma^{\alpha\beta}}\int _\Sigma R^{(2)} 
Q X^0(\sigma,\tau) $. Noticing that only contributions from the second derivatives of the world-sheet metric in $R^{(2)}$ survive this procedure, 
we obtain: $\int _\Sigma d^2\sigma ' 
\sqrt{\gamma} \partial_z^2 \delta^{(2)} (\sigma - \sigma') Q t (\sigma ')
= -\int _\Sigma \delta^{(2)} (\sigma- \sigma') Q \partial_{z'} t = -Q\partial_z^2 t $, where partial integration 
has been made in order to arrive at the last equality} {\bf ]}.  

From (\ref{stresscosmol}) it is straightforward to 
compute the conformal anomaly $c$. From basic 
conformal field theory we recall that the latter is given by~\cite{green}):
\begin{equation} 
{\rm lim}_{z \to 0} 2z^4 \langle T_{zz}(\sigma) T_{zz}(0) \rangle = c
\label{confanom}
\end{equation} 
Regulating the short-distance behaviour of the theory
by replacing $z \to 0$ by $z \to a$, where $a$ is a 
short-distance cutoff scale, and 
using the free-field contractions for two-point correlators on the 
world-sheet (\ref{freefield}) (with $\xi (x_1)  \leftarrow\rightarrow 
X^M (\sigma)$ ) it is straightforward to derive~\cite{aben}:
\begin{equation} 
c = D - 12 Q^2 
\label{ldca}
\end{equation}
where $D$ is the dimensionality of the target-space time. 

This is an important result. In the conformal field theory 
of 
a non-trivial linear dilaton background, and flat $\sigma$-model
target spacetime, the conformal anomaly is {\it no longer} 
given by the target-space dimensionality $D$ alone, which was the 
case of Minkowski space times, as we have seen in Lecture 1. 

The cancellation of the Weyl anomaly implies $c=26$ (for bosonic strings,
which we restrict ourselves from now on for definiteness, unless
otherwise stated). This, therefore, means that the critical
dimensionality of the string is $D > 26$. This string theory
is termed {\it supercritical}~\cite{aben}. 

The non-trivial issue is to demonstrate the mathematical consistency 
of such string theories, by demonstrating {\it unitarity} 
of the physical spectrum, and {\it modular invariance}, associated
with string loops~\cite{green}. We note that 
both of these properties have been shown to be valid
for the linear dilaton background (\ref{ldb}). 
We shall not demonstrate them here due to 
lack of time. We refer the interested reader to the literature~\cite{aben}
for a detailed verification. 

We next proceed to discuss the target-space time interpretation 
of the linear dilaton background (\ref{ldb}).
As we have mentioned previously, the `physical' metric, appropriate for cosmological considerations in theories with non-trivial dilaton fields, 
is provided by the Einstein frame target-space metric tensor (\ref{einstein}), 
corespondiong to  a low-energy effective action (\ref{esea})
with canonically-normalized Einstein curvature term, without dilaton 
conformal factors. For the background (\ref{ldb}), therefore, 
the Einstein-metric invariant line element reads:
\begin{equation} 
ds_E^2 = e^{\frac{4QX^0}{D-2}}\eta_{MN}dX^M dX^N 
\label{eie}
\end{equation} 
Upon redefining the time $X^0 \to t$:
\begin{equation} 
       t = \frac{D-2}{2Q}e^{\frac{2Q}{D-2}X^0} 
\label{phystime}
\end{equation}
we observe that the Einstein (``physical'') metric may be 
cast into a Robertson-Walker (RW) 
form~\cite{aben}:
\begin{equation} 
ds_E^2 = -(dt)^2 + t^2 dX^i dX^j \delta_{ij} 
\label{rwform}
\end{equation}
with a {\it linearly expanding} in time $t$, {\it non accelerating} 
scale factor $a(t) = t, \quad {\ddot a(t)} \equiv d^2 a(t)/dt^2 = 0$.
The RW Universe (\ref{rwform}) has {\it zero spatial curvature}, i.e. is 
{\it flat}.

In these coordinates the dilaton field has a logarithmic dependence
on time:
\begin{equation} 
   \Phi (t) = (2-D){\rm ln}\frac{2Qt}{D-2}
\label{logdil} 
\end{equation}

One may accommodate more general RW backgrounds 
with non-trivial spatial curvature in the above framework, 
by including non-trivial antisymmetric tensor backgrounds~\cite{aben}.
This is what we shall discuss below.

\subsubsection{The antisymmetric tensor field and 
More General Cosmological Backgrounds}

First of all we 
concentrate our attention to (\ref{ldca}). We assume  
that in our model there are $d=4$ ``large'' (non-compact) target-space 
time dimensions, one of which is the Minkowski time.  
The rest of the target dimensions (6 in the case of superstring, or 22
in the case of Bosonic strings) are replaced by
an appropriate ``internal'' conformal field theory with 
a central charge $c_I$:
\begin{equation} 
 c = d + c_I - 12Q^2 = 4 + c_I - 12Q^2 
\label{compactca}
\end{equation} 
Notice that the total central charge $c$ is required to equal 
its critical value (26 for Bosonic strings, 10 for superstrings)  
so as to ensure target-space diffeomorphism invariance (i.e. to cancel 
the Fadeedv-Popov reparametrization ghost contributions to the 
conformal anomaly), and also conformal invariance of the complete theory. 
These two requirements are essential for giving string theory a
 space-time interpretation. Then (\ref{compactca}) 
leads to (for Bosonic strings for brevity):
\begin{equation}
c_I = 22 + 12Q^2 \equiv 22 + \delta c
\label{deficit} 
\end{equation} 
where $\delta c$ is known as the {\it central charge deficit}. For a critical
dimension string theory, $\delta c =0$.

From now on we shall ignore the details of the compact internal theory,
and simply assume it is there to ensure the above properties, and hence
consistency, of the string theory at hand. One can show that
non-trivial internal conformal field theories can indeed be constructed
with the desired properties~\cite{aben}. 
We therefore consider $d=4$-dimensional backgrounds $g_{\mu\nu}(x)$,
$B_{\mu\nu}(x)$, $\Phi (x)$, where $\mu, \nu = 0, \dots 3$, and $x^\mu$ are
four-dimensional spacetime coordinates. 

In four space time dimensions the 
antisymmetric tensor field strength 
may be written in terms of a pseudoscalar field $b(x)$ 
to be identified with the axion field~\cite{green,aben}:
\begin{equation} 
H_{\mu\nu}^\lambda  = e^{2\Phi}\epsilon^\lambda_{\mu\nu\rho}\partial^\rho b 
\label{axion}
\end{equation} 
The conformal invariance conditions (\ref{diffbeta}), then, 
corresponding to the 
four-dimensional Weyl anomaly coefficients (\ref{gravwac}),
(\ref{antisymwac}) read:
\begin{eqnarray} 
&~& {\rm graviton:} \qquad R_{\mu\nu} = \frac{1}{2}
\partial_\mu \Phi \partial_\nu \Phi + \frac{1}{2}g_{\mu\nu} 
\nabla^2 \Phi + 
\frac{1}{2}e^{2\Phi}[\partial_\mu b \partial_\nu 
b- g_{\mu\nu} (\partial b)^2 ], \nonumber \\
&~& {\rm antisym.~tensor:} \qquad 
\nabla^2 b + 2 \partial_\lambda b \partial^\lambda \Phi = 0 
\label{eqsgravantis}
\end{eqnarray}
The fact that the total central charge is 26 (for Bosonic strings, or 
10 for superstrings) implies the dilaton equation~\cite{aben}:
\begin{eqnarray} 
\delta c = 12 Q^2 = -3e^{-\Phi}\left[ -R + \nabla^2 \Phi 
+ \frac{1}{2} (\partial \Phi)^2 - \frac{1}{2}e^{2\Phi}(\partial b)^2 \right]
\label{dileq}
\end{eqnarray} 
(in units $\alpha ' =1$), where $\delta c$ is defined in (\ref{deficit}). 
We stress again that in the case of critical strings
$\delta c$ would vanish. Here, as a result of the Bianchi identity 
$\nabla^\mu R_{\mu\nu} =\frac{1}{2}\nabla_\nu R$, the equations 
(\ref{eqsgravantis}) imply the consistency of 
(\ref{dileq}), i.e. that the right-hand-side is a constant,
consistent with $\delta c$=const. This consistency is nothing other
than the Curci-Paffuti equation (\ref{cpeq}), stemming from 
renormalizability of the world-sheet $\sigma$-model theory. 

The four-dimensional effective low-energy action 
obtained from (\ref{eqsgravantis}),(\ref{dileq}), in the 
Einstein frame, is:
\begin{equation} 
I_{\rm eff} = \int d^4x \sqrt{-g^E}\left[R -\frac{1}{2}(\partial \Phi)^2
-\frac{1}{2}e^{2\Phi}(\partial b)^2 -\frac{1}{3}e^\Phi\delta c \right]
\label{effaction}
\end{equation}
Note that, as a result of the non-trivial central-charge deficit
$\delta c \ne 0$, there is a non-vanishing  
potential for the ``internal'' fields, which implies a 
{\it non-trivial }  
vacuum energy term for a four-dimensional observer. 

We note now that 
the linear dilaton background (\ref{ldb}) 
is indeed a special case of the equations (\ref{eqsgravantis}),(\ref{dileq}),
leading in the Einstein frame, to spatially flat RW linearly expanding 
Universes,as we have seen above, .
For non trivial axion fields $b$ one has more general RW backgrounds,
with spatial curvature. Indeed,
it can be shown that the equations (\ref{eqsgravantis}),(\ref{dileq})
admit as solution~\cite{aben} 
\begin{equation} 
ds_E^2 = -dt^2 + a(t)^2 {\tilde g}_{ij} dx^i dx^j~, i,j =1,2,3 
\label{rwspatialcurv}
\end{equation} 
where ${\tilde g}_{ij}$ is a three-dimensional maximally symmetric 
metric:
\begin{equation} 
{\tilde g}_{ij}dx^i dx^j = \frac{dr^2}{1 - kr^2} + r^2 (d\theta^2 + {\rm sin}^2\theta 
d\phi^2)
\label{scrw}
\end{equation}
where $t$ is the phhysical time (\ref{phystime}), and 
the RW parameter $k$, related to spatial curvature,
is to de determined below.

The Hubble parameter is given by:  
$H(t) \equiv \frac{{\dot a}(t)}{a(t)}$, with the dot denoting 
derivative w.r.t. $t$. With the ansatz (\ref{rwspatialcurv}), (\ref{scrw})
the antisymmetric tensor/axion equation in (\ref{eqsgravantis})
is solved by~\cite{aben}:
\begin{equation} 
{\dot b}= b_0\frac{e^{-2\Phi}}{a(t)^3}~, \qquad b_0={\rm const.}
\label{axioneq}
\end{equation} 
and the dilaton equation (\ref{dileq}) implies for the central-charge deficit:
\begin{equation} 
\delta c = 6e^{-\Phi}\left( {\dot H} + 3H^2 + \frac{2k}{a(t)^2}\right)
\label{ccd}
\end{equation}
The graviton equations have in principle two independent
components:
\begin{eqnarray} 
&~& \mu=\nu=t: \qquad -6({\dot H} + H^2)={\ddot \Phi} + 3H{\dot \Phi}
+ ({\dot \Phi})^2~, \nonumber \\
&~& \mu, \nu= i, j : -2({\dot H} + 3H^2 + \frac{2k}{a(t)^2})
= {\ddot \Phi} + 3H {\dot \Phi} - \left(\frac{b_0^2}{a(t)^6}\right)e^{-2\Phi}
\label{geq}
\end{eqnarray} 
However, since the dilaton equation (\ref{dileq}) is an identity
(up to an irrelevant constant), one observes actually that there is 
only one independent equation for the graviton. 
Indeed, solving (\ref{ccd}) for the dilaton and substituting into 
(\ref{geq}), and subtracting these two equations we obtain:
\begin{eqnarray} 
&~& \left(\frac{{\ddot H} + 6 {\dot H}H - (4k/a(t)^2)H}{{\dot H}
+ 3H^2 + 2k/a(t)^2}\right)^2= \nonumber \\ 
&~& -4{\dot H} + \frac{4k}{a(t)^2}
-\frac{(\delta c)^2b_0^2}{36a(t)^6}\cdot \frac{1}{\left({\dot H}
+ 3H^2 + (2k/a(t)^2)\right)^2}
\label{finaleq}
\end{eqnarray}
This equation can in principle be solved, yielding the Hubble 
parameter for the string Universe under consideration. 

\noindent {\bf Asymptotic Solutions of (\ref{finaleq}):}
There are two kinds of asymptotic solutions, of (\ref{finaleq}),
which can be obtained analytically: 

\begin{itemize} 

\item{(I)} $H \to 0$, as $t \to +\infty$, $\Phi =\phi_0=$constant, 
${\dot b}=b_0e^{-2\phi_0}$ and the space curvature 
obeys
\begin{equation} 
         k = \frac{1}{4}b_0^2 e^{-2\phi_0} \ge 0
\label{nonnegativek}
\end{equation}
and thus is non negative.  This Universe is therefore {\it closed}.
The central charge deficit, in this case is determined 
via the dilaton equation (\ref{dileq}) to be : $\delta c = 
12e^{-\phi_0}k$. This asymptotic Universe is therefore 
a {\it static Einstein Universe} with non-negative spatial curvature.

\item{(II)} A linearly expanding Universe $a(t)=t$ 
with metric: 
\begin{equation} 
ds_E^2 = -(dt)^2 + t^2\left[\frac{dr^2}{1-kr^2}
+ r^2 (d\theta^2 + {\rm sin}^2\theta d\phi^2)\right]
\end{equation}
with Hubble parameter relaxing to zero as $t \to \infty$ 
as: $ H(t) \sim 1/t$, and hence one has: 
\begin{equation}
\Phi = -2{\rm ln}t + \phi_0~, \qquad b=2e^{-\phi_0}\sqrt{k}t 
\end{equation}
with $k$ again non-negative. The four-dimensional curvature
is $R=6(1 + k)/t^2$, and the central charge deficit 
is $\delta c = 12e^{-\phi_0}(1 + k)$. 

\end{itemize}  

\noindent {\bf Conformal Field Theories corresponding to the asymptotic
solutions (I) and (II)}

The asymptotic solutions, found above to leading order in $\alpha '$,
can become {\it exact solutions} if one manages to 
construct explicitly the corresponding 
conformal fied theories (CFT) on the world shweet.

This has been done in some detail in \cite{aben}.
Below we only describe the main results.
For the static Einstein Universe the corresponding CFT
is a two-dimensional Wess-Zumino model on a $O(3)$ 
group manifold, with a time coordinate which is a free
world-sheet field. The corresponding central charge 
is:
\begin{equation} 
c=1 + \frac{3{\tilde \kappa}}{{\tilde \kappa} + 1}=4-\frac{3}{{\tilde \kappa}+1}
\label{03wz}
\end{equation}
where ${\tilde \kappa}$ is the Wess-Zumino level parameter 
of the $O(3)$ Kac-Moody algebra. 
The central charge deficit (\ref{deficit}) is in this case:
\begin{equation} 
\delta c = \frac{3}{{\tilde \kappa} +1}=12e^{-\phi_0}k + \dots
\label{dclevel}
\end{equation} 
where $\dots$ denote higher orders. The important point to notice
is that the level parameter ${\tilde \kappa}$ is an {\it integer}
for topological reasons (equivalently, this result
follows from unitarity of the spectrum and modular invariance
of the underlying string theory~\cite{aben}). 
Thus, (\ref{dclevel}) implies that the central charge 
deficit is {\it quantized}. 

The conformal field theory corresponding to the second asymptotic
solution (II) of (\ref{finaleq}), that of a linearly expanding 
Unviverse, can be found most conveniently 
if we go back to the $\sigma$-model frame: $g_{\mu\nu} = e^\Phi g_{\mu\nu}^E$
and the $\sigma$-model coordinate time $X^0$ (\ref{phystime}), in wich the 
dilaton is linear: 
\begin{equation}
\label{lindilsmodelframe}
\Phi = -2e^{-\phi_0/2}X^0 + \phi_0 \equiv -2QX^0 + \phi_0
\end{equation} 
Thus we observe that $Q \equiv e^{-\phi_0/2}$ plays the r\^ole of a 
``charge at infinity'' in similar spirit to the Coulomb-gas 
conformal models~\cite{green}, an analogy prompted by the 
form of the corresponding stress tensor (\ref{stresscosmol}). 
 
The corresponding world-sheet conformnal field theory is again a 
Wess-Zumino model on a group manifold, in which $g_{ij}$ 
is the metric, and $H_{ij\ell} = \nabla_{[i}B_{j\ell]}$ is the 
volume element. The model has again a time coordinate
but with a charge $Q$ at infinity, as we have just mentioned.
The (two-dimensional) Lagrangian of the model is: 
\begin{equation} 
{\cal L}^{(2)} = -(\partial X^0)^2 - QX^0R^{(2)}  + {\cal L}_{WZW}(O(3))
\end{equation} 
where $R^{(2)}$ is the world-sheet curvature. 
The central charge is: $c=1 - 12Q^2 + c_{WZW} $ 
with the level parameter
${\tilde \kappa}$ being related to the 
spatial curvature $k$ as fllows:
\begin{equation} 
k = \frac{1}{2Q^2{\tilde \kappa}}
\label{lpsc}
\end{equation}
Since ${\tilde \kappa} \in Z^{+} U \{0\}$ for topological 
(or, equivalently  unitarity and modular invariance) reasons,
then $k >0$ and the four-dimensional Universe is again closed.
The 4-d curvature is found again to be $R=6(1 + k)/t^2$.

\subsubsection{The spectrum of the Linear-Dilaton Strings: Mass Shifts}

Consider the conformal invariant solution (\ref{ldb}). The corresponding 
Virasoro operators, i.e. the moments of the 
world-sheet stress tensor, as we have discussed in the 
first Lecture, are~\cite{aben}: 
\begin{equation} 
L_n = \frac{1}{2}\sum_{j} \eta_{\mu\nu} x_{n-j}^\mu x_j^\nu + iQ (n+1)x_n^0
\label{virgen}
\end{equation}
where $x_n^\mu$ are moments of the world-sheet operators 
$i\partial_z x^\mu$, satisfying:
\begin{eqnarray} 
&~& \left[x_m^\mu, x_n^\nu \right] = m \eta^{\mu\nu}
\delta_{m+n,0}~,\nonumber \\
&~& {x_n^\mu}^\dagger = x_n^\mu + 2iQ\eta^{\mu 0}\delta_{n,0}~\qquad ({\rm since}~~L_n^\dagger = L_{-n})
\end{eqnarray}
This implies that the $0$-th (time) component of 
Minkowski space-time momentum has a fixed {\it imaginary} part~\cite{aben}
\begin{equation} 
p^0 = E + iQ
\end{equation} 
where the real part $E$ corresponds to ``energy''. 

Consider for definiteness a bosonic scalar mode, e.g. a tachyon,
which is the lowest lying energy state of a Bosonic string 
(ground state): 
\begin{equation} 
|p\rangle = e^{-p_\mu x^\mu (0)}|0\rangle 
\end{equation}
annihilated by all $x_n^\mu $ ($n > 0$). The corresponding 
mass-shell condition is:
\begin{equation} 
\frac{1}{2}p_\mu p^\mu + iQp^0 =-\frac{1}{2}(E^2 + Q^2 - {\vec p}^2)=1
\label{masshell}
\end{equation}
where $\vec A$ denotes three-dimensional vectors. 
Thus, from (\ref{masshell}) one observes that there is a shifted mass 
for the tachyonic mode:
\begin{equation}
\delta m^2_T = m^2 - Q^2 
\label{shift} 
\end{equation} 
in such linear dilaton backgrounds. 

{}From a target-spacetime view point, this can be easily understood 
considering the Lagrangian for 
a scalar mode $\varphi$ in the background
(\ref{ldb}) in the Einstein frame (\ref{einstein}):
\begin{equation}
{\cal L}_\varphi= e^{2Qx^0}\left(-\eta^{\mu\nu} 
\partial_\mu \varphi \partial_\nu \varphi - m^2 \varphi^2\right)
\end{equation}
Indeed, rescaling 
the field $\varphi \to {\tilde \varphi} = e^{Qx^0}\varphi$,
so as to 
have a canonical kinetic term, one obtains a mode
that obeys a free scalar-field wave equation, in flat space time,
with shifted mass (\ref{shift}). 

This result can be extended to include 
all the other bosonic modes~\cite{aben}, including graviton and dilaton. 
All such bosonic modes therefore will have a mass shift of tachyonic type 
in supercritical strings:
\begin{equation} 
\delta m_B^2 = -Q^2 < 0 
\label{bosonshift}
\end{equation} 

For target-space Supersymemtric strings, including the 
phenomenologically relevant Heterotic string~\cite{green}, 
in linear-dilaton backgrounds, 
one observes that there are {\it no mass shifts} for the 
fermionic target space time modes~\cite{aben}. 

This can be readily seen, for instance, by noting first that 
the anomaly condition for superstrings becomes:
\begin{equation}
c_I + d -8Q^2 = 10
\end{equation} 
This is due to the 
additional stress tensor contributions 
on the world-sheet pertaining to fermionic backgrounds
$T_F = -\psi_\mu \partial _z x^\mu + 2Q \partial_z \psi^0$.

The lowest-lying fermionic excitations are massless,
since superstrings do not have tachyonic instabilities. 
Consider for simplicity the case $c_I =0$ and concentrate on the 
lowest-energy Ramond state. 
Consider the moments of $T_F$:
\begin{equation} 
G_n = i \sum_{n} \psi_{n-m}^\mu x_{\mu,n} - 2Q(n+1)\psi_n^0 
\end{equation}
When acting on the highest-weight state, one has:
\begin{equation}
G_0 = -i\left(\gamma_0 E - {\vec \gamma}\cdot {\vec p}\right)
\end{equation}
This is precisely the massless Dirac operator in flat space. 
Thus one observes that there is {\it no mass shift} for the 
fermionic modes.

{}From a field-theoretic view point this can be seen 
from the quadratic part of the target-space Lagrangian 
for fermionic modes $\Psi$,
in the background (\ref{ldb}):
\begin{equation} 
{\cal L}_{\rm fermion} = e^{2Qx^0}\left({\overline \Psi}\partial_\mu 
\gamma^\mu \Psi + m{\overline \Psi}\Psi + \dots \right)
\label{fermlagr}
\end{equation}
It is easily seen that 
the rescaled fermion field ${\tilde \Psi} = e^{Qx^0}\Psi$ 
obeys the free Dirac equation {\it without a mass shift}. 

Thus, in a linear dilaton background, which leads to a
linearly expanding Universe in Einstein frame, 
there is no target-space fermionic-mode mass shift:
\begin{equation} 
\delta m_F^2 =0 
\label{fermionshift}
\end{equation} 

So far, our considerations pertain to tree-level world-sheet
$\sigma$-models, i.e. world-sheets with the topology of a disc (open strings)
or sphere (closed strings). 
String loop corrections do affect the $\beta$-functions of the theory,
and actually they do result in the appearance of 
non-trivial dilaton potentials $\delta V(\Phi)$, whose effects  
we now come to discuss, from the point of view of 
Cosmological backgrounds, which are of interest to us in the context 
of these lectures. 

\subsection{String Loop Corrections and De Sitter (Inflationary) Space Times}

\begin{figure}[htb]
\epsfxsize=4in
\begin{center}
\epsffile{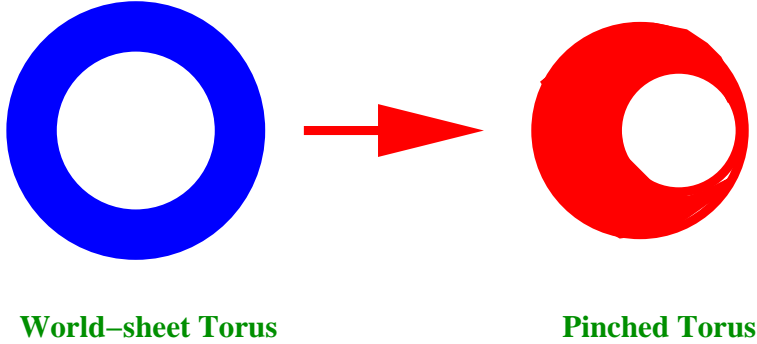}
\end{center} 
\caption[]{Extra world-sheet partition function divergencies 
arising from pinched tori. Regularizing such pinched surfaces modifies
the $\beta$-function of the theory at lower genera, since it 
introduces new types of $\sigma$-model 
counterterms. This is the essence of the 
Fischler-Susskind mechanism.}
\label{fig:pinched}
\end{figure}

The string loop corrections, i.e. effects coming from higher world-sheet
topologies, are non trivial and they do modify the tree-level 
$\beta$-functions of the theory through the so-called
Fischler-Susskind mechanism~\cite{fs}. 
To understand {\it qualitatively} the r\^ole of such effects 
let us consider the indicative example of a $\sigma$-model 
partition function on a world-sheet {\it torus}. 
As one sums up over tori geometries, with handles of variable size,
one encounters {\it extra divergencies}, as compared to the case of 
world-sheet spheres, arising from {\it pinched} tori, as indicated in fig.
\ref{fig:pinched}.

Such infinities (modular) are equivalent to considering tori with
handles of sizes below the ultraviolet cutoff on the world sheet.
Such degenerate higher-genus surfaces cannot be distinguished from 
those of spherical topology. Thus, in a regularization procedure
the effect of the presence of 
these surfaces is to induce new types of counterterms
for the spherical topology regularized $\sigma$-model action,
which result in the string-loop modifications of the $\sigma$-model 
$\beta$-functions. For technical details, the interested reader
is referred to the literature~\cite{fs,green}.

For our purposes here, we note that these string-loop corrections
induce a dilaton potential $\delta V(\Phi)$ 
in the four-dimensional 
string effective action, whose contributions to the 
conformal invariance
conditions (\ref{diffbeta}), for the $\sigma$-model (\ref{smodelcosmol}),   
can be summarized as follows: 
\begin{eqnarray} 
&~& R_{\mu\nu} = R_{\mu\nu}^{\rm old}    
+ \frac{1}{2}g_{\mu\nu}\left[\delta V(\Phi) - \delta V'(\Phi)\right]~, 
\nonumber \\
&~& \delta c = \delta c^{\rm old} - 3e^{-\Phi}\left[2\delta V(\Phi) 
- \delta V'(\Phi) \right]
\label{loopcorr} 
\end{eqnarray} 
where the suffix ``old'' denotes the right-hand-sides of the 
tree-level 
``graviton'' equation in (\ref{eqsgravantis}), and that of the dilaton
equation (\ref{dileq}), and the prime denotes differentiation
with respect to the dilaton field $\Phi$, $\delta V'(\Phi) \equiv 
\frac{\delta }{\delta \Phi}\delta V(\Phi)$.

From (\ref{loopcorr}) we observe again that $\delta c$ is a $c$-number 
(constant), as required  by consistency, for arbitrarty 
dilaton potential $\delta V(\Phi)$. 
In string-loop perturbation theory the dilaton potential
can be computed order by order, and has the generic form:
\begin{equation} 
\delta V(\Phi) =\sum_{n \ge 1} a_n e^{(n+1)\Phi}
\label{looppot}
\end{equation}
where we remind the reader that $g_s=e^{\Phi/2}$ is the string coupling 
constant, which is a 
string-loop counting parameter, as explained in Lecture 1.

An important physical consequence of the presence of a dilaton potential
due to string loop corrections is the possibility of having 
{\it De Sitter} ({\it inflationary}) solutions in string theory,
i.e. solutions  in which the Hubble parameter is constant
in time $H(t)={\rm cons}$. This implies an {\it exponentially expanding
Universe}, with scale factor 
\begin{equation} 
a(t) \sim e^{Ht}
\label{infl}
\end{equation} 
The constancy of $H$ can be achieved by selecting constant values
for the dilaton and axion fields $\phi_0={\rm const}$, $b=b_0={\rm const}$,
and non-trivial values for the 
dilaton potential, induced by string loops:
\begin{eqnarray}
&~& R=12H^2=2\left[\delta V(\Phi)-\delta V'(\Phi)\right]~, \nonumber \\
&~& \delta c = -3\delta V'(\Phi)e^{-\Phi}
\end{eqnarray} 
One should emphasize 
the crucial r\^ole of the constant value of the  
dilaton field in determining {\it both} 
the value of the 
Hubble constant during the inflationary period of the string Universe,
{\it and} 
the string coupling $g_s=e^{\phi_0/2}$. 

The physically interesting issue is how one can {\it exit}
from the inflationary phase in string Universes.
In the simplified background considered above this cannot be 
possible in a smooth continuous way. 
The rest of the lectures will be therefore devoted to a rather
brief, but to the point, discussion of more complicated 
string backgrounds and scenaria that might achieve 
such a {\it graceful exit} from the inflationary period.
We shall also point out some essential problems
that an 
{\it eternal} de Sitter Universe poses for critical string theory in general,
namely for a proper definition of scattering amplitudes
which is an essential feature of any critical string theory.

\subsection{De Sitter Universes and pre-Big Bang scenaria: the crucial r\^ole of the Dilaton Field}

\subsubsection{Life before the Big Bang in string theory?} 

As we have seen above, a non-trivial dilaton field $\Phi$ 
is an important ingredient for providing 
inflationary, and in general expanding, Universes in string theory.
As argued by Veneziano and collaborators~\cite{veneziano},
the presence of a non-trivial dlaton potential 
may  result in scenaria for expanding Universes in which 
{\it there is no initial singularity} (Big Bang), since 
in such cases the ``singularity'' is replaced by a 
(yet not fully known) {\it non-perturbative} strongly-coupled
region of string theory, in which $g_s =e^{\Phi/2} \gg 1$.
This is the so-called {\bf Pre Big-Bang scenario (PBB)} 
of the string Universe, which we now proceed to discuss
in general terms. For details we refer the interested
reader in the relevant literature~\cite{veneziano}.

\begin{figure}[htb]
\epsfxsize=4in
\begin{center}
\epsffile{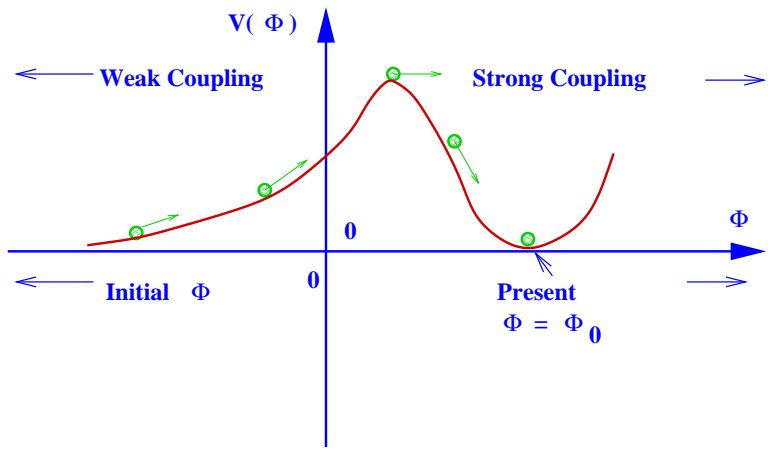}
\end{center} 
\caption[]{A typical dilaton potential encountered in 
pre-Big-Bang scenaria for string Universe. In such scenaria
the initial (Big-Bang) singularity of standard cosmology
is absent, and is replaced by a non-perturbative 
region of string theory, in which the string coupling, being 
given by the exponential of the dilaton field, is very strong.
The arrows indicate flow of cosmic time. The dilaton today 
is at (or close to) the minimum of its potential. 
At present the rigorous derivation of such potentials
from exact string theory models is lacking}
\label{fig:dilpot}
\end{figure}

In PBB scenaria one is typically encountering 
the situation for a dilaton potential depicted in figure \ref{fig:dilpot}. 
In generic PBB models the string 
Universe has a (weak string coupling) ``life'', before 
one reaches the ``big bang'', which is not a singularity, but 
a potential barrier separating 
the weak phase from that at which 
the string theory becomes strongly coupled.
The weakly coupled string-theory (pre Big Bang) region can 
be treated analytically by means of 
Einstein-type low-energy effective actions, of the form 
(\ref{esea}). 
In this region one considers 
homogeneous Bianchi type I solutions of the equations
of motion obtained from the string-effective action~\cite{veneziano}.

\begin{figure}[htb]
\epsfxsize=4in
\begin{center}
\epsffile{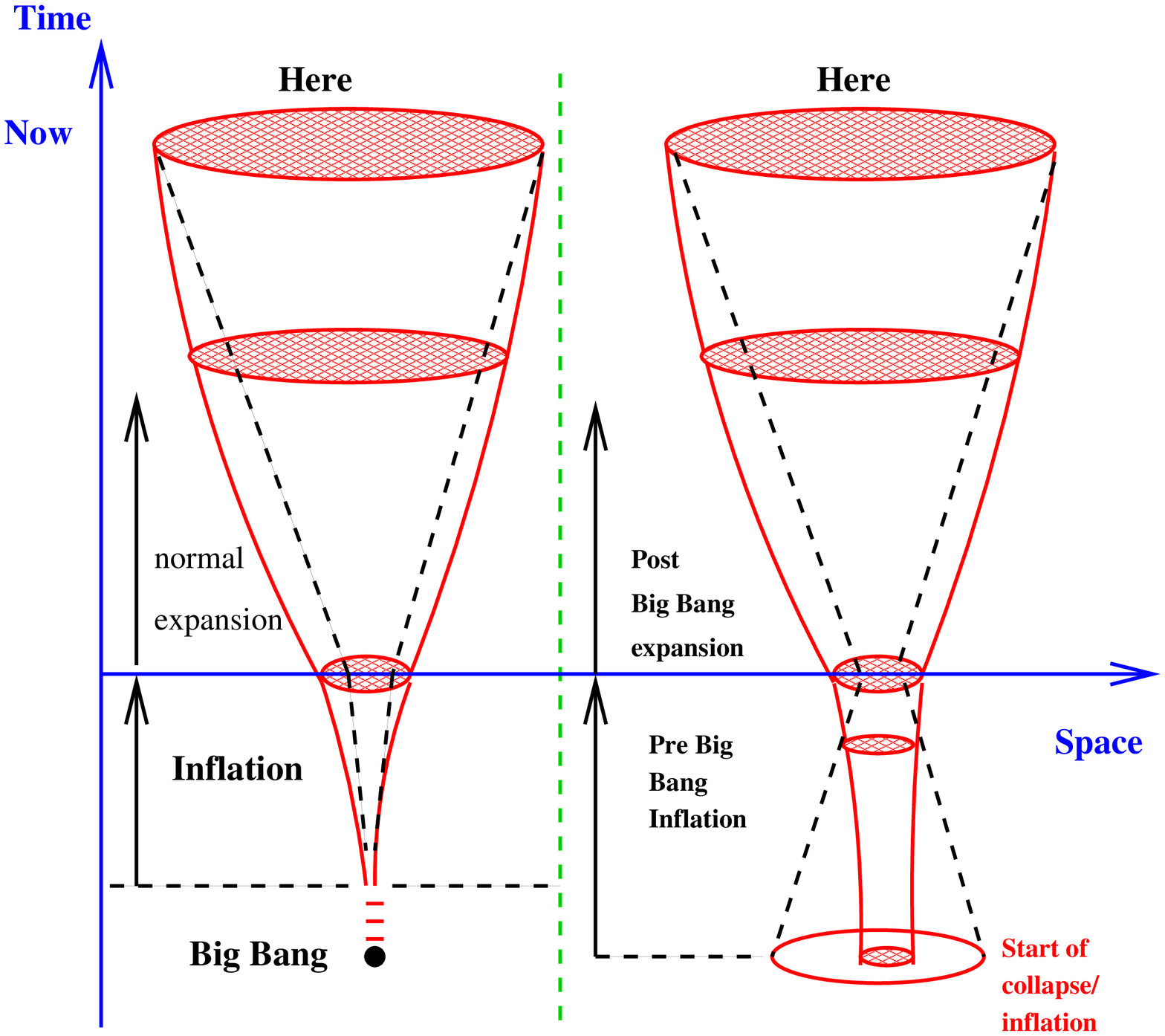}
\end{center} 
\caption[]{A typical space-time ``wine glass'' diagram
for the PBB scenario (right figure), and the 
corresponding diagram for conventional Big Bang 
Cosmology (left figure). The eras of pre big bang life of the Universe
and dilaton driven
inflation, in the PBB scenario, are indicated (original figure 
in ref. \cite{veneziano})} 
\label{fig:wineglass}
\end{figure}

\begin{figure}[htb]
\epsfxsize=3in
\begin{center}
\epsffile{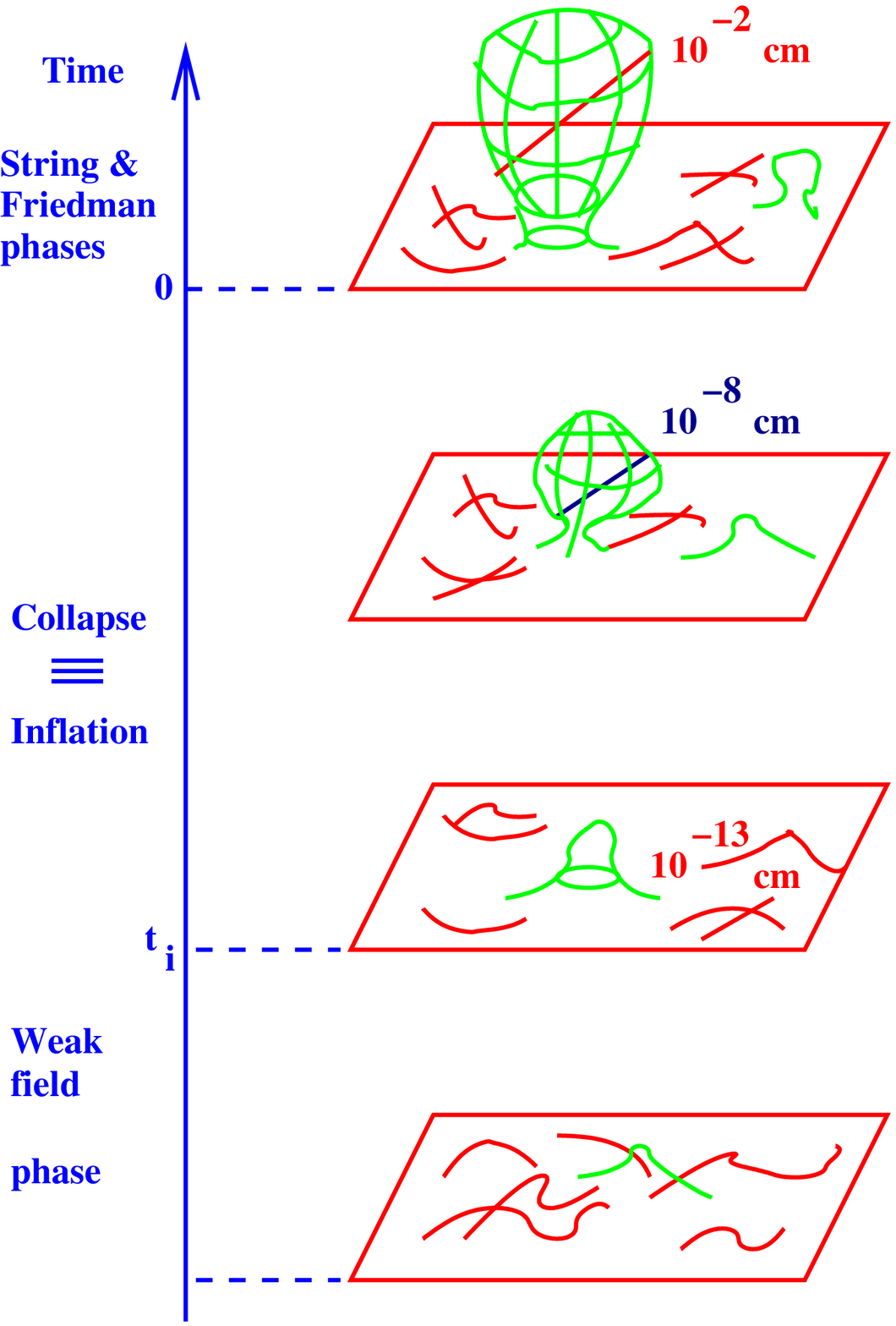}
\end{center} 
\caption[]{A physical representation of the 
PBB dilaton-driven inflationary phase 
of the string Universe. The figure is self explanatory
(original figure in ref. \cite{veneziano}).} 
\label{fig:physicalPBB}
\end{figure}

Let $t=0$ be the ``Big-Bang'' time moment, i.e. the time moment
for which the dilaton potential has its 
maximum height (see figure \ref{fig:dilpot}). 
The pre Big Bang (weakly coupled) solution occurs for $t<0$, and 
has the form~\cite{veneziano}:
\begin{eqnarray} 
&~& ds^2_E= -(dt)^2 + \sum_{i} (-t)^{2a_i}dx^i dx^j \eta_{ij}~, \nonumber \\
&~& \Phi = -(1 - \sum_{i} a_i ){\rm ln}(-t)~, \qquad \sum_{i}a_i^2 =1~, t <0
\label{pbbsol}
\end{eqnarray} 
It is customary~\cite{veneziano} to use a redefined dilaton field, 
{\it shifted} by the logarithm of the determinant of the spatial part
of the metric, 
\begin{equation}
{\overline \Phi} \equiv \Phi - \frac{1}{2}{\rm ln}{\rm det}(g_{ij})=
-{\rm ln}(-t)
\label{shifted} 
\end{equation}
Notice that in PBB scenaria it is the early times regime 
that is characterized by a weakly coupled string theory, 
and dilaton potential which asymptotes
to zero. This has to be contrasted with the situation in ref. \cite{aben},
where it is the late times region which has these features, as we have seen
in the previous subsection. 

Inhomogeneities are introduced in a straightforward manner~\cite{veneziano}:
\begin{eqnarray} 
&~& ds^2_E = -(dt)^2 + \sum_{b} e^b_i(x)e^b_j(x)(-t)^{2a_b(x)}dx^i dx^j~,
\nonumber \\
&~& \Phi = -(1 - \sum_{i} a_i ){\rm ln}(-t)~, \qquad \sum_{i}a_i^2 =1~,~~t <0
\label{inhompbbsol}
\end{eqnarray}

\subsubsection{Stringy Dilaton Driven Inflation in PBB scenaria} 

In a PBB scenario, like the one depicted in 
figure \ref{fig:dilpot}, the dilaton continues to grow (as time evolves)
in such a way that the string coupling $g_s=e^\Phi$ becomes strong
and, hence, perturbative solutions like (\ref{pbbsol}),(\ref{inhompbbsol}) 
are no longer possible.
In strong string coupling situations the resummation of string 
world-sheet genera has to be performed, something which at present
is not feasible. Moreover, many physicists believe that in such 
strong string-coupling situations  even the concept 
of a $\sigma$-model breaks down, and one encounters
a fully non-perturbative stringy situation which
is far from being understood at present. 
It is in this regime that non-perturbative 
concepts like branes, M-theory {\it etc.}, 
are applicable, and one would hope to find 
appropriate dualities which would map the strongly-coupled string theory 
to a dual theory which could be treated perturbatively in an analytic way.
At prsent, despite effort, this issue is still open in our opinion,
and this prevents one from providing analytic arguments
in support of the crossing of the potential ``Big-Bang'' barrier.

However, the lack of analytic treatment does not prevent one from 
making a {\it qualitative} description as to how the situation
is expected to be~\cite{veneziano}. 
After crossing the barrier one expects to have an {\it inflationary} 
phase, driven by the dilaton field, and eventually a graceful exit from it,
so as to reach the present era of our power-law expanding Universe. 
Schematically, the PBB scenario and its post big-bang inflationary phase
is represented by means of ``wine-glass'' space-time diagrams~\cite{veneziano}.
In figure \ref{fig:wineglass}, which is a crude 
reproduction of the original figure
suggested by Veneziano~\cite{veneziano}, the 
PBB scenario for a string Universe,
together
with its post BB evolution, is compared, 
in terms of the corresponding space time diagrams, 
with that of a standard Big-Bang 
Cosmology. A physical picture of what it is envisaged in a PBB situation,
including the dilaton-driven inflation 
is given in figure \ref{fig:physicalPBB}. 
Our Universe starts as a small (Planckian) fluctuation
of the string vacuum, and then turns into a bubble that grows
to enter the post Big-Bang era of normal Friedman expansion
we witness today. The creation of another bubble 
cannot be excluded in such scenaria. This would 
bear similarities to stochastic inflationary scenaria.
At present a rigorous derivation of such a picture
from specific string models is still far from being 
complete, at least in our opinion.

Before closing this subsection it is worthy of pointing out 
that in the Einstein-frame PBB scenaria 
the issue of {\it dilaton-driven inflation} 
becomes equivalent to that of studying {\it gravitational 
collapse}~\cite{veneziano}, 
in the sense of the Einstein-metric spatial volume 
element being shrunk to zero size
at a certain moment, 
as time goes backwards. The reason is simple: in this frame, 
one observes from (\ref{esea}) that the dynamics of the 
problem are those of a minimally coupled scalar field $\Phi$ 
to Einstein gravity. Such a situation is characterized by 
positive pressure, as can be trivially verified, and thus
it cannot lead to inflation. However, at these singularities
the dilaton also blows up, and one can verify that in PBB scenaria
the stringy metric, related to the Einstein one via 
(\ref{einstein}), {\it also blows up} there, leading to 
stringy inflation. Such a situation is depicted in 
figs. \ref{fig:wineglass}, \ref{fig:physicalPBB}. 
For more details
on such issues in the context of PBB scenaria we refer the 
reader to the literature~\cite{veneziano}.

\subsection{Some Phenomenological Implications of String Cosmology}

The string cosmologies we have discussed so far have a far richer 
spectrum of physical excitations, as compared with 
standard cosmologies. The quantum fluctuations of these 
stringy excitations are expected to 
undergo amplification under inflation, which is expected to 
lead to a rich unconventional phenomenlogy, not characterizing the 
case of conventional cosmologies.

In PBB scenaria one can actually show~\cite{veneziano} 
that some ``pump'' fields, a terminology to be defined immediately below,
tend to grow during the PBB inflation
in contrast to the situation encountered in 
standard (conventional, field-theoretic) inflationary scenaria, 
where they tend to 
decay. 

Consider a generic perturbation $\Psi$ 
in the low-energy limit of string theory with action 
(\ref{sea}) in the $\sigma$-model frame 
(e.g. metric, dilaton, axion fluctuation {\it etc}). 
We assume the theory has been appropriately compactified 
to four space-time dimensions. As mentioned previously,
in the context of our generic discussion in this lectures, 
we shall not bother
with explicit details of the internal dimensions. 
The effective action of this
perturbation has the generic form: 
\begin{equation} 
I_{\rm eff, pert} = \int d\eta d^3x s(\eta) \left[ {\Psi '}^2 - (\nabla \Psi)^2\right]
\label{effpert}
\end{equation} 
where $\eta$ is the conformal time, defined by $d\eta = dt/a(t)$, 
with $a(t)$ the scale factor of the Universe (in the 
$\sigma$-model frame), and the prime denotes
differentiation with repsect to $\eta$, $\partial/\partial \eta$. 
The function $s(\eta)$ is a function of the scale factor $a(\eta)$
and other scalar fields (dilaton, moduli-i.e. fields related to the internal dimensions {\it etc}), which characterize the string background under 
consideration. The function $s(\eta)$ is called a ``pump'' field,
since a  $s(\eta) \ne {\rm const}$  couples non-trivially 
to the perturbation $\Psi$ and leads to the production of pairs of quanta
of this perturbation.

The pump fields are crucial in determining the evolution of the 
perturbation. Let $\Psi _{\vec k}$ be a Fourier component 
of such a perturbation. Then one may define: 
${\widehat \Psi}_{\vec k} \equiv s^{1/2}(\eta)\Psi_{\vec k}$, which can be shown to satisfy a Sch\"odinger type equation~\cite{veneziano}:
\begin{equation} 
{\widehat \Psi}_{\vec k}'' + \left[ k^2 -
(s^{1/2})''s^{-1/2}\right]{\widehat \Psi}_{\vec k}=0
\label{pumpevol}
\end{equation} 
where the prime denotes differentiation
w.r.t. the conformal time $\eta$. 

In string cosmology, and in particular PBB scenaria, 
the most interesting perturbations correspond to the following 
pump fields~\cite{veneziano}:
\begin{eqnarray} 
&~& {\rm Gravity~waves,~~dilaton}: \qquad s(\eta)=a^2e^{-\Phi}~, \nonumber \\
&~& {\rm Heterotic~gauge~bosons}: \qquad s(\eta)= e^{-\Phi}~, \nonumber \\
&~& B_{\mu\nu}~{\rm Kalb-Ramond~field,~(axion)}: \qquad s(\eta)=a^{-2}e^{-\Phi}~.
\label{pumpfields}
\end{eqnarray}
where $a$ is the RW scale factor in the $\sigma$-model frame, 
related to the scale factor in the Einstein frame $a_E$ by $a_E=ae^{-\Phi/2}$.
These are found easily by looking at the corresponding terms of the
low-energy string effective action (in these lectures we only exhibited
explicitly the gravitational part of the effective action (\ref{sea}), 
(\ref{esea}) (or (\ref{effaction})), 
and not the gauge and other parts. The interested reader
is referred to the literature for explicit 
forms of such background fields~\cite{green}). For example, 
looking at the axion term in the Einstein frame effective 
action (\ref{effaction}) 
it is immediate to see that the axion $b$ perturbations will have a pump field 
$a^{2}e^\Phi$. On the other hand, when expressed in terms of 
the field strength of the Kalb-Ramond field $B_{MN}$, 
$H_{\mu\nu\rho} = e^{2\Phi} \epsilon_{\mu\nu\rho\lambda} \partial^\lambda b $, 
such axion terms lead to effective action $H$-terms of the form (\ref{esea}),
and therefore to 
the Kalb-Ramond 
pump field
indicated in (\ref{pumpfields}).

After amplification during PBB inflation, such perturbations may lead to 
{\it observable effects}. Below we shall briefly catalogue the 
claimed effects. The interested reader may find more detailed discussion
in the literature~\cite{veneziano}. 

\begin{itemize} 

\item{Tensor Perturbations:} such perturbations 
are associated with gravitational field perturbations, 
and may have effects in the observable cosmic gravitational 
radiation 
background (gravity waves). Such effects are though extremely tiny,
due to the weakness of the itneraction. 
Conventional models of inflation
also have such perturbations, and it will be very difficult
to disentangle the stringy situations from the conventional
ones, as far as tensor perturbations are concerned, even if the 
gravitational radiation is observed.  

\item{Dilaton Perturbations:} since the dilaton plays the 
r\^ole of the {\it inflaton} in string cosmology, 
as it drives string inflation, as discussed above, 
it is the natural source for adiabatic 
scalar perturbations. One would expect it to lead quite naturally
to a {\it quasi} scale invariant Harrison-Zeldovich spectrum
of adiabatic perturbations. This would be desirable in 
explaining the {\it observed} 
cosmic microwave bacground (CMB) anisotropies.
Unfortunately, however, detailed studies in the PBB scenaria~\cite{veneziano}
have revealed that both scalar and tensor perturbations remain exceedingly 
small at large  scales, 
so CMB data cannot be explained by the 
dilaton inflation-amplified perturbations.

\item{Gauge-Field Perturbations:} in standard cosmology there is 
{\it no amplification} of vacuum fluctuations of gauge fields. 
This is due to the fact that the inflaton in such cases makes the 
metric conformally flat, and in such metrics, the gauge fields 
{\it decouple} from geometry
in $D=3+1$ 
dimensions. In contrast, in PBB stringy scenaria, 
the effective gauge coupling, being related proportionally to 
the string coupling $g_s=e^{\Phi/2}$, grows together with the inflated 
space. This is an exclusive feature of stringy models.
In this sense, one would expect~\cite{veneziano} that 
PBB, or in general stringy inflationary scenaria, could provide 
an explanation for the origin of primordial seeds of the 
{\it observed} galactic magnetic fields. 
This, however, still remains a theoretically unsolved problem.
Gauge perturbations interact considerably with the 
hot plasma of the early post big Bang Universe, 
and hence covnerting the primordial seeds into those that may have
existed in the era of galaxy formation is a non-trivial 
and still unresolved task. 

\item{Axion Perturbations:} As we have discussed above, in four space-time
dimensions, the field strength  of the antisymmetric tensor field 
of the $\sigma$-model is related to the axion field $b$:
$H_{\mu\nu\rho} = e^{2\Phi} \epsilon_{\mu\nu\rho\lambda} \partial^\lambda b $. 
It must be stressed that 
the spectrum of the axion field perturbations is very sensitive to the 
cosmological behaviour of the {\it internal } (compactified) dimensions
during the string inflationary era, thereby making axions a window
to extra dimensions. On the other hand, the axion spectrum is flat even
red (tilted towards large scales). 

\end{itemize} 

With these brief comments we finish our discussion on 
the string cosmological scenaria. We only glazed the surface
of a huge subject here, and the interested reader is strongly advised
to seek further details in the literature. As we have seen, there are 
many issues that need further exploration, both theoretical and 
experimental ones. There are important differences from standard
cosmology. However, de Sitter Universes in string theory 
pose serious theoretical challenges as well, which we did not 
discuss so far. This, and ways of incorporating such backgrounds 
in a mathematically consistent string-theory 
framework, will be the (speculative) 
topic of the third Lecture, which we now turn to.

\section{Lecture 3: Challenges in String Cosmology and Speculations on 
their Treatment}

\subsection{Exit from Inflationary Phase: a theoretical challenge 
for String Theory}

An important, and still unresolved issue, in stringy inflationary cosmology
is the {\it graceful exit} from the De Sitter (inflation) phase. As we 
have seen previously, 
an important ingredient for 
inflation is the existence of a dilaton potential, which in
critical (conformal) string theories is absent at tree world-sheet
level, and can only be generated by resumming string loops (higher
genera).
In PBB scenaria, during the inflationary period   
one is dealing with a 
strongly coupled phase of string theory, and hence analytic arguments
on such a resummation 
cannot be provided. Exit mechanisms have been proposed though 
at a qualitative level by many groups~\cite{veneziano,exitB,nonlocal,tunel}; 
some of them involve 
non-local dilaton potentials~\cite{nonlocal}, 
which however lack a good motivation
within the framework of string theory; others impliment
exit via quantum tunnelling~\cite{tunel} through the dilaton 
potential barrier (see figure \ref{fig:dilpot}), 
which exploits the associated Wheeler-de-Witt
equation, without modification of the low-energy string effective actions.
Unfortunately, although in such tunnelling scenaria the 
quantum probability of a classically forbidden exit turns out
to be suppressed by $e^{-\Phi}$ factor, 
and {\it a priori} looks to be a promising scenario, 
however this suppression 
exist throughout the three-space, and thus in such scenaria only 
{\it tiny} regions have a reasonable chance of tunnelling. 

Exit from inflationary phase is 
a generic challenge for critical string theory,
not only of PBB scenaria. This problem becomes even more serious today,
where there seems to be experimental evidence~\cite{evidence} 
(from high-redshift supernovae Ia data, supported by complementary 
observations of CMB data~\cite{cmb}) that 
our Universe today is in an {\it accelerating phase}, ${\ddot a}(t) > 0$, 
which, within the
Friedman cosmological model, implies 
also a non-trivial positive cosmological constant $\Lambda > 0$.
In fact there is evidence that $70\%$ of the total available
energy density is {\it dark energy component}, not matter, which could 
be an honest cosmological constant, or, even, a relaxing to zero 
time-dependent energy component of a quintessence field~\cite{carroll}. 
This may be said differently: our Universe is still in a de Sitter
phase, which if true may imply {\it eternal acceleration}, given that
in such a phase, with a non-zero positive cosmological constant,
eventually the vacuum energy due to $\Lambda$ will 
become dominant over the matter, whose density 
decays with the scale factor as
$a(t)^{-3}$. In such a vacuum-dominated Universe,  
the scale factor of a Friedman model varies exponentially  
with the Robertson-Walker time $t$:  
\begin{equation}
       a(t) \sim e^{\sqrt{\frac{8\pi G_N}{3}\Lambda}t}~,\qquad 
\Lambda ={\rm const} > 0 
\label{accel}
\end{equation}
Such eternally accelerating Universes are plagued by the presence
of finite {\it cosmic horizons} $\delta_H$:
\begin{equation} 
   \delta_H \propto \int_{t_0}^\infty \frac{dt}{a(t)} < \infty 
\label{horizon} 
\end{equation} 
If the Universe does not exit from the inflationary (de Sitter)  
phase, then the inevitable existence of horizons 
will imply the {\it impossibility} of defining properly 
{\it asymptotic states} , and hence a scattering matrix $S=e^{-iH t}$,
where $H$ the Hamiltonian operator of the Universe. 

The situation is somewhat {\it analogous} to that of having 
space-time boundaries, e.g. due to the existence of 
microscopic black hole fluctuations in certain scenaria 
of quantum gravity. There is ``information'' loss in such a situation
for asymptotic observers, due to modes crossing these boundaries,
which implies that an asymptotic observer {\it cannot} define pure 
quantum states $| \psi \rangle$, but {\it only} {\it mixed} 
states, defined by a density matrix 
$\rho = {\rm Tr}_M |\psi \rangle \langle \psi |$, obtained by tracing over unobserved degrees of freedom $\{ M \}$.

\begin{figure}[htb]
\epsfxsize=3in
\begin{center}
\epsffile{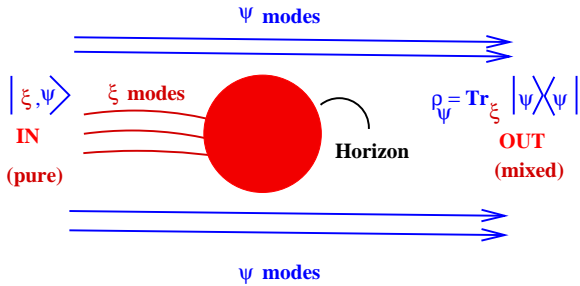}
\end{center} 
\caption[]{Evolution of pure quantum mechanical 
states to mixed ones in the presence of a space-time boundary.
An asymptotic future observer $O'$ has to trace over 
modes $\xi$ that cross the horizon, and hence are unobserved by him/her.
The situation is common for both local(black hole) and global (cosmic)
horizon boundaries.}
\label{fig:mixed}
\end{figure}

We recall that in a unitary quantum theory, 
the S-matrix connects asymptotic {\it in} states to asymptotic 
{\it out} states:
\begin{equation} 
      |out \rangle = S~| in \rangle 
\label{smatrix} 
\end{equation}
On the other hand, if one encouters mixed states, as is the case of 
{\it open quantum mechanical systems}, non-equilibrium systems,
or systems with space time boundaries, such as gravitational 
theories with local or global (cosmological) horizons, 
then the concepts of ``in'' and ``out states'' should
be  replaced 
by those of ``in'' and ``out density matrices'', given that pure 
states evolve to mixed ones, 
as depicted in figure \ref{fig:mixed}.

In such a case, as suggested first by Hawking~\cite{hawk},
one can still link the in and out density matrices
using not the $S$-matrix, but another object,
called  $\nd{S}$ matrix:
\begin{equation} 
   \rho_{\rm out} = \nd{S} \rho_{\rm in} 
\label{dollar}
\end{equation} 
The operator $\nd{S}$ factorizes into a product $S S^\dagger$ 
{\it only} in pure state quantum mechanics without unobserved degrees 
of freedom, in which 
$\rho = |\psi \rangle \langle \psi |$. In general, however, 
once there are unobserved degrees of freedm, thereby opening up
the system, as is the case of horizons (local or global) (see figure
\ref{fig:mixed}), the factorization property of \nd{S} {\it is lost}:
\begin{equation} 
           \nd{S} \ne S~S^\dagger
\label{nonfactor}
\end{equation}
In such systems one cannot define on-shell scattering amplitudes. 

This is a serious theoretical {\it challenge for string theory}. 
As we have discussed in Lecture 1, string theory is 
{\it by construction} a theory of on-shell $S$-matrix, 
based on scattering 
amplitudes which are reproduced by the appropriate
conformal invariance conditions. Thus, de-Sitter
Universes (with eternal acceleration) pose a challenge which 
needs to be resolved~\cite{challenge,emnaccel}. 

In the remaining of these lectures I shall present some 
speculations as to how this problem can be tackled.
A straightforward possibility would be to demonstrate 
the existence of string backgrounds which allow
graceful exit both from the de-Sitter as well as the accelerating 
phases. Within critical (conformal) string theory 
such a scenario has not yet been achieved. 
However, I will discuss an alternative possibility to
such critical string theory models, known as {\bf Liouville 
Strings}~\cite{ddk},                        ,
where graceful exit may be a realistic possibility.
Such string theories are supposed to be
mathematically consistent attempts to formulate $\sigma$-models
away from their world-sheet renormalization-group fixed (conformal)
points, i.e. $\sigma$-models for which the conditions (\ref{diffbeta})
are {\it not valid}. 
The topic, however, is by no means as well established as
critical strings, and therefore the treatment requires
extreme caution. Nevertheless, as I will try to argue in this 
part of the Lectures, Liouville strings have some nice and 
quite interesting features which certainly support further studies
and are worthy of discussion. 

\subsection{Cosmological Backgrounds in String Theory and 
World-Sheet Renormalization-Group Flow}

We shall introduce the reader into the topic of Liouville 
strings by first elaborating further on time-dependent (cosmological)
backgrounds of the $\sigma$-model theory. 
Consider for definiteness 
a $(d+1)$-dimensional target space time $\sigma$-model,
describing propagation 
of a Bosonic closed string on a 
background consisting of the massless string multiplet
of graviton $g_{MN}({\vec x}, t)$, antisymmetric tensor $B_{MN}({\vec x},t)$
and dilaton $\Phi ({\vec x}, t)$ fields. Here ${\vec x}$ span 
a $d$-dimensional Euclidean {\it space} ($x^i, i=1, \dots d$) 
and $t$ is the time.  Assume the $(d+1)$-dimensional 
$\sigma$-model at {\it its conformal point}, at which the conformal 
invariance conditions (\ref{betaddim}) are satisfied.

Target-space diffeomorphism invariance and the Abelian gauge symmetry 
associated with $B_{MN}$, discussed in Lecture 1, 
can be used to ensure:
\begin{equation} 
       B_{0i}=0~, \qquad G_{00}=-1~, \qquad G_{0i}=0~, \qquad i=1, \dots d
\label{conditions}
\end{equation} 

A $(d+1)$-dimensional string solution can be represented as a 
{\it trajectory} $f^A(t)$ in the space of $d$-dimensional $\sigma$-model 
fields $x^i$, with the time $t$ being 
a {\it parameter} along the trajectory:
\begin{equation} 
f^A = \{ g_{ij}({\vec x}(t)), B_{ij}({\vec x}(t)), \Phi ({\vec x}(t))\}~, i,j =1, \dots d
\label{traject}
\end{equation}
[{\bf NB5}: {\it in this section we shall use the notation $M,N, \dots $ for $(d+1)$-dimensional spacetimes, and $i,j \dots$ for $d$-dimensional space} ]

The set of fields $f^A$ can be viewed as couplings of a $\sigma$-model 
in $d$-dimensional target {\it space}~\cite{tseytl} 
\begin{equation} 
{\cal S}_\sigma = \frac{1}{4\pi \alpha '}\int d^2\sigma 
\left[ \sqrt{\gamma} g_{ij}(\vec x)\partial_\alpha x^i \partial^\alpha x^j
+ i \epsilon^{\alpha\beta}B_{ij}(\vec x)\partial_\alpha x^i \partial_\beta x^j
+ \alpha ' \sqrt{\gamma} R^{(2)} \Phi (\vec x) \right]
\label{straj}
\end{equation} 
As we shall discuss now, the orbits $f^A$ resemble standard 
world sheet renormalization group (RG) trajectories in the space 
of couplings $g^i$ of the two-dimensional theory (\ref{straj}).
This is a very important feature which goes beyond a simple analogy~\cite{emn},
as we shall discuss later on in the lectures.

For the moment we note that the string theory {\ref{straj}) 
lives necessarily in a {\it non-critical} dimension, since the 
$(d+1)$-dimensional theory has been assumed {\it critical}. 
Thus, the couplings of (\ref{straj}) lie away from 
their fixed point, and hence must have non-trivial 
RG flows (and therfore non-trivial Weyl anomaly coefficients). 
Their flows are to be identified with the flow in the real time $t$,
as we shall discuss now~\cite{emn}. 

Consider the ${\cal O}(\alpha ')$ $\widehat \beta$-functions
of the $d$-dimensional $\sigma$-model theory:
\begin{eqnarray} 
{\widehat \beta}_{ij}^{g(d)} &= &\alpha ' \left( R_{ij} + 2 \nabla_i \partial_j \Phi - 
\frac{1}{4} H_{imn}H_{j}^{mn} 
\right)~, \nonumber \\
{\widehat \beta}_{ij}^{B(d)} &=& \alpha ' \left( -\frac{1}{2}\nabla_m H^m_{ij} 
+ H^m_{ij}\partial_m \Phi \right)~, \nonumber \\
{\tilde {\widehat \beta}}^{\Phi(d)}  &=& \beta^{\Phi(d)} - \frac{1}{4}
g^{ij}{\widehat \beta}^{g(d)}_{ij} = \frac{1}{6}\left[ c^{(d)}(\vec x) -26 \right]~, \nonumber \\
 c^{(d)}(\vec x) &=& d - \frac{3\alpha '}{2}\left[R - \frac{1}{12}H^2 
- 4(\nabla \Phi)^2 + 4 \nabla ^2 \Phi \right]
\label{betaddim}
\end{eqnarray} 
Above the superscript $(d)$ denotes $d$-dimensional quantities, 
and $c^{(d)}(\vec x)$ is the Zamolodchikov running central charge~\cite{zam}
of the {\it non-critical} theiory (\ref{straj}). This determines the 
$d$-dimensional  target-space effective action~\cite{mms,osborn}:
\begin{equation} 
I_{\rm eff,off-shell}^{(d)} = \int d^d x \sqrt{g}e^{-2\Phi (\vec x)}
\left[c^{(d)}(\vec x) - 26 \right]
\label{offshellaction}
\end{equation}
The {\it off-shell} variations of (\ref{offshellaction}) yield 
the $\widehat \beta$-functions (\ref{betaddim}).
It must be stressed that the non-criticality of the $d$-dimensional 
$\sigma$-model (\ref{straj}) implies that the ${\widehat \beta}$-functions
in (\ref{betaddim}) are non-vanishing (so the conformal invariance conditions
(\ref{diffbeta}) are not satisfied for the $d$-dimensional theory).

Let us now consider the corresponding $(d+1)$-dimensional 
${\widehat \beta}^i$, which should be 
set to {\it zero}, on account of the 
criticality assumption of the $(d+1)$-dimensional theory.
Let us then split the equations into temporal and 
spatial ($d$-dimensional) parts. The result is facilitated
if one uses a shifted dilaton: 
\begin{equation} 
\varphi \equiv 2\Phi - {\rm ln}\sqrt{g}
\label{shifteddil}
\end{equation}
Then we have~\cite{tseytl}:
\begin{eqnarray} 
0 &=& {\widehat \beta}_{00}^{g(d+1)} = 
2{\ddot \varphi} - \frac{1}{2}g^{ik}G^{j\ell} \left({\dot g}_{ij} {\dot g}_{k\ell} + {\dot B}_{ij}{\dot B}_{k\ell}\right)~,
\nonumber \\
0 &=& {\widehat \beta}_{ij}^{g(d+1)} = {\widehat \beta}^{g(d)}_{ij} - g^{00}\left[{\ddot g}_{ij} - {\dot \varphi}{\dot g}_{ij} - g^{mn}\left({\dot g}_{im}{\dot g}_{jn} - {\dot B}_{im}{\dot B}_{jn} \right)\right]~, \nonumber \\
0 &=& {\widehat \beta}_{ij}^{B(d+1)} = {\widehat \beta}_{ij}^{B(d)} 
- g^{00}\left({\ddot B}_{ij} - {\dot \varphi }{\dot B}_{ij} - 2
g^{k\ell}{\dot g}_{k[i}{\dot B}_{j]\ell}\right)~, \nonumber \\
0 &=& c^{(d+1)} -26 = c^{(d)} -25 - 3g^{00}\left({\ddot \varphi }
-({\dot \varphi })^2\right)
\label{d+1eq}
\end{eqnarray} 
where the superscript $(d+1)$ denotes critical dimension $d+1=26$
Bosonic string quantities, and the indices $0$ denote temporal compontents. 

It is straightforward to 
show that these equations are derived from the action~\cite{tseytl}:
\begin{eqnarray}
&~& I_{\rm eff}^{(d+1)} = 
\int dt d^dx \sqrt{|g^{00}|}e^{-\varphi}\left(
c^{(d)}(\vec x) - 25 + \right. \nonumber \\
&~& \left. 3g^{00}\left[{{\dot \varphi}}^2 
-\frac{1}{4}g^{ik}g^{j\ell}({\dot g}_{ij}{\dot g}_{k\ell} + 
{\dot B}_{ij}{\dot B}_{k\ell}) \right]\right)
\label{d+1action}
\end{eqnarray}
In adition, one also should satisfy the 
conditions (\ref{conditions}), which imply the constraints:
\begin{eqnarray} 
0 &=& {\beta}_{0i}^{g(d+1)} = \nabla_k \left(g^{k\ell}
{\dot g}_{\ell i}\right)-\frac{1}{2}{\dot B}_{k\ell}H_{i}^{k\ell}
+ 2 \partial_i {\dot \varphi } - g^{k\ell}{\dot g}_{\ell i}\partial_k \varphi~,
\nonumber \\
0 &=& {\beta}_{0i}^{B(d+1)} = -g_{ik}\partial_j \left(g^{k\ell}
g^{jn}{\dot B}_{n\ell}\right) + 2{\dot B}_{ij}\partial^j \varphi
\label{constraints}
\end{eqnarray}
which, to ${\cal O}(\alpha ')$,
can be shown not to have any 
important consequences other than  restricting 
the initial values of the fields and their 
derivatives~\cite{tseytl}. 

To proceed with our cosmological solutions one should define 
quantities integrated over spatial coordinates $\vec x$,
which thus have only a time dependence.
In this spirit we define~\cite{tseytl}: 
\begin{equation} 
\varphi _0 (t) \equiv -{\rm ln}\left(\int d^dx e^{-\varphi ({\vec x}, t)}
\right)
\end{equation}
This allows a splitting of the dilaton field $\varphi ({\vec x},t)$
into $\vec x $-dependent and $\vec x $-independent parts:
\begin{equation}
\varphi ({\vec x}, t) = \varphi _0(t) + {\tilde \varphi}(\vec x, t)
\end{equation}
From (\ref{shifteddil}), then, we have:
\begin{equation}
\varphi_0 (t) = -{\rm ln}\left[\int d^d x 
\sqrt{g}e^{-2\Phi ({\vec x},t)} \right]
\equiv -{\rm ln}V^{(d)} 
\end{equation}
where $V^{(d)}$ is the proper volume of the $d$-dimensional space. 

One can also define the space-aveage of a function $f({\vec x}, \dots )$
as~\cite{tseytl}:
\begin{equation}
\langle\langle f({\vec x}, \dots \rangle\rangle =
\frac{\int d^d x f({\vec x}, \dots) e^{-\varphi ({\vec x}, \dots)}}
{\int d^dx  e^{-\varphi ({\vec x}, \dots)}}
\label{spaceav} 
\end{equation} 
From this we observe that 
\begin{equation} 
{\dot \varphi}_0 (t) \equiv -Q(t) =  \langle\langle {\dot \varphi}({\vec x}, t)\rangle\rangle 
\label{qdef}
\end{equation}
The time-dependent function $Q(t)$ is related to the central charge deficit,
and hence to the $Q$ of the linear dilaton background (\ref{ldb}),
of Lecture 2, as follows: 
At the ``fixed points'' of the $t$-flow: ${\dot g}_{ij}={\dot B}_{ij}=
{\ddot \varphi}={\ddot \Phi}=0 $, it follows that 
$Q=Q_0$=constant, and 
in fact:
\begin{equation}
\varphi _0 (t) = -\frac{1}{2}Q_0~t + {\rm const} 
\end{equation}
which is a linear dilaton background, analogous to that examined
in (\ref{ldb}), corresponding to  
conformal field theory models
with central-charge deficits $Q_0$. 

In general, however, away from the ``fixed-pointas'' of the $t$-flow,
$Q(t)$ is a function of $t$. Integrating the dilaton equation in 
(last of) (\ref{d+1eq}), and taking the space average (\ref{spaceav}), 
we obtain~\cite{tseytl}:
\begin{eqnarray} 
&~& {\dot Q}(t) + Q^2(t) =-\frac{1}{3}g^{00}({\overline c} - 25 )~,
\nonumber \\
&~& {\overline c} = {\overline c}(g,B,\varphi) \equiv \langle\langle
c^{(d)} (\vec x) \rangle\rangle 
= \frac{{\cal I}_{\rm eff,off-shell}^{(d)}}{V^{(d)}} + 26
\label{averagecc}
\end{eqnarray} 
The function ${\overline c}$ plays the r\^ole of the 
`running central charge' of a non-conformal world-sheet 
field theory away from the fixed points, in the presence of 
non-constant dilatons. Notice that in case ${\dot Q}=0$ 
then (\ref{averagecc}) becomes just the definition of the 
central charge deficit appearing in the linear dilaton background
(\ref{ldb}), or in standard non-critical strings~\cite{aben,ddk}.

Using the first of equations (\ref{d+1eq}) one may compute ${\dot Q}$:
\begin{eqnarray} 
&~& {\dot Q}(t) = \langle\langle -{\ddot \varphi} + {\dot \varphi}^2 
- Q^2 \rangle\rangle = \langle\langle ({\dot \varphi} - \langle\langle 
{\dot \varphi} \rangle\rangle )^2 - {\ddot \varphi} \rangle\rangle = 
\nonumber \\
&~& \langle\langle ({\dot \varphi }-\langle\langle {\dot \varphi} 
\rangle\rangle )^2 -\frac{1}{4}g^{ik}g^{j\ell}({\dot g}_{ij}
{\dot g}_{k\ell} + {\dot B}_{ij}{\dot B}_{k\ell})\rangle\rangle 
\end{eqnarray} 

We also notice that the first three of (\ref{d+1eq}) 
can be written in a compact form~\cite{tseytl}: 
\begin{eqnarray}
&~& {\ddot {\vec g}} + Q(t){\dot{\vec g}} =g^{00}{\widehat{\vec \beta}}^{\vec g}
+ {\cal O}({\dot{\vec g}}^2)~, \nonumber \\
&~& {\vec g} = \{ g_{ij}(\vec x)~,~B_{ij}(\vec x)\}~, 
\label{liouveq}
\end{eqnarray}
with
\begin{eqnarray} 
&~& Q^2(t) =-\frac{1}{3}g^{00}\left[{\overline c}(\vec g) -25\right]
+ \frac{1}{4}{\dot {\vec g}}^2~, \nonumber \\
&~& g^{00} =-1 \qquad {\rm if} ~~~ {\overline c}({\vec g}) > 25~, \nonumber \\
&~& g^{00} =+1 \qquad {\rm if} ~~~ {\overline c}({\vec g}) < 25~.
\label{signspace}
\end{eqnarray} 
The equations (\ref{liouveq}) are sufficient to describe the 
theory in {\it the vicinity} of fixed-points (with respect to
the $t$-flow) in the space of couplings $\{ \vec g \}$
of the $\sigma$-model (\ref{straj}). 
Notice the ``{\it friction}'' form of these equations, due to the 
presence of a non-trivial dilaton (\ref{qdef}).

\subsection{Liouville Strings and Time as a world-sheet RG flow parameter}

The similarity of the $t$-flow with the two-dimensional 
renormalization-group flow is more than a mere analogy, and if made~\cite{emn},
it results in some important consequences for the 
underlying physics~\footnote{It should be stressed, though, 
that this is not the interpretation
adopted by the authors of ref. \cite{tseytl}.}. 
In that case, i.e. after identifying the target time $t$ 
with a renormalization-group flow parameter on the world sheet 
of the $\sigma$-model (\ref{straj}), the $t$-dependence of $Q(t)$ 
is identified with the RG scale dependence of the running Zamolodchikov
central charge~\cite{zam} of this two-dimensional non-conformal
theory. 

Notice that the equations (\ref{liouveq}) 
refer to couplings of a {\it non-conformal} $\sigma$-model,
in a $d$-dimensional target space, which however can become 
{\it conformal} in {\it one target-space dimension higher}, i.e. by 
making the trajectory parameter $t$ a fully-fledged quantum field
in the $\sigma$-model. 
In this sense, the equations (\ref{liouveq}) may be thought of as a 
{\it generalization } of the conformal invariance conditions 
${\widehat \beta}^i=0$ (\ref{diffbeta}) of a critical (fixed point) theory.
This is precisely the principle of {\bf Liouville Strings}~\cite{ddk}.

{}From this point of view the equations (\ref{liouveq}) 
stem from the following fact: 
As just said, Liouville theory~\cite{ddk} restores conformal 
invariance of $\sigma$-models which are away from their fixed points,
by coupling them with an extra fully fledged world-sheet
quantum field $\rho (\sigma, \tau)$, the Liouville mode. 
If a vertex deformation $V_i$ is not a conformal (marginal) 
operator of the $\sigma$-model, then the``Liouville-dressed'' operator :
\begin{equation} 
V_i^L \equiv e^{\alpha_i \rho(\sigma,\tau)}V_i
\label{ldop}
\end{equation} 
is a marginal operator, in the two-dimensional renormalization 
group sense. 
The quantity $\alpha_i$ is known as the `gravitational anomalous
dimension'~\cite{ddk}, and it 
satisfies the equation (for ${\overline c} \ge 25$ we are interested in here):
\begin{equation}
\alpha_i (\alpha_i + Q) = \Delta_i~\qquad {\rm no~sum~over~i}
\label{liouvdress}
\end{equation} 
where $Q$ is a `charge at infinity', with $Q^2$ 
denoting the central charge
deficit, and $\Delta_i=h_i -2$ 
is the anomalous dimension of the operator $V_i$,
with $h_i$ its conformal dimension.
We repeat that 
eq. (\ref{liouvdress}) is nothing other than the condition 
that the Liouville dressed operator $V_i^L$ have vanishing anomalous
dimension~\cite{ddk}.  

Consider, now, a Liouville-dressed deformation 
of the $\sigma$-model (\ref{ldop}). The gravitationally-renormalized 
couplings can be read off directly from this expression as:
$g_L^i \equiv g^ie^{\alpha_i \rho }$. Considering the 
second derivative of $g_L^i$ 
with respect to the world-sheet
zero mode of the Liouville field, $\rho_0$, 
and using (\ref{liouvdress}),
one can arrive~\cite{emn} at equations of the form (\ref{liouveq}),
with the overdot denoting differentiation with respect to $\rho_0$.
In such equations 
the ${\cal O}({\dot g}^2)$ terms stem from 
possible $\rho_0$ dependence of $Q$, as in our case.

The Liouville mode $\rho (\sigma,\tau)$ is nothing other than 
a dynamical $\sigma$-model field mode, which appears in the 
sum over geometries of a non-conformal $\sigma$-model through, e.g. 
the conformal gauge fixing (\ref{fiducialflat}).
In a conformal field theory the Liouville mode decouples
from the world-sheet path integral. This is not the case, however, 
in a non-conformal $\sigma$-model, and this is what we demonstrated above
with our simplified example of stringy cosmology. 
In such non-conformal cases, the Liouville mode becomes a fully fledged
$\sigma$-model field in order to {\it restore  the lost 
conformal invariance} of the $\sigma$-model.
From a physical point of view the reader's attention is drawn to
the property (\ref{signspace}) of the central charge deficit
in a Liouville theory. As we have seen above, the $(d+1)$-dimensional 
target-space time (after taking into account the Liouville field
as a time coordinate)
has a Minkowskian signature for {\it supercritical strings}, i.e. 
${\overline c} > 25$~\cite{aben,emn}, and Euclidean signature 
for {\it subcritical strings}, i.e. ${\overline c} < 25$.

In other words, the above-described ``Liouville dressing'' procedure
implies a temporal signature  for the Liouville field, which can thus
be identified with the time $t$, only in the case where the 
central-charge deficit of the non-conformal $\sigma$-model theory
is supercritical~\cite{aben,emn}. 
By construction (\ref{fiducialflat}), the Liouville mode
may be viewed~\cite{emn} 
as a {\it local world-sheet renormalization-group scale},
since it enters the expression  of a covariant cutoff distance in space,
necessary for regulating ultraviolet divergencies in curved 
space in a way compatible with two-dimensional 
general covariance~\cite{osborn}. The target-time then is nothing other
than the {\it world-sheet zero mode} of the Liouville field~\cite{emn}.

In this interpretation of target time as a world-sheet
renormalization group scale there is hidden an {\it important
property}, which makes the Liouville coordinate different from the 
rest of $\sigma$-model coordinates.  That 
of \textbf{its irreversibility}~\cite{emn}. 
This stems from the fact that a world-sheet RG flow encodes
information loss due to the presence of an ultraviolate
cutoff in the theory, and as such is irreversible. This irreversbility
can be expressed in terms of the irreversibility of the 
flow of the running central 
charge of the non-conformal cut-off theory~\cite{zam} (Zamolodchikov's 
$c$-theorem), ${\dot c} \le 0$ towards a non-trivial infrared fixed point. 
We shall come back to this important point later on. 

Notice that the central charge has been argued to count 
physical target-space 
degrees of freedom in the case of a stringy 
$\sigma$-model~\cite{kutasov}, and hence its decrease along a RG trajectory
is in perfect agreement with the loss of degrees of freedom 
in a cosmological situation with horizons as the time (RG scale)  
evolves. 
It is for this reason that Liouville strings with the 
time identified with a world-sheet RG scale
are viewed as sort of {\it non-equilibrium} string theories,
with the conformal strings corresponding to {\it equilibrium}
points~\cite{emn}. 
What we shall do in the remainder of the lecture, then, is to 
discuss some important physical
features of Liouville strings, such as time-dependent 
vacuum energy for the Liouville Universe, as well as 
the impossibility of defining a proper on-shell scattering matrix
for a Liouville string. 
We shall also revisit the de Sitter string Universes from this point of view,
and present various 
possibilities for a graceful exit from the de Sitter, or
in general, the accelerating phase in the context of string theory. 

\subsection{Liouville String Universe and time-dependent Vacuum Energy} 

The presence of a time dependent central charge deficit 
$Q(t)$ in Liouville strings on cosmological backgrounds, 
with the time identified with the 
world-sheet RG scale~\cite{emn}, 
implies - from the point of view
of the corresponding effective target-space action (\ref{d+1action})-
a time-dependent dilaton pontential, {\it already at tree level world-sheet
topologies}~\cite{emn}:
\begin{equation}
I_{\rm eff}^{(d+1)} \ni \int dt d^dx \sqrt{g(\vec x, t)}e^{-2\Phi ({\vec x},t)}
\left[-g^{00}Q^2(t)\right] 
\label{vacuumenergy}
\end{equation}
One should compare this term with the corresponding term 
(\ref{effaction}) of the model of \cite{aben} 
(after appropriate metric redefinitions to 
go to the Einstein frame). In that case, $\delta c$ came from the 
internal conformal field theory (Wess-Zumino model), 
and this is why it turned out to be constant.  
In contrast, in (\ref{vacuumenergy}), which 
represents a more general situation, the deficit depends on the RG scale $t$,
since the underlying $\sigma$-model theory is considered away from its
fixed point (unlike the situation in \cite{aben}). 

One may construct consistent examples 
of string theories, compactified appropriately to 
four-dimensional cosmological backgrounds~\cite{georgal}, 
in which 
the theory flows to a linear dlaton conformal field theory background
of \cite{aben} 
asymptotically, as $t \to \infty$
(which here plays the r\^ole of the infrared fixed point).
The non-conformality of the original theory is then attributed to 
some sort of fluctuations of the geometry, which result in
departure from equilibrium of the corresponding string theory.

Such non-critical string theories allow for relaxing to zero 
vacuum energies, asymptotically in time. 
Indeed, in the Einstein frame, the respective vacuum energy densities 
have the form~\cite{georgal}: 
\begin{equation} 
\sqrt{g({\vec x}, t)}\Lambda ^E = \sqrt{g({\vec x}, t)}e^{2\Phi ({\vec x}, t)}
Q^2 (t) \to \sqrt{g({\vec x}, t)}\frac{Q_0^2}{t^2}~, \quad t \to \infty~,   
\label{vacenergquint} 
\end{equation}
which is a consequence of the fact that, as $t \to \infty$,
the theory flows to a conformal field theory of ref. \cite{aben},
i.e. $Q^2(t) \to Q_0^2$=constant, and $\Phi \to -{\rm ln}t $
in the Einstein frame, with $t$ the Robertson-Walker time, discussed
previously in the Lectures.  
Such vacuum energies are compatible with recent 
observations~\cite{evidence,cmb}, and in fact there is a 
similarity here with {\it quintessence models}~\cite{carroll},
where the r\^ole of the quintessence field is played by the 
dilaton~\cite{emn,georgal}.

\subsection{No Scattering Matrix for Liouville Strings}

When consider a Liouville string, which as discussed above 
represents a mathematically consistent description of a 
string theory away from its conformal point, the concept of a 
string scattering amplitude breaks down. 
Below, I shall not give a detailed discussion of this important 
issue, but I would rather sketch the main reason behind it 
in a simple way. For details the interested reader is 
referred to the literature~\cite{emn,emnaccel}.

Consider a generic correlation function among $n$ vertex operators
$V_i$ of a Liouville string. In a critical string theory,
this can be associated with appropriate on-shell scattering amplitudes.
In Liouville strings, though, with the target-time identified as 
the Liouville (RG) mode, 
this association cannot be made. Let us see briefly why.
In such a case the correlator reads:
\begin{equation} 
\langle V_{i_1} \dots V_{i_n} \rangle_g = \int {\cal D}\rho 
 {\cal D}X e^{-S^* -g^i\int _\Sigma d^2\sigma V_i + Q^2\partial \rho 
{\overline \partial} \rho - Q^2\rho \int_\Sigma d^2\sigma \rho R^{(2)}}
  V_{i_1} \dots V_{i_n}
\label{liouvcor} 
\end{equation} 
where $\rho$ is the Liouville mode, and $Q^2$ denotes the central charge
deficit, quantifying the departure 
of the non-conformal theory from criticality~\cite{ddk}.

\begin{figure}[htb]
\epsfxsize=3in
\begin{center}
\epsffile{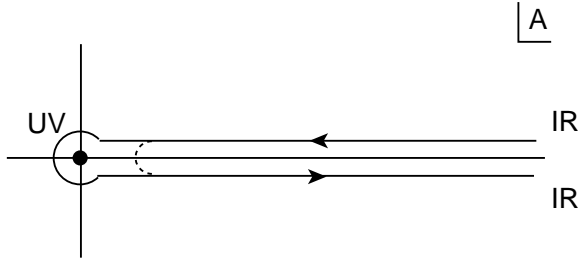}
\end{center} 
\caption[]{Contour
of integration for a proper definition of Liouville field path
integration.
The quantity A denotes the (complex) world-sheet area, which is identified with the logarithm of the Liouville (world-sheet) zero mode.
This is known in the literature as the Saalschutz contour,
and has been used in
conventional quantum field theory to relate dimensional
regularization to the Bogoliubov-Parasiuk-Hepp-Zimmermann
renormalization method. Upon the interpetation of the
Liouville field as target time, this curve
resembles closed-time-paths in non-equilibrium field theories.}
\label{fig:contour}
\end{figure}

A detailed analysis~\cite{emn} shows that, upon  
performing the world-sheet zero-mode $\rho_0$ integration of the Liouville 
mode $\rho$ in (\ref{liouvcor}), one obtains that the dominant contributions
to the path integral can be represented by a steepest-descent
contour of $\rho_0$ as indicated in fig. \ref{fig:contour}.
The interpretation of the Liouville zero mode as the target time
implies a direct analogy of this contour with {\it closed time like paths}
in {\it non-equilibrium} field theories~\cite{ctp}.

When consider infinitesimal Weyl shifts of the world-sheet metric
of the correlators (\ref{liouvcor}), $\delta _w \langle V_{i_1} \dots V_{i_n} \rangle$, then a straightforward but rather tedious world-sheet analysis 
shows that~\cite{emn}:
\begin{equation}
\delta _w \langle V_{i_1} \dots V_{i_n} \rangle \propto 
{\cal O}\left(\frac{s}{A}\right)\langle V_{i_1} \dots V_{i_n} \rangle
+ {\rm A-independent~terms}
\label{weylshifts} 
\end{equation} 
where $s=\sum_{i}\alpha_i/Q$ is the sum of the corresponding 
Liouville anomalous dimensions of the vertex operators $V_i$~\cite{ddk},
and $Q^2$ is the corresponding central charge deficit.
The $\alpha_i$ are defined such that, if $V_i$ is not a conformal (marginal) 
operator, then the ``Liouville-dressed'' operator 
 $V_i^L \equiv e^{\alpha_i \rho(\sigma,\tau)}V_i$ is a marginal 
operator. 

In the scenario of \cite{emn}, the identification of the world-sheet
area (covariant scale) $A$ with $e^{-t}$, where $t$ is the target time, 
implies therefore, on account of (\ref{weylshifts}),   
that these correlators do exhibit time-dependence, and as such 
cannot be associated with on-shell S-matrix elements. 
Such an association can only be made at the infrared fixed point
of the world-sheet flow, $A \to \infty$, where the string reaches
its equilibrium position. It should be mentioned though that 
the definition of the correlators (\ref{liouvcor}) on the 
closed-time-like contour of fig \ref{fig:contour} implies 
that they represent $\nd{S}$ elements, associated with 
density matrices. 

To understand better this last point, it suffices to mention that 
the world sheet partition function $Z$ of a conformal $\sigma$-model,
resummed (in general) over world-sheet topologies, 
is related to the 
wavefunctional $\Psi [g]$ of the underlying string theory:
\begin{equation} 
Z[{\vec g}] \equiv e^{-I_{\rm eff}[{\vec g}]} \leftarrow\rightarrow 
\Psi [{\vec g}]
\label{wavefunctional}
\end{equation} 
where $I_{\rm eff}[{\vec g}]=\int dt d{\vec X}{\cal L}[{\vec g}]$,
with $t$ the time, and ${\vec X}$ spatial coordinates,  
is the target-space effective action
of the backgrounds ${\vec g}$, which is the appropriate Legendre transform 
of the generating functional of connected correlators in target space.

In the non-critical string approach of \cite{emn}, discussed here, 
the time $t$ is nothing other but the world-sheet zero mode 
of the Liouville field $\rho(\sigma, \tau)$. As we have discussed above, 
the proper definition of Liouville correlators necessitates 
an integration of this time variable over the closed-time-like path 
of fig. \ref{fig:contour}. Due to the different sense of the two branches
of this contour, it is then straightforward to see that, upon analytic 
continuation to the target-space Minkowski formalism,  
the middle side of (\ref{wavefunctional}) becomes ``almost'' the product 
of $\Psi \Psi^\dagger $ (with $\Psi (\Psi^\dagger)$ 
associated with, say, the lower (upper) branch of the 
curve of fig. \ref{fig:contour}). 
We say ``almost'', because, as discussed in 
some detail in \cite{emn}, there are world-sheet infinities around the 
turning (ultraviolet) point of the curve ($A \sim 0$), whose
regularization (dashed curve in fig. \ref{fig:contour}) prevents 
such a complete factorisation. In this sense, the world-sheet
Liouville correlation functions are associated with $\nd{S}$-matrix
elements, linking 
density-matrices instead of pure quantum states.

In this respect, one might conjecture~\cite{emnaccel} 
that an eternally accelerating Universe
can be represented by a (non-equilibrium) Liouville
rather than critical string, with the target time
variable being identified with the world-sheet zero mode 
of the Liouville field. This is consistent with the previous discussion
in the beginning of this section, 
on the impossibility of constructing a proper $S$-matrix
in such situation, but rather a $\nd{S}$ matrix, non factorizable
in $S S^\dagger$. 

\subsection{Graceful Exit from Inflation in Liouville Strings} 

In the previous section we have argued on the equivalence
of a Liouville string theory with a non-equilibrium dynamical system,
for which asymptotic states cannot be defined properly. 
From a physical point of view, one of the most interesting applications
are the eternally accelerating Universes, characterized by cosmic 
(global) horizons
beyond which an observer cannot ``see'', and hence the system is open. 

Another intereting possibility, however, can arise in the context 
of non-critical strings, namely that of a {\it graceful} exit
from the de Sitter or in general the accelerating phase. 
Such a possibility has been discussed in detail in \cite{georgal},
in the context of a specific 
cosmological model based on the so-called
type $0$-string theory~\cite{type0}. Such models involve three-dimensional
branes worlds (appropriate stringy domain walls), 
playing the r\^ole of our observable Universe. 
We shall not discuss details here, but outline the main results
of that work. 
Due to the specific choice of a 
background flux field characterizing the 
type $0$ strings~\cite{type0}, 
the internal dimensions freeze out after inflation 
in different sizes in such a way that one dimension 
(along the chosen flux background) freezes out to a much larger
size than the others, thereby implying an effectively
five-dimensional model. 
In such a model the departure from criticality 
is provided by quantum fluctuations of the three-dimensional
brane worlds.

\begin{figure}[htb]
\epsfxsize=3in
\begin{center}
\epsffile{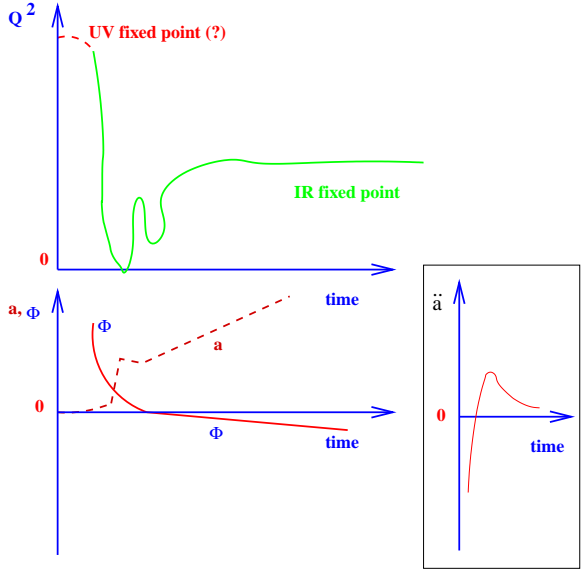}
\end{center} 
\caption[]{The behaviour of the central charge deficit (upper)
and the dilaton (continuous line) and scale factor (dashed line) 
(lower), in the Einstein frame,  
during the various evolutionary phases of the 
cosmological non-critical type-$0$ string theory of ref. \cite{georgal}.
The central charge relaxes asymptotically to a constant value,
when the model asymptotes, for large times, 
to a conformal field theory of 
the type of ref. \cite{aben}, describing a non-accelerating
Universe with a negative (logarithmically divergent) dilaton. 
The diagram inside the box on the right shows the cosmic acceleration
for late Einstein times, indicating the passage from a decelarating 
phase after inflation, to an
accelerating one, with asymptotic exit from it.}
\label{fig:centralcharge}
\end{figure}

The model has an {\it inflationary} (de Sitter type) phase,
characterized by a positive dilaton potential, 
and then a smooth {\it exit from it}. It is crucial, for consistency
of the theory that the central charge deficit, 
quantifying the departure from criticality, 
{\it depends on time}. Immediately after the inflationary period
the Universe enters a {\it decelerating} phase, which is 
succeeded by an {\it accelerating} one~\cite{georgal}. 
The important feature of this model is that, asymptotically, for 
large times, it tends to a {\it non-accelerating} conformal field theory
with a {\it linear dilaton} in the $\sigma$-model frame~\cite{aben}
(or, equivalently logarithmic dilaton in the Einstein frame, depicted
in fig. \ref{fig:centralcharge}). 
Asymptotically, the dilaton potential, which plays the r\^ole of 
an (equilibrium) vacuum energy, relaxes to zero as a
quintessence like field (\ref{vacenergquint}), the r\^ole 
of the quintessence field being provided by the dilaton. 
However, we stress again, here one encounters a non-eternally accelerating 
quintesssence model. 
During such phases the behaviour of the central charge $Q^2(t)$ 
is as indicated in figure \ref{fig:centralcharge} in the 
Einstein frame. 
Notably, due to the Minkowski signature of the target time 
(``non-unitary'' $\sigma$-model field) there is some oscillation
of the central charge before relaxing into its asymptotic 
infrared fixed-point value. 
There is a {\it conformal metastable point} at which momentarily the 
theory becomes critical ($Q^2=0$), and after this there is some oscillatory
behaviour until the theory settles in its final infrared fixed point. 
The existence of the conformal metastable point is a result
of the fact that the theory asymptotes to that of a linear dilaton.
In such a case the dilaton equation forces $Q$ 
to change sign at a certain stage of the evolution~\cite{georgal}. 
Despite the oscillatory behavior,
however, 
there is an {\it overall} decrease of the central charge
as it flows 
from the Gaussian (UV) fixed point value (Big-Bang? Early Universe)
to the infrared one (far future).
Unfortunately, the perturbative ${\cal O}(\alpha ')$ calculations 
of \cite{georgal} (solid line in fig. \ref{fig:centralcharge}) 
cannot give sufficient information
on the value of the UV fixed point (dashed line) at present,
but we conjectured in \cite{georgal} that the initial 
fixed point (constant) value of $Q^2$ 
is also finite, corresponding to a 
given conformal field theory. 

It is interesting to remark that in this model, 
at late stages of the evolution,
the string coupling $g_s=e^\Phi \ll 1$, and thus perturbation theory
applies. This is due to the fact that the dilaton asymptotes to 
$-\infty$ for large times. This situation has to be contrasted
with the pre-Big-Bang scenario~\cite{veneziano} where the 
weak field regime occurs for early (pre-Big-Bang) Universes. 

The ${\cal O}(\alpha ')$ analysis of \cite{georgal} implies
{\it initial singularities} (Big-Bang type), but, as mentioned
already, this may be 
an artifact of the lowest-order truncation. 
Summing up higher orders of $\alpha'$ corrections,
as well as world-sheet topologies, in other words
going to a fully {\it non perturbative string level},  
may indeed
lead to the removal of such singularities. For instance, 
this is known to be the case in some stringy cosmological models
with curvature-squared corrections of ${\cal O}({\alpha '}^2)$,
in the 
string effective action~\cite{art}. The latter effects 
are known to be induced by string loops.

The asymptotic exit from the accelerating phase, and the absence of 
cosmic horizons in the model of \cite{georgal} 
is a very welcome feature
from the point of view of the possibility of defining
asymptotic states~\cite{challenge,emnaccel}, 
and hence a proper $S$-matrix (for this, however,
a resolution of the initial singularities will be desirable, if not 
essential). 
In this respect, our work  is somewhat similar in spirit to 
the arguments of \cite{banks}, where eternal 
quintessence was argued not to occur in perturbative string theory,
which thus was conjectured to exhibit exit from de Sitter phase, 
and have a proper S-matrix, calculated though by purely 
non perturbative
methods. 

The basic argument of \cite{banks},
which however, we stress, should not be considered as a rigorous proof, 
can be summarized as follows: 
in perturbative string cosmology, like 
the case examined in \cite{aben,georgal}, but {\it not} in 
PBB scenaria~\cite{veneziano} (see fig. \ref{fig:dilpot}), 
the dilaton 
potential $V_{\rm dil}$ 
vanishes asymptotically in time, together with the energy 
$E$ of the dilaton field $\Phi$, 
which, in this context, plays the r\^ole of a quintessence
field. 
In the framework of (low-energy) perturbative string-inspired 
Friedmann-Robertson-Walker 
Cosmologies, invovling the (minimal) coupling of the dilaton field  
to gravity, it can be shown that 
the existence of cosmic horizons (\ref{horizon}) 
depends on how fast $V_{\rm dil}$ approaches zero
as compared with $E$. In critical strings, as we have discussed in 
Lecture 2, a non trivial 
dilaton potential is generated through string loops 
via the Fischler-Susskind 
mechanism~\cite{fs} (dilaton tadpoles), and as such 
it is given by infinite sums of the form (\ref{looppot}),
being proportional to various powers of the string
coupling $g_s \sim e^\Phi$.  
In the case of a non-perturbative string potential, then, one expects 
such resummations to exponentiate, and in this case 
$V_{\rm dil} $ would be the exponential function of an exponential 
of the dilaton field $\Phi$. On the other hand, general 
arguments~\cite{banks} can be given in 
support of the fact that in perturbative string theory, i.e. in regimes where
the string coupling is weak, so that $\sigma$-model perturbation 
theory is valid, $E$ has at most a power-law dependence on $g_s$.
Thus, as $\Phi \to -\infty$, one has that $E \gg V_{\rm dil}$ and,
therefore, there will be no {\it cosmic horizon}, in the sense that
the integral (\ref{horizon}) would diverge in the limit $t \to \infty$.

Notice, however, one important difference of the non-critical 
string approach of \cite{georgal} from that of \cite{banks}. 
As just mentioned, in standard critical string theory, 
a positive cosmological constant
in the effective action, as required by the de Sitter phase, 
is obtained through string loops. 
In contrast, as we have discussed in this Lecture, 
the non-criticality of the 
stringy model of \cite{georgal} 
introduces a vacuum energy (dilaton potential) already at 
a tree $\sigma$-model level.

There are many open issues that are left undiscussed in the 
non-critical string approach, 
regarding the phase after inflation, 
such as reheating
{\it etc}. These are open issues for future work.
I must stress though that, although the non-critical 
string approach to cosmology appears promising, and already gave 
physically interesting results, such as the possibility 
of graceful exit from de Sitter (and in general accelerating)
Universe phase, nevertheless it is still very far from 
being considered as well established. 
So far we have treated the departure from criticality 
at a ``phenomenological'' level, by treating the 
time dependence of the central charge deficit 
as being determined by consistency with the rest of the 
Liouville conditions (\ref{liouveq}), which replace the 
conformal invariance (\ref{diffbeta}) 
conditions of the critical strings. 
To be complete one should discuss explicitly the 
internal conformal field theory (pertaining to the extra dimensions),
whose `flow' between fixed points results in the $Q(t)$ 
under consideration. Moreover, from the physical viewpoint one should also
examine the r\^ole of supersymmetric 
target-spaces in cosmological 
scenaria. Note that even in the case of 
type-$0$ strings, with explicitly broken 
supersymmetry, fermionic target-space 
backgrounds do exist, given that the original underlying 
theory is a superstring~\cite{type0}. These issues present important 
theoretical challenges, awaiting further studies, which, in my personal 
opinion, is something 
that should be done.

\section{Conclusions}

In these lectures I have tried to give a brief account of 
interesting cosmological scenaria from the point of view of string theory.
As we have seen, there are amusing possibilities, such as a pre-Big-Bang 
life of the Universe, 
graceful exit from accelerating Universe phases {\it etc}., 
which do not seem to be characterizing conventional cosmological models.

Recent experimental developments in the field of astrophysics,
concerning for instance the possibility for the current era
of the Universe to be an accelerating phase, present
important theoretical challenges for string theory, which 
probably necessitate a fresher look at string cosmology.
One such possibility might be the representation of a cosmological
(time-dependent) background of string theory as a 
non-critical (non-conformal), non-equilibrium situation.
Although speculative,  such a possibility seems, at least to the me,
a mathematically viable one, if the non-conformal 
nature of the background is seen from the point of view
of a renormalization-group flow between fixed (equilibrium) 
points in string theory space.

In this context it should be mentioned that 
there are many explicit 
models one can construct, which exhibit graceful exit from de Sitter,
or, in general, accelerating phases. One of them was presented
in \cite{georgal}, and analysed briefly in these lectures. 
Additional non-critical string models with such exit 
properties 
can be found 
in toy two-dimensional non-critical stringy cosmologies~\cite{grace}, 
where the non-criticality is induced by initial fluctuations
of matter backgrounds. 
Moreover, in higher-dimensional theories, 
one encounters such graceful exit properties 
in cases of intersecting brane
cosmologies. For instance, it can be shown that if 
one represents our universe as a three-brane domain wall,
punctured with D-particles (point-like solitonic defects)~\cite{gravanis},
then recoil of these D-particle during scattering with macroscopic numbers
of closed string states propagating on the brane can also lead to 
space-time back reaction, which 
is sufficient to induce exit from an accelerating 
phase, so that the final equilibrium theory 
will again asymptote to a conformal
field theory of the type of ref. \cite{aben}.

In general, there are many issues in the context of string cosmology 
that remain open, apart from the exit problem. 
Issues like reheating after the inflationary phase, 
the r\^ole of supersymmetry 
in inflationary scenaria {\it etc}, 
are some of them. We have not touched such 
issues here, but we believe that we have presented enough material 
in this admitedly brief and by far not complete exposition, 
which would motivate the interested reader to do further research in 
the exciting directions opened up by string cosmology.

\section*{Acknowledgements}

\noindent I would like to thank the organizers of the First Aegean School
on Cosmology for creating an excellently organized, 
very successful and thought stimulating 
school, in a very pleasant and relaxed atmosphere.

\end{document}